\documentclass[12pt,preprint]{aastex}

 \shorttitle{Catalogue of XPEs Observed by the YOHKOH SXT}
 \shortauthors{Tomczak |& Chmielewska}

\begin{document}


\title{A Catalogue of Solar X-ray Plasma Ejections \\
observed by the Soft X-ray Telescope onboard YOHKOH}


\author{M. Tomczak, \& E. Chmielewska}
\affil{Astronomical Institute, University of Wroc{\l }aw, \\ ul.
Kopernika 11, PL-51-622 Wroc{\l }aw, Poland}
\email{tomczak@astro.uni.wroc.pl; chmielewska@astro.uni.wroc.pl}

%\author{C. D. Biemesderfer\altaffilmark{4,5}}
%\affil{National Optical Astronomy Observatories, Tucson, AZ 85719}
%\email{aastex-help@aas.org}

%\and

%\author{R. J. Hanisch\altaffilmark{5}}
%\affil{Space Telescope Science Institute, Baltimore, MD 21218}


%\altaffiltext{1}{Visiting Astronomer, Cerro Tololo Inter-American Observatory.
%CTIO is operated by AURA, Inc.\ under contract to the National Science
%\altaffiltext{2}{Society of Fellows, Harvard University.}
%\altaffiltext{3}{present address: Center for Astrophysics,
%    60 Garden Street, Cambridge, MA 02138}
%\altaffiltext{4}{Visiting Programmer, Space Telescope Science Institute}
%\altaffiltext{5}{Patron, Alonso's Bar and Grill}


\begin{abstract}
A catalogue of X-ray Plasma Ejections (XPEs) observed by the Soft
X-ray Telescope onboard the YOHKOH satellite has been recently
developed in the Astronomical Institute of the University of Wroc{\l
}aw. The catalogue contains records of 368 events observed in years
1991-2001 including movies and crossreferences to associated events
like flares and Coronal Mass Ejections (CMEs). 163 XPEs from 368 in
the catalogue were not reported until now. A new classification
scheme of XPEs is proposed in which morphology, kinematics, and
recurrence are considered. The relation between individual
subclasses of XPEs and the associated events was investigated. The
results confirm that XPEs are strongly inhomogeneous, responding to
different processes that occur in the solar corona. A subclass of
erupting loop-like XPEs is a promising candidate to be a
high-temperature precursor of CMEs.
\end{abstract}





\section{Introduction}
\label{intro}

X-ray Plasma Ejections (XPEs) are sudden expulsions of hot
magnetized plasma in the solar corona seen in X-rays. They establish
a wide range of macroscopic motions showing different morphology,
kinematics and physical conditions. XPEs occur usually during the
impulsive phase of flares, but their connection with other
solar-activity phenomena like: Coronal Mass Ejections (CMEs),
prominences, radio bursts, coronal dimmings, global waves is also
known. There are some restrictions in calling any motions in the
corona around the flare times as XPEs. The restrictions regard the
size, duration, brightness, speed, etc. and are introduced mainly by
spatial, temporal and spectral resolutions of imaging instruments
and their operational schemes.

XPEs have been systematically observed since 1991 when {\sl Yohkoh}
satellite began to operate. They became commonly known since the
paper written by \citet{shi95} was published. However, we note
earlier articles on essentially the same phenomena from the {\sl
Solar Maximum Mission} \citep{har85} and from {\sl Yohkoh}
\citep{kli94}. Until now images recorded by the {\sl Yohkoh} Soft
X-ray Telescope, SXT \citep{tsu91} are the largest database of XPEs,
even though the newer solar X-ray imaging instruments operate, e.g.,
{\sl GOES} Solar X-ray Imager, {\sl Reuven Ramaty High-Energy Solar
Spectroscopic Imager} ({\sl RHESSI}), {\sl Hinode} X-Ray Telescope.

Detailed analyses of individual XPEs were performed first by
\citet{tsu97} and \citet{o+s97,o+s98}. In these papers the authors
determined values of physical parameters describing an XPE using
temperature and emission measure maps obtained from SXT images. The
maps allowed them to investigate overall magnetic configuration
including a flare loop and a reconnection region. They also used
hard X-ray light curves, derived by the {\sl Yohkoh} Hard X-ray
Telescope, HXT \citep{kos91}, for a detailed description of
reconnection timing.

\citet{n+a99} made the first attempt to correlate XPEs and CMEs. For
17 well-observed limb flares they found that flares associated with
CMEs show XPEs and opposite -- flares not associated with CMEs also
lack XPEs. A more extensive investigation of association between
XPEs and flares was performed by \citet{o+s00}. For 57 well-observed
limb flares they found that almost 70\% show XPEs. They also
reported dependence on X-ray class, namely the association is larger
for stronger flares, but it could be caused by observational biases.

To investigate interesting examples of XPEs other {\sl Yohkoh}
instruments also have been used, namely the HXT \citep{hud01} and
the Bragg Crystal Spectrometer, BCS \citep{tom05}. In both papers a
special location of investigated events has been chosen. These XPEs
occurred far behind the solar limb and due to their fast expansion
they came into the view of an instrument before brighter flares,
which expand slower. It is virtually the only way for using full-Sun
instruments like the BCS to resolve faint soft X-ray emission of
XPEs. The behind-the-limb location also protects against strong
emission of footpoint hard X-ray sources of flares, which usually
dominate fainter coronal emission. The obtained results proved that
an XPE can conatin energetic non-thermal electrons \citep{hud01} and
superhot thermal plasma \citep{tom05}.

An important progress in investigation of XPEs gave a trilogy made
by \citet{kim04,kim05a,kim05b}. They investigated systematically SXT
observations obtained during a two-year interval and found 137 XPEs.
The events were a subject of multipurpose analysis -- the authors
introduced a morphological classification of XPEs, investigated
their kinematics, specified the association with flares and CMEs.
The present name of XPEs also comes from these papers. We
recapitulate the results of Kim et al. in details in further
sections, where we compare them to our results.

More recently, an association between XPEs and radio events and
prominences has been investigated. In statistical surveys
\citet{sha06} studied type II radio bursts, whereas \citet{kol07}
studied drifting pulsating structures (DPS). Both surveys suggest a
kind of connection between XPEs and radio events but further
examinations are needed to establish the connection. The
relationship between hot (XPEs) and cold (prominences) ejections was
discussed by \citet{o+s08,kim09} for single events, which was
followed up with statistical studies \citep{c+t12}.

Finally, in our short review illustrating a research progress we
would like to recall the two following papers. Firstly, the results
of a quantitative analysis of SXT images describing time evolution
of basic physical parameters for 12 XPEs were given by
\citet{t+r07}. Secondly, after extensive analysis of a complex XPE
that consisted of several recurrent episodes, \citet{nis10} reported
a close connection between sequential ejections and successive hard
X-ray bursts.

The most commonly accepted physical explanation of XPEs connects
these phenomena directly with flare magnetic reconnection.
\citet{shi95} regarded XPEs as a proof of the presence of plasmoids
driven by magnetic reconnection occurring above a soft X-ray loop in
short-duration, compact-loop flares similar to the {\sl canonical}
2D CSHKP model \citep[and references therein]{s+c92}, which was
proposed for long-term, two-ribbon flares. In this way,
\citet{shi95} postulated a unification of two observationally
distinct classes of flares, i.e. two-ribbon flares and compact-loop
flares, by a single mechanism of magnetic reconnection called the
{\sl plasmoid-induced-reconnection model}.

The first qualitative studies of individual events
\citep{o+s97,o+s98} reported that the measured velocities of XPEs
are much smaller than the velocity of reconnection outflow expected
from the model to be about the Alfv$\rm{\acute{e}}$n speed. To
reconcile this discrepancy the authors suggested: (1) the high
density of the XPEs, (2) the time evolution effect (i.e., the
plasmoid should be accelerated as it propagates, thus the
investigated XPEs have not yet reached the maximum velocity), or (3)
an interaction with coronal magnetic fields overlying the XPEs.

Although more recently, 2D resistive-MHD numerical simulations of
the reconnection explain kinematical properties of various
observational features attributed to the current-sheet plasmoids
\citep{bar08}, it has been expected that 3D reconnection renders a
more realistic description of eruptive phenomena. For example,
\citet{nit10} suggested 3D quadrupolar reconnection of two loop
systems that appear to exchange their footpoints as a result of
loop-loop interaction \citep{asc99}.

On the other hand, in some cases the XPEs seem to play the same role
as phenomena called precursors of CMEs \citep{che11}. This opinion
is supported observationally by common kinematical evolution of XPEs
and CMEs \citep{gal03,dau06,bak11} as well as their morphological
resemblance \citep{kim05a}. If so, loss of equilibrium or MHD
instability, commonly accepted as one of the CME triggering
mechanism \citep{for00}, also should be taken seriously into
consideration as a cause of XPEs.

Reports concerning XPE observations in the SXT database were
scattered until now across many different sources: refereed
articles, conference communications, electronic bulletins, etc. An
exception was the survey given by \citet{kim05a}, which includes
almost all the XPEs associated with limb flares for a two-years
interval. Our motivation was to ingest all available reports in one
catalogue and organize them in a uniform way for an easy usage. We
have examined SXT images in those time intervals, in which any
systematic searches of XPEs did not perform.

Knowledge about XPEs has so far been shaped by a limited number of
events that have repeatedly appeared in the literature. Our
catalogue is meant to serve as a convenient tool for every scientist
who wants to better understand the nature of XPEs.

\section{Description of the catalogue}
\label{catal}

\subsection{General contents}
\label{cont}

The catalogue contains the all XPEs we know that were observed by
the SXT during the entire {\it Yohkoh} operations, i.e. between 1991
October 1 and 2001 December 14. There are three main surveys of
events that we used in our catalogue:
\begin{enumerate}
\item \citet{kim05a}, which contains 137 limb events, observed between
1999 April and 2001 March.
\item Ohyama (2009, private communication) with 53 limb events that occurred between 1991
October and 1998 August. The survey was prepared for the aim of
statistical research \citep{o+s00}, but it was not published.
\item \citet{chm10}, which reports 113 events, observed basically within two
time intervals: 1998 September -- 1999 March and 2001 April -- 2001
December that were not systematically searched before.
\end{enumerate}
We also incorporated 65 XPEs reported in other scientific papers as
well as in the electronic bulletin {\it YOHKOH SXT Science
Nuggets}\footnote{\url{http://www.lmsal.com/YPOP/Nuggets/}}.

Keeping in mind the examination of SXT images made by different
authors, we can conclude that the list of XPEs associated with limb
flares (defined as $|\lambda| > 60^{\circ}$, where $\lambda$ is
heliographic longitude) is almost complete. On the other hand, the
list of XPEs associated with disk flares is largely incomplete, with
the exception of time intervals examined by \citet{chm10}.
Occasional reports of XPEs not associated with any flares
\citep{kli94} teach us that the SXT images made without any flares
should be also examined and this work still awaits to be done.

\begin{figure}
\includegraphics[angle=-90,scale=.65]{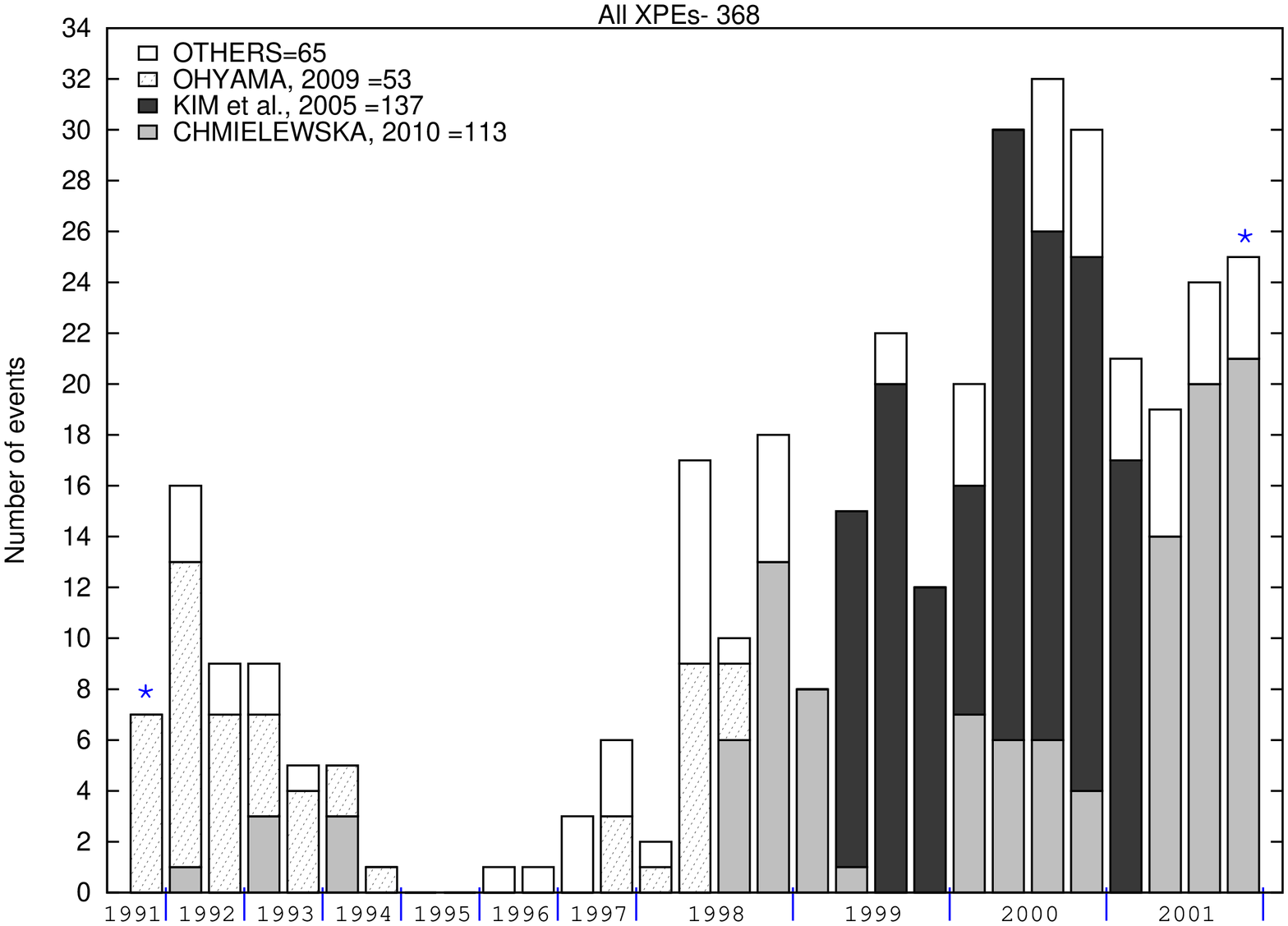}
\caption{Frequency of XPEs for each year during the {\it Yohkoh}
mission. Bin sizes are 6 and 3 months for years 1991-1997 and
1998-2001, respectively. The bins marked with stars refer to partial
years. Each survey mentioned in Section~3.1 is indicated
differently.\label{occ}}
\end{figure}

 In summary, our catalogue contains 368 events. Time
frequency of XPEs occurrence during the {\it Yohkoh} mission is
given in Fig.~\ref{occ}, where sizes of bins are 6 or 3 months for
years 1991-1997 and 1998-2001, respectively. This traces variability
of general solar activity. A larger occurrence rate in cycle 23 in
comparison with cycle 22 may be attributable to the revision of the
SXT flare mode observing sequences. Indeed, the ratio of the number
of XPEs and that of flares taken from the {\sl Solar-Geophysical
Data} (SGD) is 3.6 times greater for solar cycle 23 than for cycle
22.

Before 1997, a routine scheme of observations during the flare mode
was dominated by images in which the exposure time and the position
of the field of view (typically 2.5$\times$2.5 arcmin$^2$) were
automatically adjusted by the signals and locations of the brightest
pixels. XPEs are distinctly fainter and located higher in the corona
than flares. Thus, they had usually too poor statistics during short
exposure times and due to a fast expansion they left immediately the
narrow field of view. Under these circumstances, XPEs were rarely
well-observed.

In 1997, the frequency of images with sufficiently long and constant
exposures and broader field-of-view (5.2 $\times$ 5.2 arcmin$^2$ and
10.5 $\times$ 10.5 arcmin$^2$) was increased to every 10-20 s
\citep{n+a99}. This observational scheme worked more favorably for
the XPEs identification, however flare structures seen in those
images often suffer from heavy saturation that manifests itself as
vertical spikes disturbing a picture of XPEs.

We registered events to the catalogue on the basis of the SXT
observations exclusively. For this reason, we omitted some X-ray
ejections from years 1991-2001 identified using observations made
with other instruments alone, like the HXT, e.g. \citet{hud01}.

\begin{figure}
\includegraphics[scale=.90]{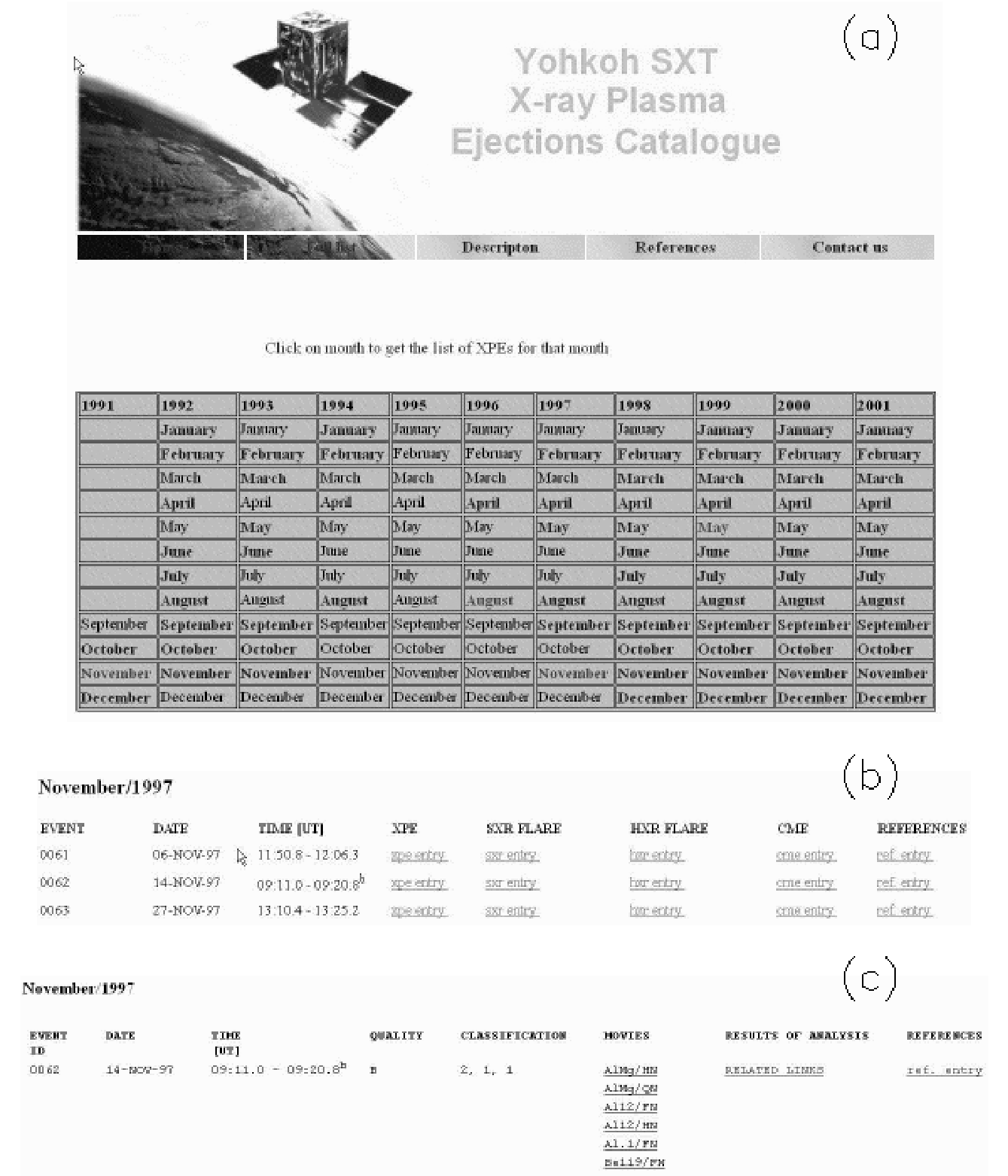}
\caption{Overview of the {\it Yohkoh} SXT XPE Catalogue, which
resides online at http://www.astro.uni.wroc.pl/XPE/catalogue.html.
(a) The main entry into the catalogue as a matrix of years and
months of observations. (b) A few of the entries in the catalogue
for 1997 November. (c) A screenshot for the XPE entry of event
No.~62 on 1997 November 14. \label{mentry}}
\end{figure}

The online catalogue resides at
\url{http://www.astro.uni.wroc.pl/XPE/catalogue.html} since 2010
October 22. It is also linked from the {\sl Yohkoh} Legacy Data
Archive\footnote{\url{http://solar.physics.montana.edu/ylegacy/}}
\citep{tak09}. The general arrangement of the catalogue as a matrix
of years and months of observation is presented in
Fig.~\ref{mentry}a. After clicking the month, each XPE is identified
by a chronological catalogue number, date, and time of occurrence.
The letter (a) added to a start time means that the XPE began
earlier than shown in the available movies. The letter (b) added to
an end time means that the XPE finished later than shown in the
available movies. The letter (c) added to an end time means a time
interval of available movies in which we cannot identify the XPE
reported earlier by other authors. Each events has links to five
entries that provide detailed information on the XPE, flare (SXR and
HXR), CME, and references on the XPE (see an example in
Fig.~\ref{mentry}b).

\subsection{The ``XPE'' entry}
\label{xpe}

This entry contains 8 columns labeled as follows: (1) event ID, (2)
date, (3) time, (4) quality, (5) classification, (6) movies, (7)
results of analysis, and (8) references (see an example in
(Fig.~\ref{mentry}c). The first three columns are replicated from
the higher entry.

In Col.~(4), we indicate the quality of available SXT observations
by assigning one letter between (A) and (D). The letter (A) means
the highest quality: an XPE is clearly seen and only slightly
disturbed by flare saturation, observations have almost full spatial
and time coverage, images are made by at least two different
filters. In conclusion, events with this letter are a good source
for any kind of quantitative analysis including plasma diagnostics
on the basis of the filter-ratio method \citep{har92}. The letter
(B) also means quite good quality of observations, but the usage of
only a one filter in some cases makes a plasma diagnostics
unavailable. Nevertheless, XPEs marked with this letter are always
good for kinematical studies. The letter (C) means poor quality for
some of the following reasons: the brightness of the XPE only
marginally above the background, short observation window,
inadequate field-of-view, or strong effect from flare saturation.
For events with this letter only limited analyses are usually
possible, e.g. a description with our 3-parameters classification.
The letter (D) is designed for XPEs, which were mentioned by other
authors but whose presence is not confirmed in the movies that we
made.

In Col.~(5) we characterize general observational features of XPEs
using a new classification scheme that we have developed in this
catalogue. In our classification we define three criteria
considering: (a) morphology of an XPE, (b) its kinematics, and (c)
recurrence. Examining each criterion we distinguish two subclasses
of events only: (a) 1 -- collimated, 2 -- loop-like; (b) 1 --
confined, 2 -- eruptive; (c) 1 -- single, 2 -- recurrent. In
consequence, our classification can resolve $2^3=8$ subclasses.

Our motivation should be commented in the context of the earlier
classification made by \citet{kim04}, who proposed 5 morphological
groups of XPEs: a loop-type, spray-type, jet-type, confined, and
other. In our opinion, the classification that is too
`hair-splitting' may be uncomfortable in practical usage because it
is easy to make a wrong assignment in case of poor quality of the
observational data or their limited coverage. We may recall attempts
of organizing properties of CMEs as observed by different
coronagraphs \citep{m+s85,how85,b+c93,gop09}. They have never worked
out any commonly accepted classification scheme for CMEs on the
basis of morphological features only.

Our morphological criterion resolves only a direction of soft X-ray
plasma movement in comparison with the direction of local magnetic
field. Roughly speaking, in the case of the subclass 1 the direction
is parallel i.e. along the already existing magnetic field lines, in
the case of the subclass 2 --- perpendicular i.e. across the already
existing lines (or strictly speaking -- together with them). XPEs
from the first morphological subclass usually take a form of a blob
or a column of matter propagating within a bundle of magnetic lines
without any serious modification of their structure. Therefore,
these events are more collimated (hence its name) and less
energetic. A direction of their motions depends on the configuration
of guiding lines. Our subclass 1 comprises the majority of events
classified by \citet{kim04} as the spray-type and jet-type events.
XPEs from the second morphological subclass take a form of a rising
loop or a system of loops. Our subclass 2 is very similar to the
loop-type events proposed by \citet{kim04}. Some events showed
features of both morphological subclasses, 1 and 2, in this case we
classified them according to a more evident feature.

For the XPE assignment into one of the kinematical subclasses we
have chosen height increase rate above the chromosphere, $\dot{h}$.
A negative value, $\dot{h} < 0$, means the subclass 1, the opposite
case, $\dot{h} \ge 0$ means the subclass 2. There are several papers
presenting plots $h(t)$ for events belonging to both subclasses
\citep{kli94,tsu97,o+s97,o+s98,n+a99,kun01,ale02,tom03,
tom04,kim05a,kim05b,o+s08,kim09,nis10}. XPEs from the first
kinematical subclass can be connected with plasma motion within
closed magnetic structures as well as with some changes in a plasma
situation or in the local magnetic field structure which do not
evacuate any mass from the Sun. In summary, XPEs from the first
kinematical subclass suggest the presence of kind of magnetic or
gravitational confinement of X-ray plasma. For XPEs from the second
kinematical subclass, an increasing velocity in the radial direction
in the field of view of the SXT allows us to anticipate further
expansion leading to irreversible changes (eruption) of the local
magnetic field. In consequence, at least a part of the plasma
escapes from the Sun.

For many weak XPEs a construction of the diagram $h$ vs. $t$ was
impossible. Therefore, we estimated $dh/dt$ qualitatively by
watching the expansion rate of XPEs in movies made with images
uniformly spaced in time. In some cases the classification was
problematic because of a limited coverage of available observations,
hence we added a question mark to the digit for this criterion.

According to our third criterion we separate disposable, unique XPEs
that occurred once in a time (subclass 1) from recurrent events for
which following expanding structures can be seen with time (subclass
2). The majority of XPEs described in the literature belongs to the
subclass 1, however samples of the subclass 2 were already presented
\citep{n+a99,tom03,nis10}. A partial time coverage of the available
observations probably introduces a bias toward single XPEs, because
a narrow observational window allows us to resolve only a single
feature even for recurrent XPEs.

Especially important is Col.~(6) in which all available movies
illustrating evolution of the XPE are collected. The movies consist
of images obtained by the SXT and are written in the MPEG format.
Images made by using particular filters and spatial resolutions are
collected in separate movies. We label the movies to indicate their
contents, e.g., AlMg/HN marks images obtained with the AlMg filter
of half resolution. We use the following standard annotations of
filters and spatial resolutions applied in the {\it Yohkoh} software
\citep{mor94}: the filter Al.1 -- the wavelength range 2.5-36 \AA\ ,
AlMg -- 2.4-32 \AA\ , Mg3 -- 2.4-23 \AA\ , Al12 -- 2.4-13 \AA\ ,
Be119 -- 2.3-10 \AA\ ; the full resolution, FN, -- 2.45 arcsec, half
resolution, HN, -- 4.9 arcsec, quarter resolution, QN, -- 9.8
arcsec. A particular resolution means a specific field of view: 2.6
$\times$ 2.6 arcmin$^2$, 5.2 $\times$ 5.2 arcmin$^2$, 10.5 $\times$
10.5 arcmin$^2$, for the FN, HN, and QN resolution, respectively.
Sometimes we divided images made with the same filter and the same
spatial resolution onto separate movies consisting of images made
with the same time exposition. In that case, the labels contain
additionally successive roman digits.

The movies consist of images that we previously processed using the
standard {\it Yohkoh} routine SXT\_PREP, allowing us to reduce an
influence of typical instrumental biases, e.g. telemetry
compression, electronic offset, dark current, straylight,
de-jittering. In the images the heliospheric coordinates are
overwritten by using the SolarSoft routine PLOT\_MAP. For better
identification of a faint features slightly above the background, we
represented a signal distribution with non-linear color tables Nos.
16 (``Haze''), 33 (``Blue-red''), or 3 (``Red temperature'')
available in the Interactive Data Language (IDL). Images that form
movies in the catalogue are sometimes non-uniformly spaced in time,
therefore it is strongly recommended to watch a time print that is
present in each image.

The XPEs for which a more detailed analysis have been already
performed show in Col.~(7) an entry with a concise report concerning
results. Inside the report, the obtained values of investigated
parameters like velocity, acceleration, temperature, emission
measure, electron density, pressure, and secondaries, as well as
references, are given. For 14 events (Nos. 29, 30, 34, 53, 62, 67,
72, 126, 144, 169, 252, 293, 303, and 330) a more complete set of
results is presented in form of plots and tables illustrating the
whole evolution \citep{ron07}.

Finally, in Col.~(8) references to all the reports (also in the
electronic form) in the chronological order are given.

\subsection{Other catalogue entries}
\label{other}

The ``SXR Flare'' entry contains a basic info about a flare that was
associated with the given XPE. Finding the associated flare for the
majority of XPEs in the catalogue was very easy. A flare is seen
usually in movies illustrating evolution of an XPE as heavy
saturation due to its much stronger soft X-ray radiation. In several
cases a flare occurred simultaneously with an XPE but in another
active region. We considered this flare as the associated event only
when some distinct magnetic loops connecting both active regions
were seen. Finally, there are several XPEs that occurred when no
flare was observed on the Sun.

Each record describes the following attributes: date, time of start,
maximum, and end defined on the basis of the {\it Geostationary
Operational Environmental Satellites (GOES)} 1--8 \AA\ light curve,
{\it GOES} class, location in heliographic coordinates, NOAA active
region number. By clicking on the {\it GOES} class, one can view the
{\it GOES} light curves in two wavelength ranges: 1--8 \AA\ (upper)
and 0.5--4 \AA\ (lower). Time span of plots is always two hours and
includes the occurrence of an XPE, which is marked by vertical
lines. The hatched area on the plot represents {\it Yohkoh} nights.

Records presented in this entry are generally adopted from the SGD,
however some clarifications and supplements were necessary. For
example, the lacking locations were completed on the basis of SXT
images as a place of flare bright loop-top kernels. The values
obtained in this way are given in parenthesis. Coordinates of events
that occurred behind the solar limb are taken basically from
\citet{tom09}.

If no flare was associated with an XPE, the tags devoted to flare
characteristics are empty. Exceptions are {\it GOES} light curves,
heliographic coordinates, and NOAA active region number. The last
two tags describe then an XPE.

The ``HXR Flare'' entry presents some attributes of hard X-rays
emitted by a flare that was associated with a given XPE. We used
data from the HXT onboard {\sl Yohkoh}. This telescope measured the
hard X-ray flux in four energy bands: 14-23 (L), 23-33 (M1), 33-53
(M2), and 53-93 keV (H). Each record contains peak time and peak
count rate (together with the background), inferred for the energy
band M1. By clicking on the peak count rate, one can view the HXT
light curves in all energy bands. Time span of plots usually
includes the maximum of hard X-ray flux and the occurrence of an
XPE, which is marked by vertical lines. Wherever available, we also
include the event ID from the {\it Yohkoh} Flare Catalogue
(HXT/SXT/SXS/HXS)\footnote{This catalogue is available as the online
material to \citet{sat06}. It also resides at
http://gedas22.stelab.nagoya-u.ac.jp/HXT/catalogue/.}

If no flare was associated with an XPE, the tags devoted to flare
characteristics are empty. In case the hard X-ray flux in the energy
band M1 was below the doubled value of the background we left the
tags describing peak time and peak time rate empty.

The ``CME'' entry contains some attributes of a CME that was
associated with a given XPE. The observations are derived by the
Large Angle and Spectrometric Coronagraph (LASCO) onboard the {\it
Solar and Heliospheric Observatory (SOHO)}. From the {\it SOHO}
LASCO CME
Catalog\footnote{\url{http://cdaw.gsfc.nasa.gov/CME\_list/}}
\citep{gop09} values of the following parameters are given: date and
time of the first appearance in the C2 coronagraph field of view,
central position angle, angular width, speed from linear fit to the
$h(t)$ measurements, acceleration inferred from the quadratic fit.
The first appearance time is the link to the beginning of the list
of events in the {\it SOHO} LASCO CME Catalog for a given year and
month. By clicking on the entry ``Related links'' one can view a
javascipt movie of the CMEs within the C2 field of view for a given
day. Movies reside at the the homepage of the {\it SOHO} LASCO CME
Catalog.

According to \citet{yas08}, we consider a pair XPE-CME as physically
connected if the XPE occurred within position angles defined by the
CME angular width increased by 10$^{\circ}$ from both sides.
Moreover, time of the XPE occurrence had to fall within
3-hours-interval centered around extrapolated time of the CME start
for $h=1 R_{\sun}$. For extrapolation we used time of the first
appearance in the LASCO/C2 field of view and linear velocity taken
from the {\it SOHO} LASCO CME Catalog. The ``CME'' entry is empty
when the XPE occurred during a LASCO gap. XPEs not associated with a
CME are labeled: 'No related event'.

In ``References'' entry the references to all reports (also in the
electronic form), known for us, that mentioned a particular XPE are
given in the chronological order.

\subsection{Statistics of the catalogue content}
\label{stat}

Some useful characteristics of XPEs included in the catalogue are
extracted in Table~\ref{tb1}. In this subsection we discuss general
statistics of the catalogue content.

In Table\,\ref{tb2} the quality of XPE observations from the
catalogue is summarized. The most frequent are events that we
categorized as (B) and (C). Contribution of remaining categories is
marginal. In Sections\,\ref{ass-fl} and \ref{ass-cme} we present
results of a statistical analysis that was performed for two
different populations of events: from (A) to (C) and from (A) to
(B). The first population is more frequent, which offers some
advantages in statistical approach, however for events categorized
as (C) observations are often not complete enough to give confidence
in our classification choices. In consequence, in the first
population an additional bias can be introduced which is
inadvisable. We expect that this problem is overcome for less
frequent, second population of XPEs which was better observed.

In Table\,\ref{tb3} we summarize heliographic longitudes of XPEs in
the catalogue. A distinct concentration of the XPEs around the solar
limb is seen. This is an artificial effect caused by observational
constrains. XPEs are easier for detection when we observe them
against the dark background sky than when we observe them between
plenty of different features seen on the solar disk. Moreover, in
two from three main surveys that we used in our catalogue
\citep[Ohyama 2009, private communication]{kim05a}, only flares that
occurred close to the solar limb ($|\lambda| > 60^{\circ}$) were
systematically reviewed.

Is it possible to estimate the actual number of XPEs, which occurred
on the Sun during the {\sl Yohkoh} years? Assuming their uniform
distribution with heliographic longitude and taking the number for
the interval $60^{\circ} \le |\lambda| \le 90^{\circ}$ as the most
representative, we obtain a value $6 \times 218 \approx 1300$.
Including a duty time of {\sl Yohkoh} to be about 0.65 (ratio of
satellite day to the total orbital period) we obtain  a number $2
\times 10^3$.

However, even this huge number could be significantly lower than
actual due to several reasons. Firstly, we do not include an
influence of worse detection conditions before 1997. Secondly, for
strong flares, especially in 2001, the conditions for detecting XPEs
were quite bad because of too long exposures. Thirdly, the estimated
number is roughly representative for flares stronger than the {\sl
GOES} class C5-C6. Only for those events the flare mode was
initiated in the {\sl Yohkoh} operation \citep{tsu91} and this mode
guarantees a sufficient time resolution of an image cadence for a
successful detection of XPEs. XPEs associated with weaker flares are
only known accidentally, since no systematic examination of images
recorded during the quiet mode of the {\sl Yohkoh} satellite has
been performed yet.

In conclusion, XPEs should be considered as very frequent events
occurring in the solar corona. XPEs described in the catalogue are
only a minor representation of a countless population of events,
which are typical for the hot solar corona.

In Table\,\ref{tb4} we present a time coverage of the observed XPEs.
Evolution of an important fraction of events (42.6\%) is illustrated
only partially. This limits its detailed investigation. Even
relatively simple activities like classification can be meaningless.
For example, XPEs classified as single and observed only partially
can be actually recurrent.

In Table\,\ref{tb5} we present a number of XPEs that were classified
onto one of eight subclasses defined under following three criteria:
morphological, kinematical, and recurrence. We organize the results
twofold: for the total population and for carefully selected events.
In the second case we omitted events categorized as (C),
untrustworthy assignments of kinematical criterion as shown in the
catalogue with the question mark, and examples classified as single
in case of partial time-coverage of observations. It reduces the
whole population almost three times and for particular subclasses
even more, but we believe that numbers less affected by
observational limits, seen in the last column of Table\,\ref{tb5},
are more representative for real conditions.

What do these numbers tell us about XPEs? At first sight, loop-like
XPEs seem to be more frequent than collimated XPEs of a factor 2.1
or 3.2 for total population and special selection, respectively.
However, it can be caused by the effect of observational selection.
On average, loop-like XPEs are more massive than collimated ones
\citep{t+r07}, thus they are easier for detection above the
background. Indeed, the exclusion of faint events categorized as (C)
increases relative contribution of loop-like XPEs. It is interesting
that in \citet{kim05a} the number of loop-type XPEs (60) only
slightly outnumbers the sum of spray-type and jet-type events (51).

For other relations, the carefully selected events seem to be less
affected by observational constraints than those of the entire
populations. Therefore we conclude that the former events adequately
characterize intrinsic features of particular subclasses. For
example, for loop-like XPEs the relation between eruptive and
confined or between recurrent and single events is distinctly
different from that for collimated XPEs. Loop-like XPEs are
dominantly eruptive (82 to 14) and recurrent (59 to 37), whereas
collimated XPEs are more frequently confined (19 to 11) and single
(20 to 10).

We would like to stress that subclasses of XPEs defined by us
resemble some types of classical prominences observed in H$\alpha$
line \citep{tan95}. Namely, a surge is a prominence that is
collimated and confined, a spray is a prominence that is collimated
and eruptive, a loop-like and confined event we call as an
activation of a prominence, and a loop-like and eruptive event is an
eruptive prominence or 'disparision brisque'. Classifications of
prominences do not distinguish the recurrence criterion,
nevertheless there are known observations, in which aforementioned
types of prominences were observed as single or recurrent events
(Rompolt 2011, private communication). This similarity between XPEs
and prominences suggests a close association between hot and cold
components of active regions. This relation was not investigated in
details so far, except for \citet{o+s08}.

\section{XPEs association with solar flare}
\label{ass-fl}

Using the classification we were able to separate several subclasses
of XPEs looking more homogeneous than the full population.
Unfortunately, we are not sure if particular subclasses of XPEs
refer to events that are physically different. A quantitative
analysis of soft X-ray images would give closer confirmation,
however for the majority of XPEs in the catalogue this kind of
analysis is practically unreliable, due to minor signal and other
observational limits. Therefore, the main motivation of this section
is to justify the presence of physically different subclasses of
XPEs by a comparison of properties of other solar-activity phenomena
associated with particular XPEs. Basic characteristics of flares and
CMEs are known well. In this Section we present the association of
XPEs with flares, in Section~\ref{ass-cme} we present the
association of XPEs with CMEs.

\subsection{Soft X-rays} \label{sxr}

It has been commonly agreed that an XPE is a consequence of a flare
occurrence. As a matter of fact, there are 5 XPEs in our catalogue,
for which we could not find any associated flare. However, this
sample is too small to justify the existence of flareless XPEs.
Moreover, three of the five XPEs occurred close enough to the solar
limb that they might have come from flares from the backside. Are
they just the tip of an iceberg? The answer may depend on extensive
and careful examination of SXT images made in the quiet mode.

\subsubsection{Time coincidence} \label{fl-tc}

\begin{figure}
\includegraphics[angle=-90,scale=.65]{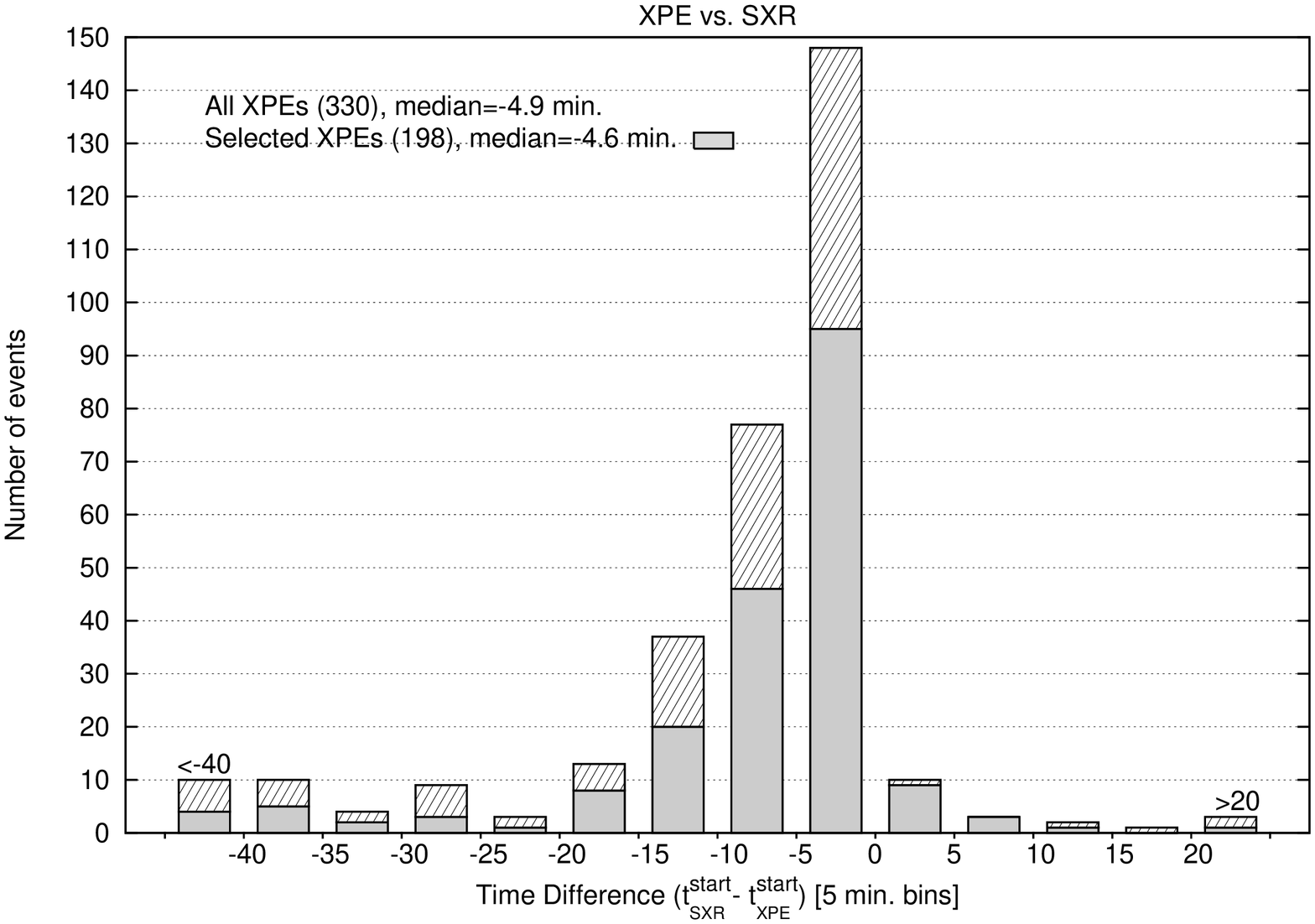}
\caption{Histogram of time differences between the SXR start of the
associated flare (from {\sl GOES} light curve) and the XPE start.
Gray and hatched bins represent the better observed XPEs (quality
A-B, so-called selected XPEs) and the rest of XPEs (quality C),
respectively. The size of bins is 5 minutes with exception of
outermost ones. Numbers of all the considered XPEs and the selected
XPEs as well as their medians are given. \label{flb-xpeb}}
\end{figure}

For better insight in time coincidence between XPEs and flares we
mark the time of an XPE on the {\sl GOES} light curve of an
associated flare. In Fig.~\ref{flb-xpeb} we present a histogram of
time differences between start times of flares and XPEs for 330
pairs of events. In 311 cases from 330 (94.2\%), an increase of soft
X-ray emission occurred earlier than the XPE. The time difference is
very often several minutes only (see maximum and median of the
histogram), however higher values also occur. Similar conclusions
can be given regarding better observed XPEs of quality A and B
(compare the gray bins in Fig.~\ref{flb-xpeb}). Similar histograms
made for particular subclasses of the XPEs introduced in our
classification do not show any important differences.

\begin{figure}
\includegraphics[angle=-90,scale=.65]{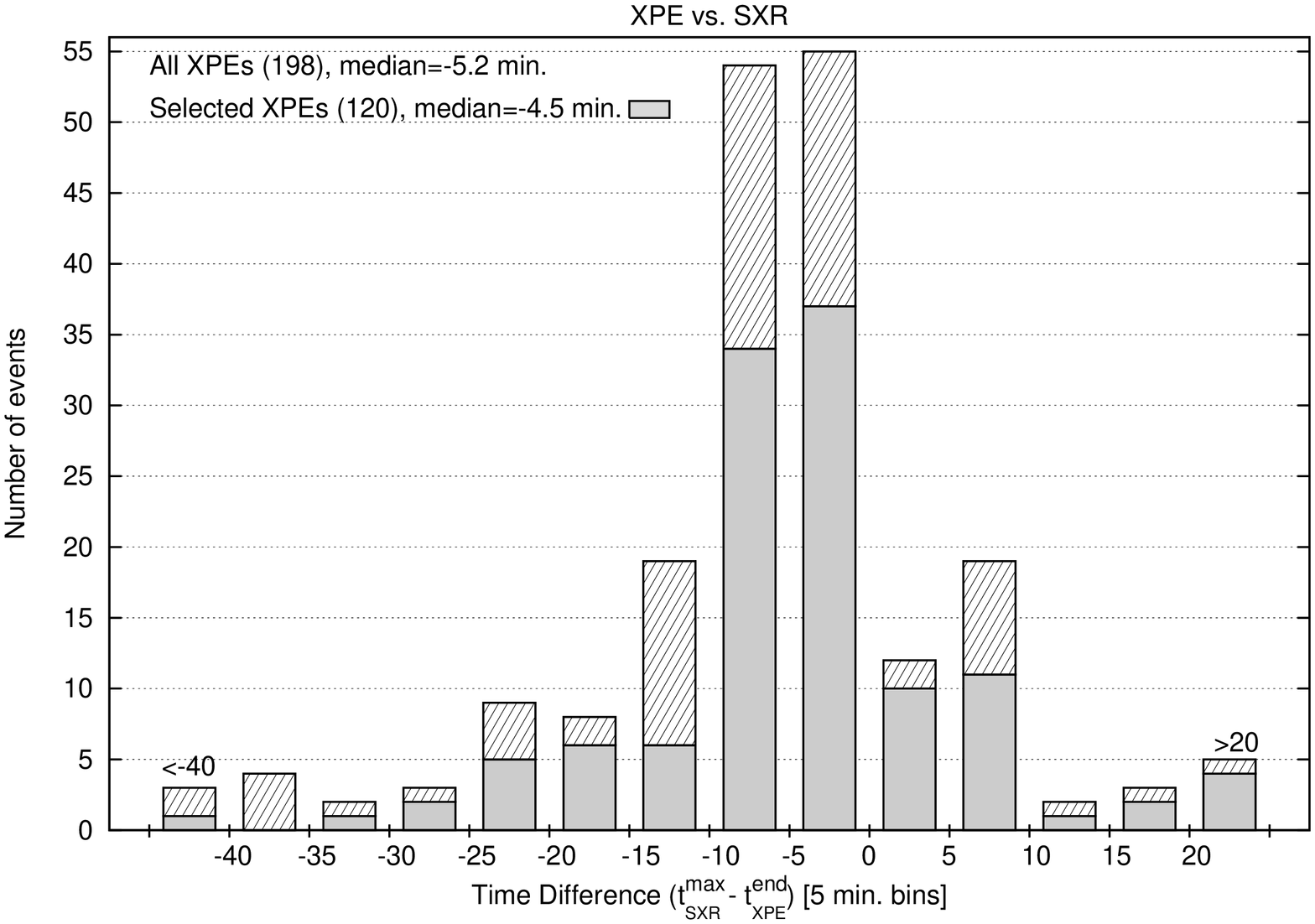}
\caption{Histogram of time differences between the SXR maximum of
the associated flare (from {\sl GOES} light curve) and the XPE end.
Gray and hatched bins represent the better observed XPEs (quality
A-B, so-called selected XPEs) and the rest of XPEs (quality C),
respectively. The size of bins is 5 minutes with exception of
outermost ones. Numbers of all the considered XPEs and the selected
XPEs as well as their medians are given. \label{flm-xpee}}
\end{figure}

In Fig.~\ref{flm-xpee} we present a histogram of time differences
between the end of XPEs and the peak of the associated flares as
determined from the {\sl GOES} light curves. In about 20\% of
investigated samples (41 from 198) any XPEs were completed before
the flare peak, i.e. within the rising phase of a flare. For almost
80\% of events the final evolution of XPEs is seen after the maximum
of soft X-ray emission, very often no longer than 10 minutes (108
examples from 198, 54.5\%).  Similar conclusions can be given
regarding better observed XPEs of quality A and B (compare the gray
bins in Fig.~\ref{flm-xpee}). Similar histograms made for particular
subclasses of XPEs introduced in our classification do not show any
important differences.

\begin{figure}
\includegraphics[angle=-90,scale=.65]{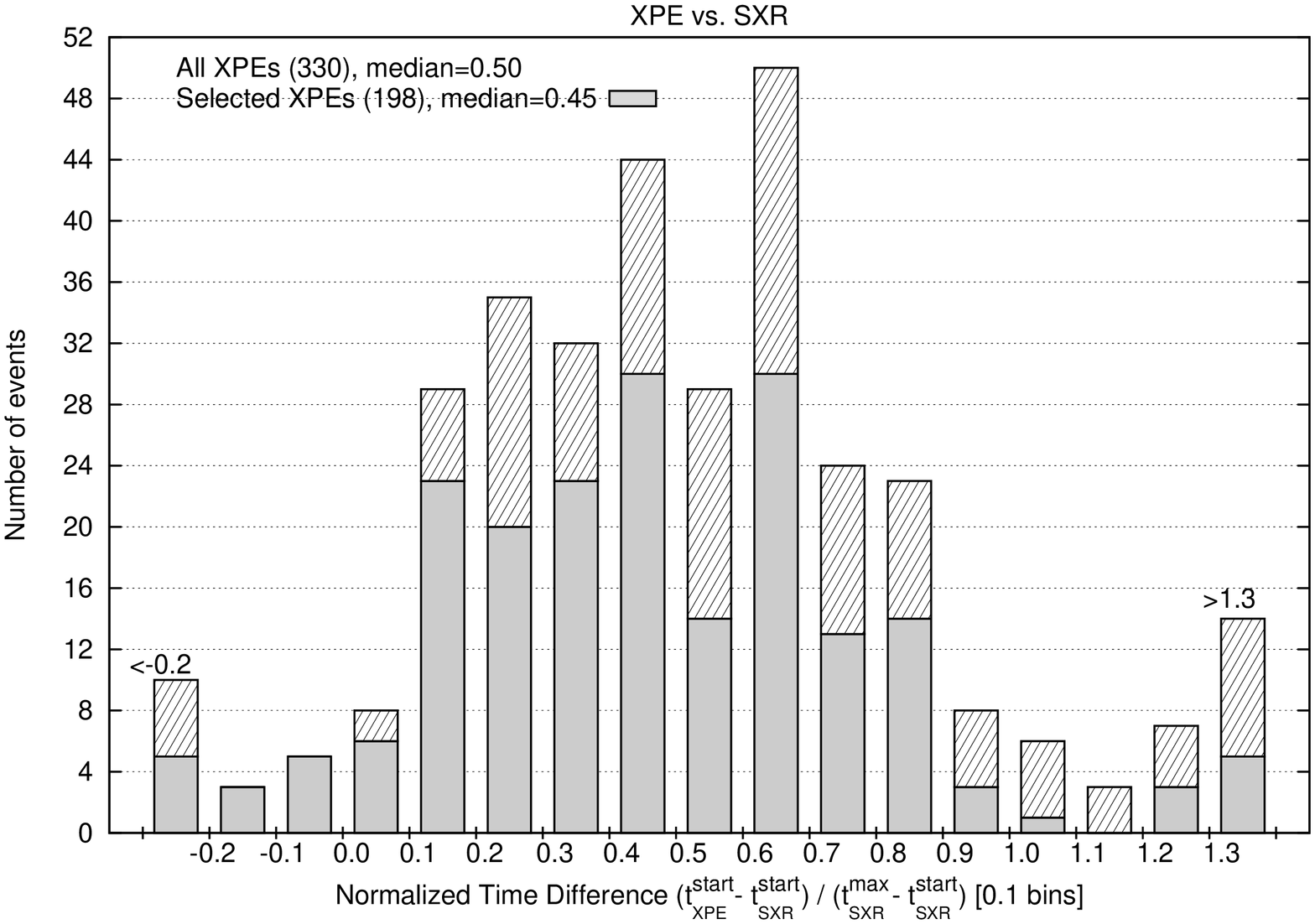}
\caption{Histogram of time differences between the SXR start of the
associated flare (from {\sl GOES} light curve) and the XPE start
normalized with the flare rising-phase duration. Gray and hatched
bins represent the better observed XPEs (quality A-B, so-called
selected XPEs) and the rest of XPEs (quality C), respectively. The
size of bins is 0.1 of the flare rising-phase duration with
exception of outermost ones. Numbers of all the considered XPEs and
the selected XPEs as well as their medians are given.
\label{flbn-xpebn}}
\end{figure}

The above results should be normalized to the time scales of flares.
Therefore, we have prepared counterparts of Figs.~\ref{flb-xpeb} and
\ref{flm-xpee} in which we normalize time differences with the flare
rising-phase duration. In Fig.~\ref{flbn-xpebn} we illustrate
occurrences of the XPE start: negative values mean that a XPE
preceded its flare, the value 0 -- simultaneous start, the value 1
-- start of a XPE at the maximum of its flare, values grater than 1
-- later XPE start. As we see, the majority of XPEs (282 from 330,
85.5\%) starts within the rising phase of flares. This rule is
fulfilled even stronger for better observed XPEs (gray bins) -- 176
from 198, 88.8\%.

\begin{figure}
\includegraphics[angle=-90,scale=.65]{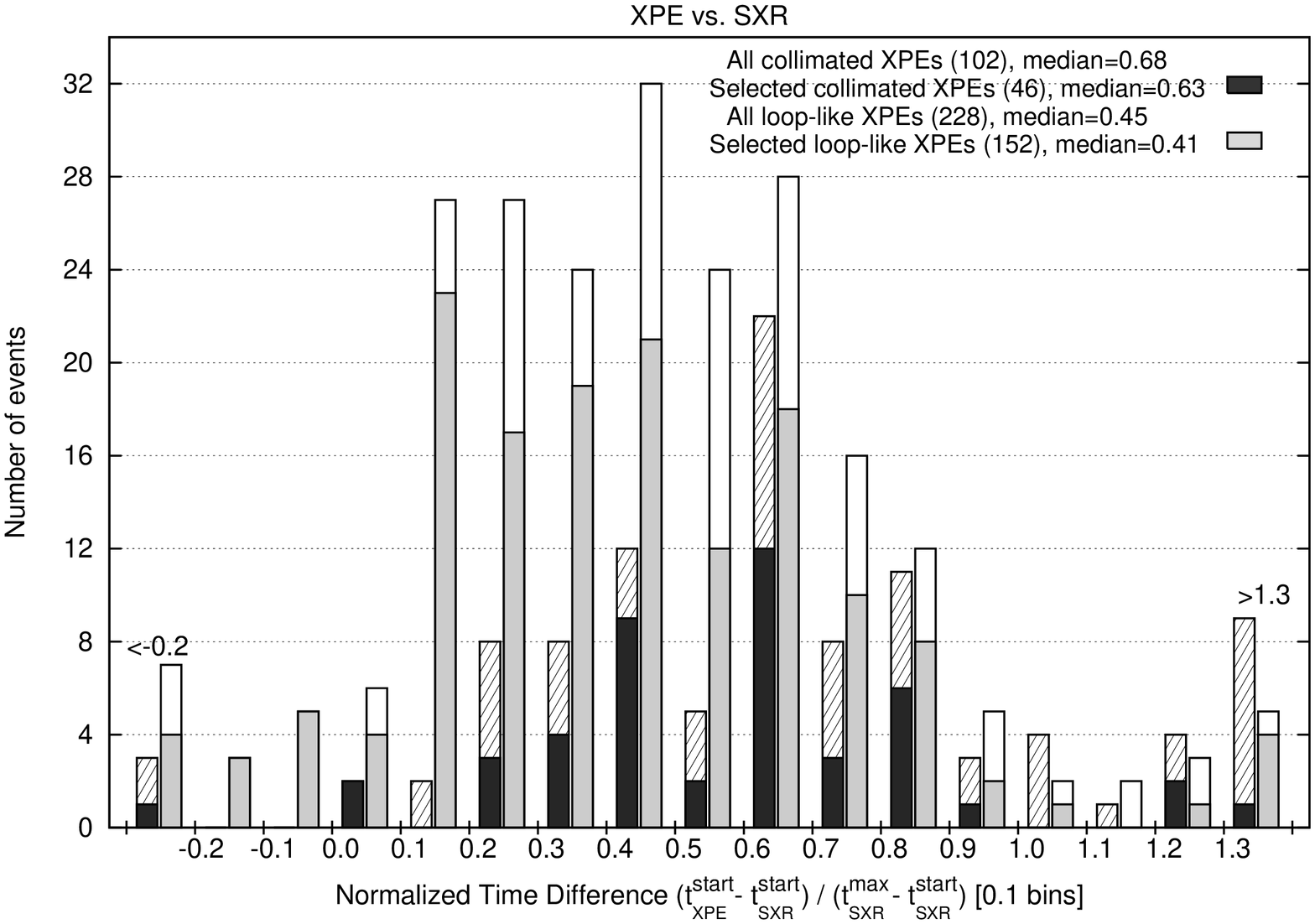}
\caption{Histogram of time differences between the SXR start of the
associated flare (from {\sl GOES} light curve) and the XPE start
normalized with the flare rising-phase duration made for collimated
and loop-like XPEs separately. Black or gray and hatched or white
bins represent the better observed XPEs (quality A-B, so-called
selected XPEs) and the rest of XPEs (quality C), respectively. The
size of bins is 0.1 of the flare rising-phase duration with
exception of outermost ones. Numbers of all the considered XPEs and
the selected XPEs as well as their medians are given.
\label{flbn-xpebn-cl}}
\end{figure}

\begin{figure}
\includegraphics[angle=-90,scale=.65]{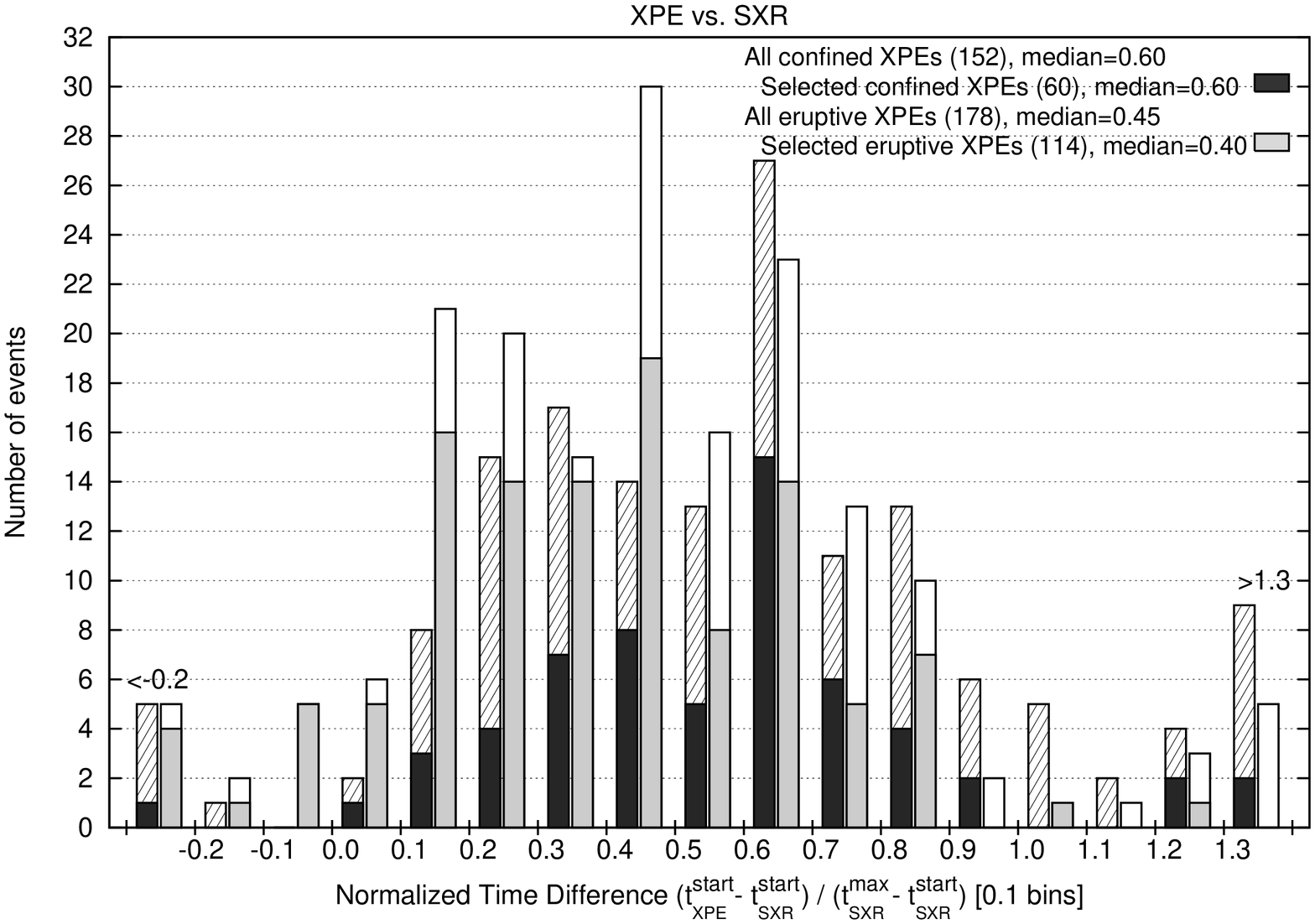}
\caption{Histogram of time differences between the SXR start of the
associated flare (from {\sl GOES} light curve) and the XPE start
normalized with the flare rising-phase duration made for confined
and eruptive XPEs separately. Black or gray and hatched or white
bins represent the better observed XPEs (quality A-B, so-called
selected XPEs) and the rest of XPEs (quality C), respectively. The
size of bins is 0.1 of the flare rising-phase duration with
exception of outermost ones. Numbers of all the considered XPEs and
the selected XPEs as well as their medians are given.
\label{flbn-xpebn-ce}}
\end{figure}

\begin{figure}
\includegraphics[angle=-90,scale=.65]{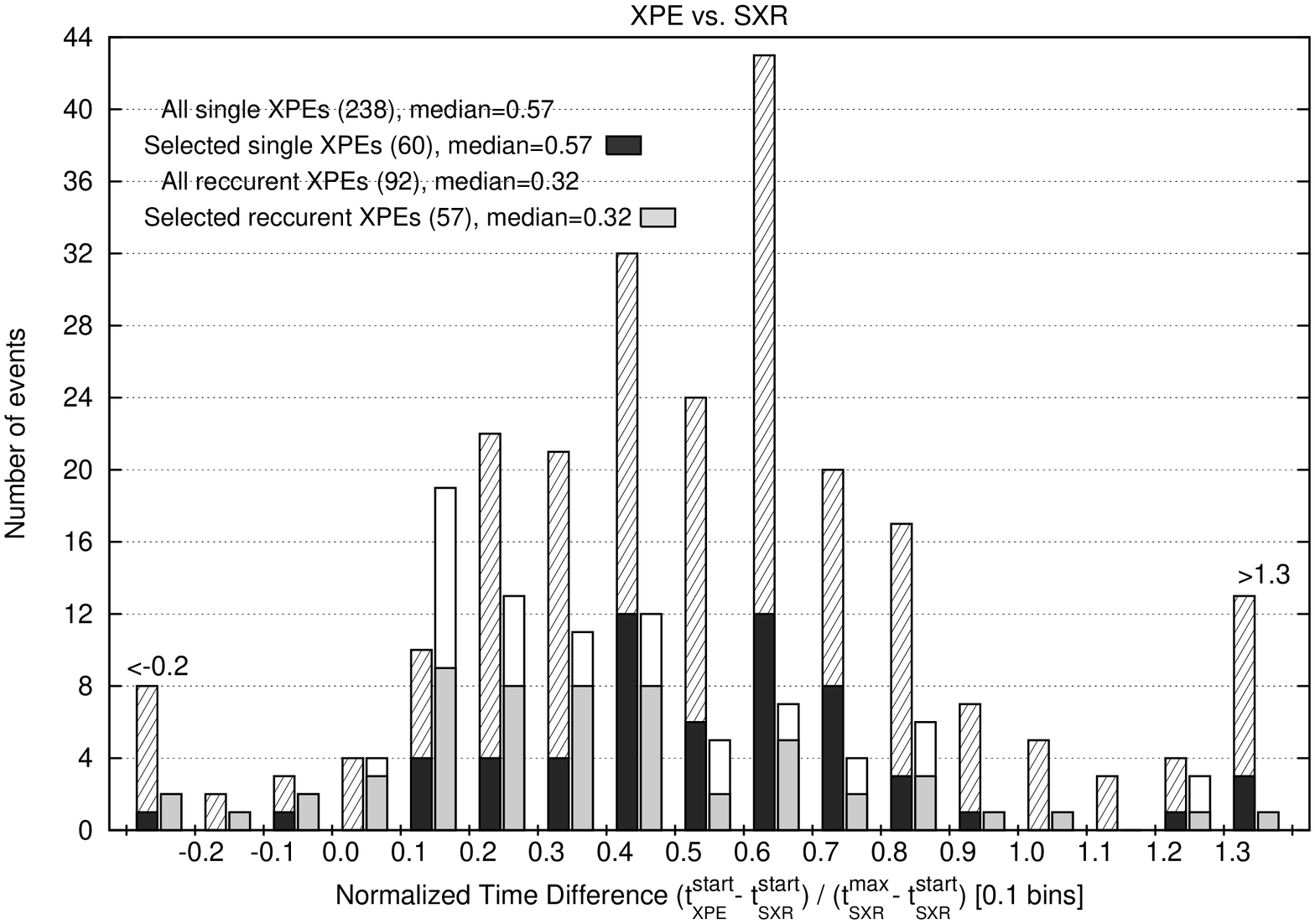}
\caption{Histogram of time differences between the SXR start of the
associated flare (from {\sl GOES} light curve) and the XPE start
normalized with the flare rising-phase duration made for single and
recurrent XPEs separately. Black or gray and hatched or white bins
represent the better XPEs (quality A-B, so-called selected XPEs) and
the rest of XPEs (quality C), respectively. The size of bins is 0.1
of the flare rising-phase duration with exception of outermost ones.
Numbers of all the considered XPEs and the selected XPEs as well as
their medians are given. \label{flbn-xpebn-sr}}
\end{figure}

Interesting results are revealed by further versions of
Fig.~\ref{flbn-xpebn}, in which particular subclasses of XPEs are
separated: collimated and loop-like, confined and eruptive, single
and recurrent, in Figs.~\ref{flbn-xpebn-cl}-\ref{flbn-xpebn-sr},
respectively. In these figures, we show side-by-side the
distributions of the XPEs that have contrasting properties. For
example, in Fig.~\ref{flbn-xpebn-cl} loop-like XPEs show tendency to
start earlier in the rising phase of flares than collimated ones:
the difference for medians is more than 0.2 of the rising-phase
duration. A similar tendency is seen in Fig.~\ref{flbn-xpebn-ce}
where eruptive XPEs start earlier in the rising phase of associated
flares than confined ones and in Fig.~\ref{flbn-xpebn-sr} where
recurrent XPEs precede, on average, single ones.

\begin{figure}
\includegraphics[angle=-90,scale=.65]{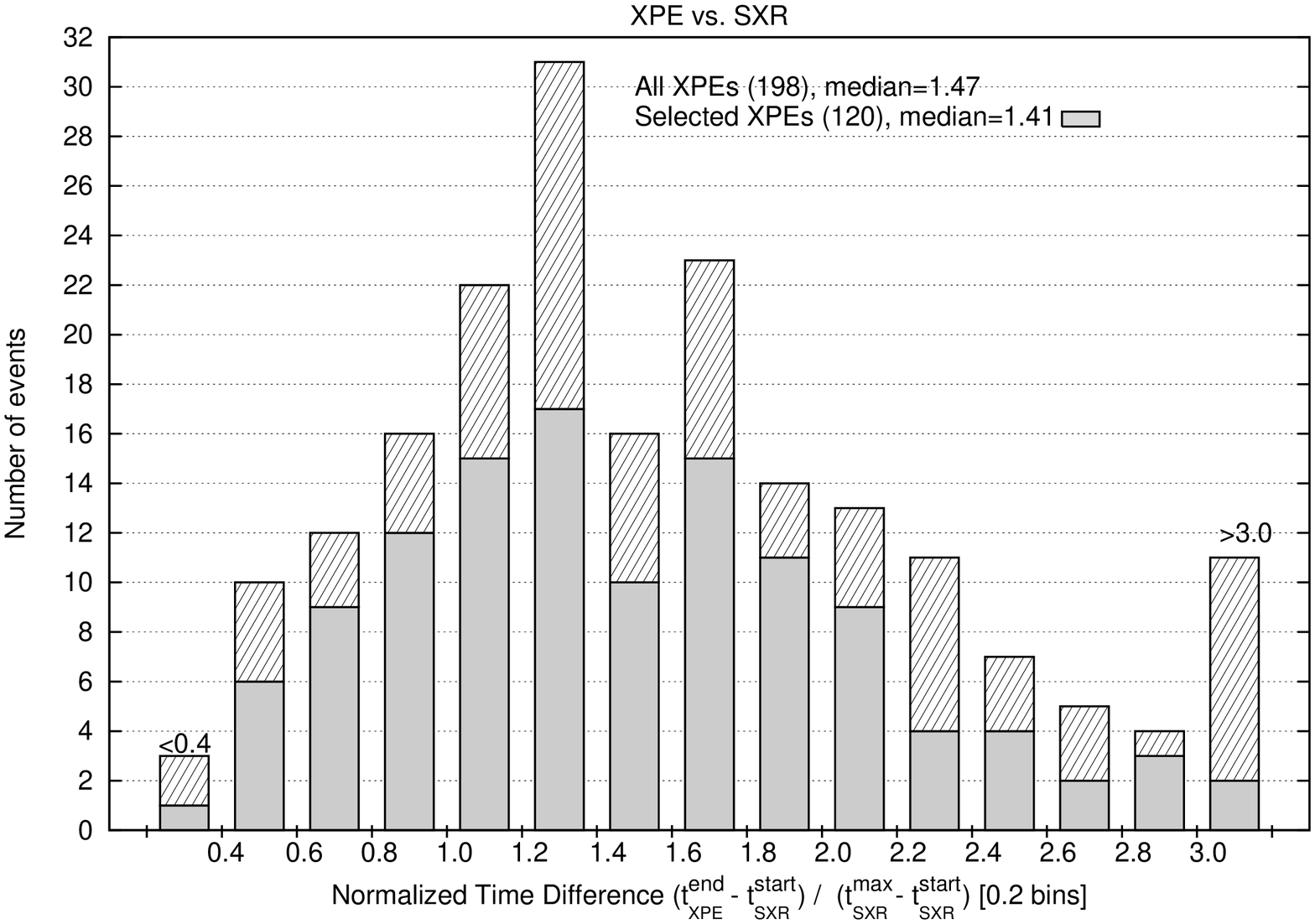}
\caption{Histogram of time differences between the SXR start of the
associated flare (from {\sl GOES} light curve) and the XPE end
normalized with the flare rising-phase duration. Gray and hatched
bins represent the better observed XPEs (quality A-B, so-called
selected XPEs) and the rest of XPEs (quality C), respectively. The
size of bins is 0.2 of the flare rising-phase duration with
exception of outermost ones. Numbers of all the considered XPEs and
the selected XPEs as well as their medians are given.
\label{flmn-xpeen}}
\end{figure}

In Fig.~\ref{flmn-xpeen} we have normalized time differences between
the XPE end and the associated flare start with the rising-phase
duration of a flare. In this scale the value 1 means that the XPE
end occurred exactly at the maximum of the associated flare. The
histogram is rather gradual with two maxima between 1.2-1.4 and
1.6-1.8. The number of XPEs lower than the first maximum and greater
than the second one decreases systematically with a marginal
contribution of those that are lower than 0.4 and greater than 3. It
means that all soft X-ray plasma motions are limited within a
relatively narrow part of total duration of associated flares.

\begin{figure}
\includegraphics[angle=-90,scale=.65]{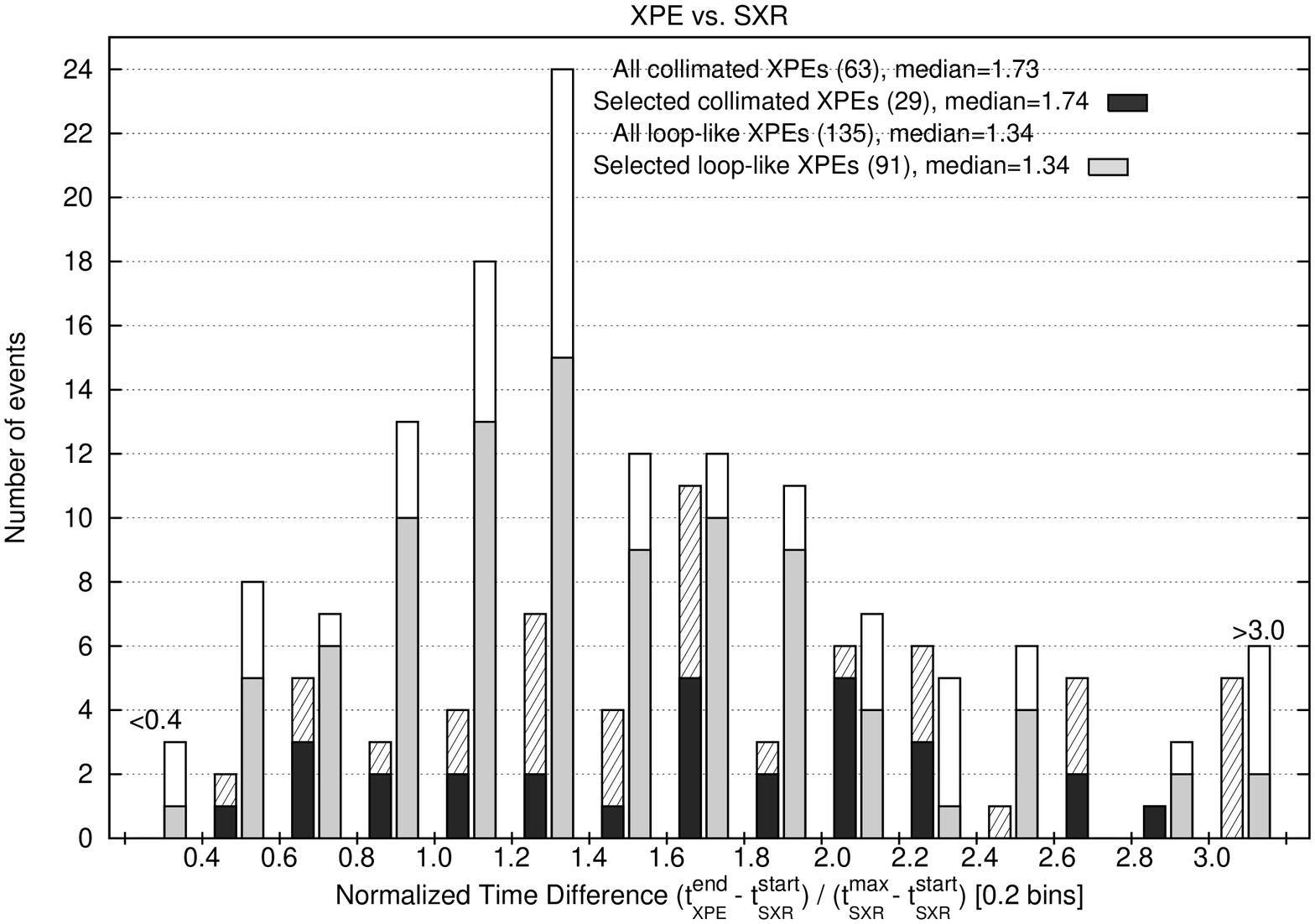}
\caption{Histogram of time differences between the SXR start of the
associated flare (from {\sl GOES} light curve) and the XPE end
normalized with the flare rising-phase duration made for collimated
and loop-like XPEs separately. Black or gray and hatched or white
bins represent the better observed XPEs (quality A-B, so-called
selected XPEs) and the rest of XPEs (quality C), respectively. The
size of bins is 0.2 of the flare rising-phase duration with
exception of outermost ones. Numbers of all the considered XPEs and
the selected XPEs as well as their medians are given.
\label{flmn-xpeen-cl}}
\end{figure}

\begin{figure}
\includegraphics[angle=-90,scale=.65]{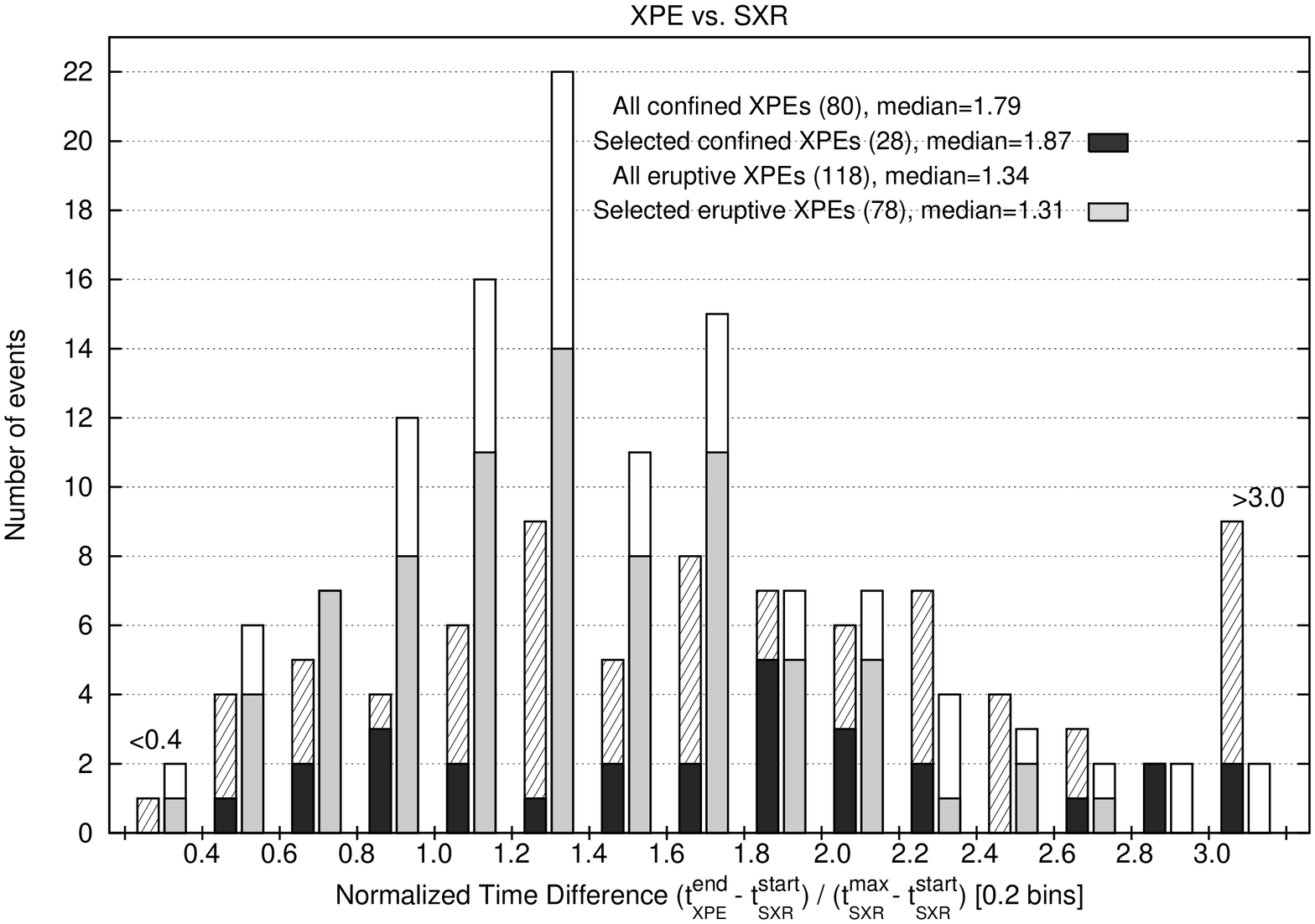}
\caption{Histogram of time differences between the SXR start of the
associated flare (from {\sl GOES} light curve) and the XPE end
normalized with the flare rising-phase duration made for confined
and eruptive XPEs separately. Black or gray and hatched or white
bins represent the better observed XPEs (quality A-B, so-called
selected XPEs) and the rest of XPEs (quality C), respectively. The
size of bins is 0.2 of the flare rising-phase duration with
exception of outermost ones. Numbers of all the considered XPEs and
the selected XPEs as well as their medians are given.
\label{flmn-xpeen-ce}}
\end{figure}

\begin{figure}
\includegraphics[angle=-90,scale=.65]{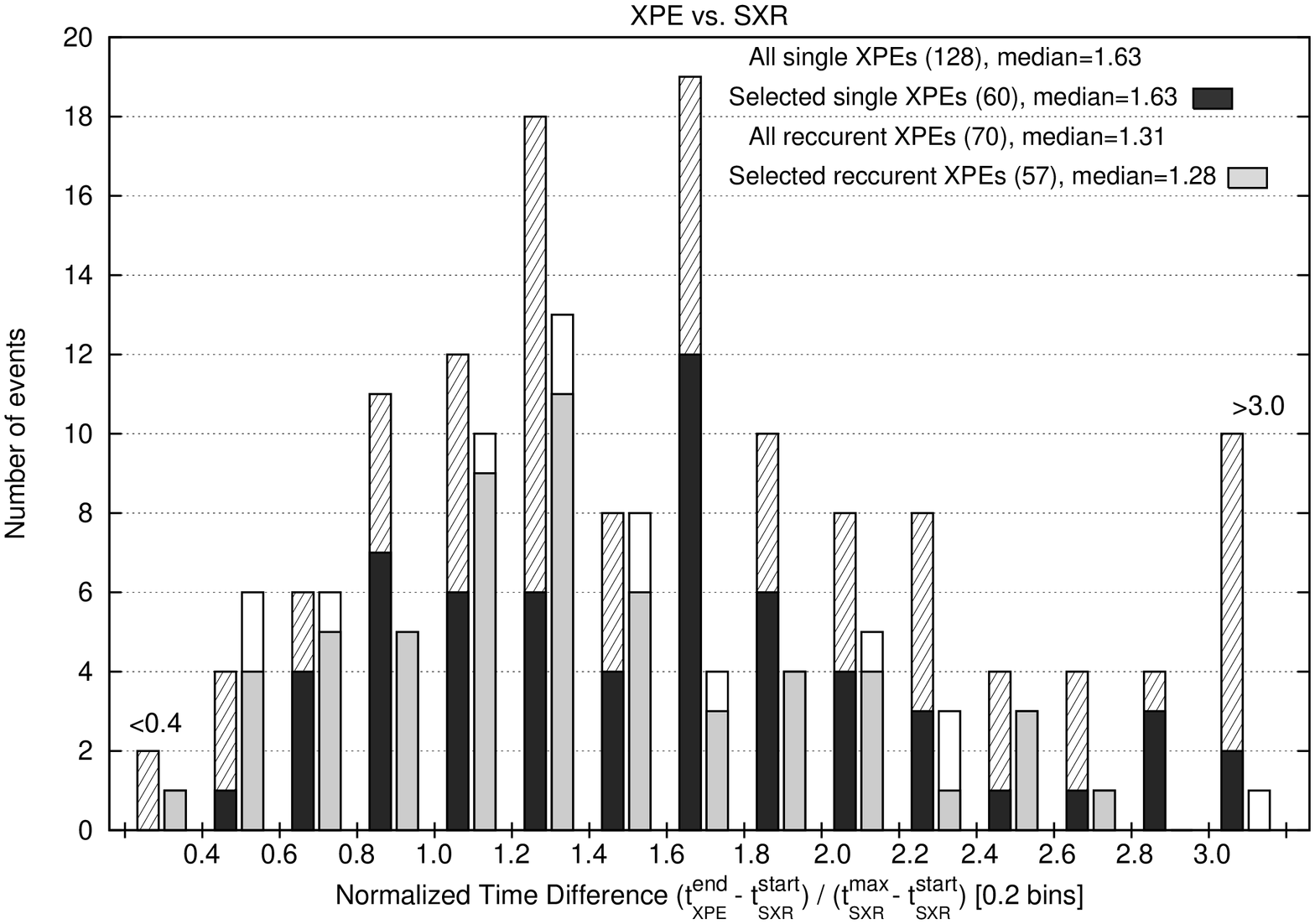}
\caption{Histogram of time differences between the SXR start of the
associated flare (from {\sl GOES} light curve) and the XPE end
normalized with the flare rising-phase duration made for single and
recurrent XPEs separately. Black or gray and hatched or white bins
represent the better observed XPEs (quality A-B, so-called selected
XPEs) and the rest of XPEs (quality C), respectively. The size of
bins is 0.2 of the flare rising-phase duration with exception of
outermost ones. Numbers of all the considered XPEs and the selected
XPEs as well as their medians are given. \label{flmn-xpeen-sr}}
\end{figure}

The variants of Fig.~\ref{flmn-xpeen}, in which particular
subclasses of XPEs are separated: collimated and loop-like, confined
and eruptive, single and recurrent, are presented in
Figs.~\ref{flmn-xpeen-cl}-\ref{flmn-xpeen-sr}, respectively. In
these figures, as in Figs.~\ref{flbn-xpebn-cl}-\ref{flbn-xpebn-sr},
we show side-by-side the distributions of the XPEs that have
contrasting properties. In Fig.~\ref{flmn-xpeen-cl} the collimated
XPEs seems to last longer, on average, than the loop-like ones:
medians of both distribution differ by 0.4 of the rising-phase
duration. Similarly, the confined XPEs seems to last longer than the
eruptive ones (Fig.~\ref{flmn-xpeen-ce}) -- medians differ about 0.6
of the rising-phase duration, in case of better observed XPEs. In
Fig.~\ref{flmn-xpeen-sr} the single XPEs last longer, on average,
than the recurrent ones -- medians differ about 0.35 of the
rising-phase duration, in case of better observed XPEs.

\subsubsection{Flare class and total duration} \label{class+dur}

For each associated flare we determined X-ray class and total
duration based on light curves recorded by {\sl GOES}, in the
wavelength range of 1-8 \AA\ . We defined the total duration as the
interval between a constant level of the solar soft X-ray flux
before and after a flare, therefore our values of this parameter are
larger than intervals between a start time and end time that are
routinely reported in the SGD. In some cases we could not estimate
the total duration properly. This is the reason why a number of
considered events in this paragraph is slightly lower than the
number of XPEs associated with flares.

\begin{figure}
\includegraphics[scale=.90]{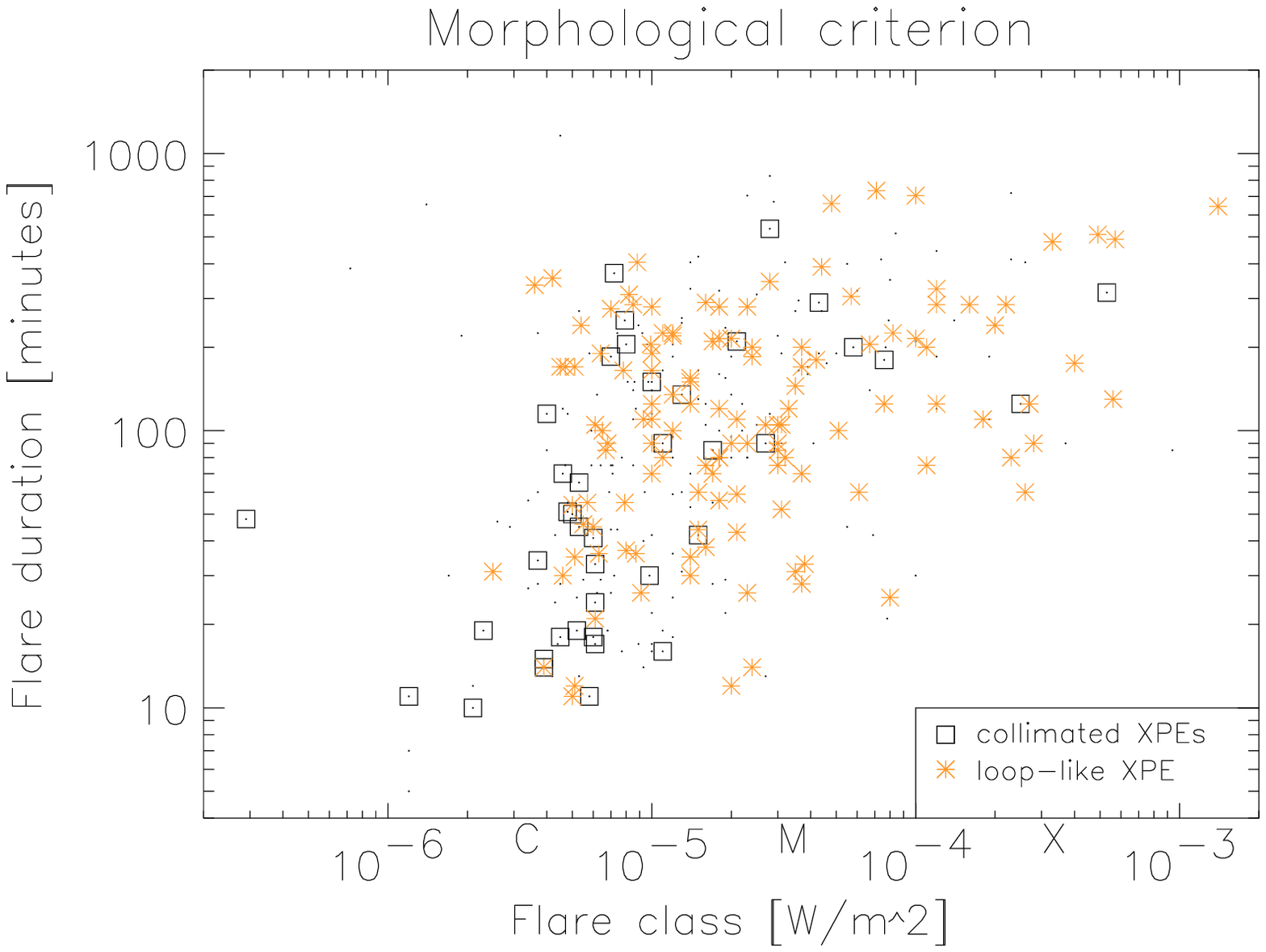}
\caption{Scatter plot of flare X-ray class versus flare total
duration. This plot compares flares associated with XPEs classified
according to the morphological criterion. All points are marked with
dots. A subset of well-observed (see text) flares are additionally
marked with boxes and stars for collimated and loop-like XPEs of
high quality (A--B), respectively.\label{xpefl1}}
\end{figure}

In Fig.~\ref{xpefl1} we present scatter plot of X-ray class versus
total duration for flares associated with morphological subclasses
of XPEs, i.e., collimated and loop-like XPEs. All points are marked
with dots. Additionally we emphasized well-observed XPEs (quality A
or B) and flares that are non-occulted by the solar disk. These
flares associated with well-observed collimated and loop-like XPEs
are marked with boxes and stars, respectively. Both groups of flares
are mixed in the plot, however some shifts toward higher X-ray class
and longer duration can be seen for flares associated with loop-like
XPEs.

\begin{figure}
\includegraphics[scale=.90]{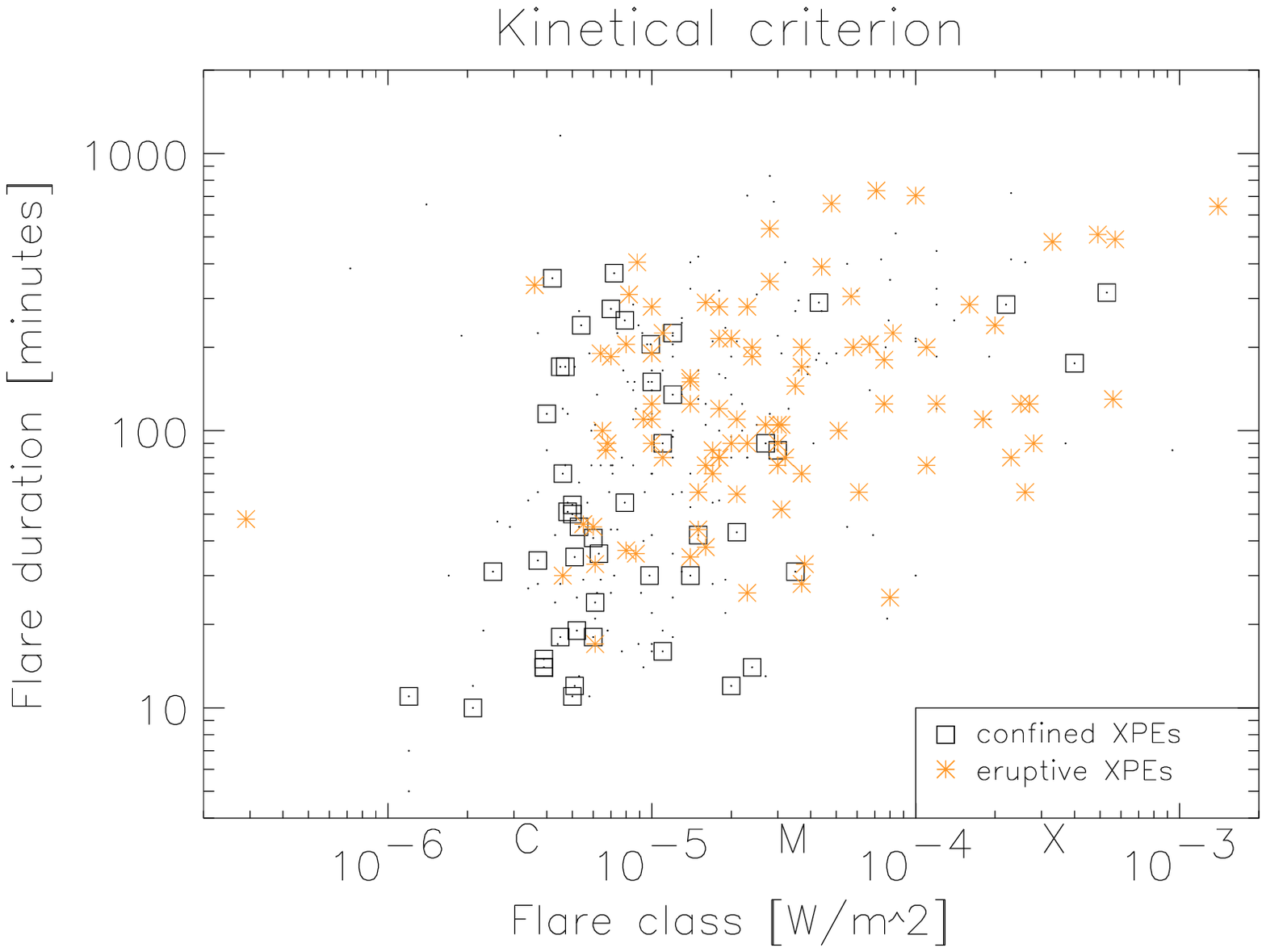}
\caption{Scatter plot of flare X-ray class versus flare total
duration. This plot compares flares associated with XPEs classified
according to the kinematical criterion. All points are marked with
dots. A subset of well-observed (see text) flares are additionally
marked with boxes and stars for confined and eruptive XPEs of high
quality (A--B), respectively. \label{xpefl2}}
\end{figure}

\begin{figure}
\includegraphics[scale=.90]{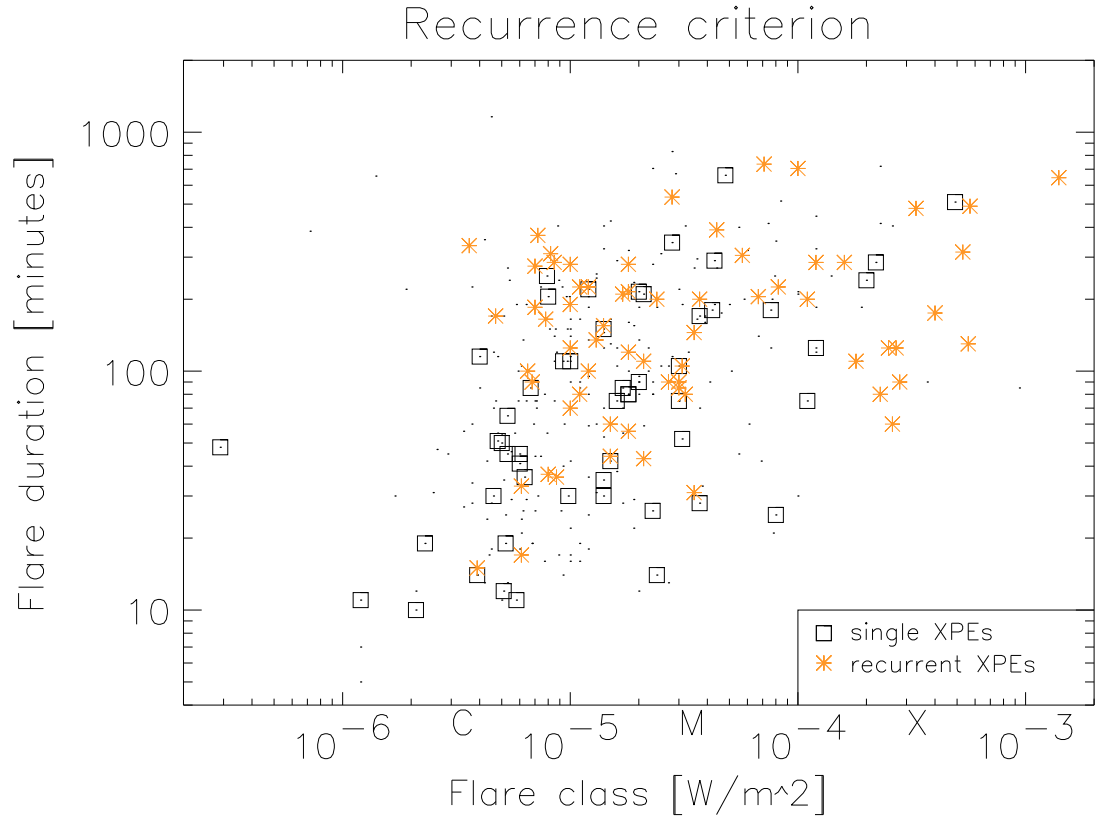}
\caption{Scatter plot of flare X-ray class versus flare total
duration. This plot compares flares associated with XPEs classified
according to the recurrence criterion. All points are marked with
dots. A subset of well-observed (see text) flares are additionally
marked with boxes and stars for single and recurrent XPEs of high
quality (A--B), respectively.\label{xpefl3}}
\end{figure}

Similar scatter plots of X-ray class versus total duration for
flares associated with kinematical and recurrence subclasses of XPEs
are given in Figs.~\ref{xpefl2}-\ref{xpefl3}. All points are marked
with dots. Again, additionally we emphasized well-observed XPEs
(quality A or B) and flares that are non-occulted by the solar disk.
Moreover, we excluded events for which the assignment of kinematical
subclasses for XPEs was uncertain (Fig.~\ref{xpefl2}) and events
associated with XPEs that were classified as single in case of
partial time-coverage of observations (Fig.~\ref{xpefl3}). In both
figures, flares associated with well-observed XPEs classified as
subclass 1 (confined and single, respectively) are marked with
boxes, whereas flares associated with XPEs classified as subclass 2
(eruptive and recurrent, respectively) are marked with stars.
Similarly to Fig.~\ref{xpefl1}, both groups of flares are mixed in
the plots and some shifts toward higher X-ray class and longer
duration is seen for flares associated with XPEs of subclasses 2.

The shifts seen in Figs.~\ref{xpefl1}-\ref{xpefl3} are confirmed by
medians calculated separately for both groups of flares for each
classification criterion. As it is seen in Table~\ref{flare}
(bold-faced columns), medians for flares associated with XPEs of
subclass 2 are 1.5--3.8 times and 2.1--2.7 times greater than
medians for flares associated with XPEs of subclass 1 for flare
X-ray class and flare duration, respectively. Higher X-ray class and
longer duration mean a more energetic flare, thus we can conclude
that more energetic XPEs are, on average, associated with more
energetic flares and less-energetic XPEs rather prefer
less-energetic flares.

\begin{figure}
\includegraphics[scale=.90]{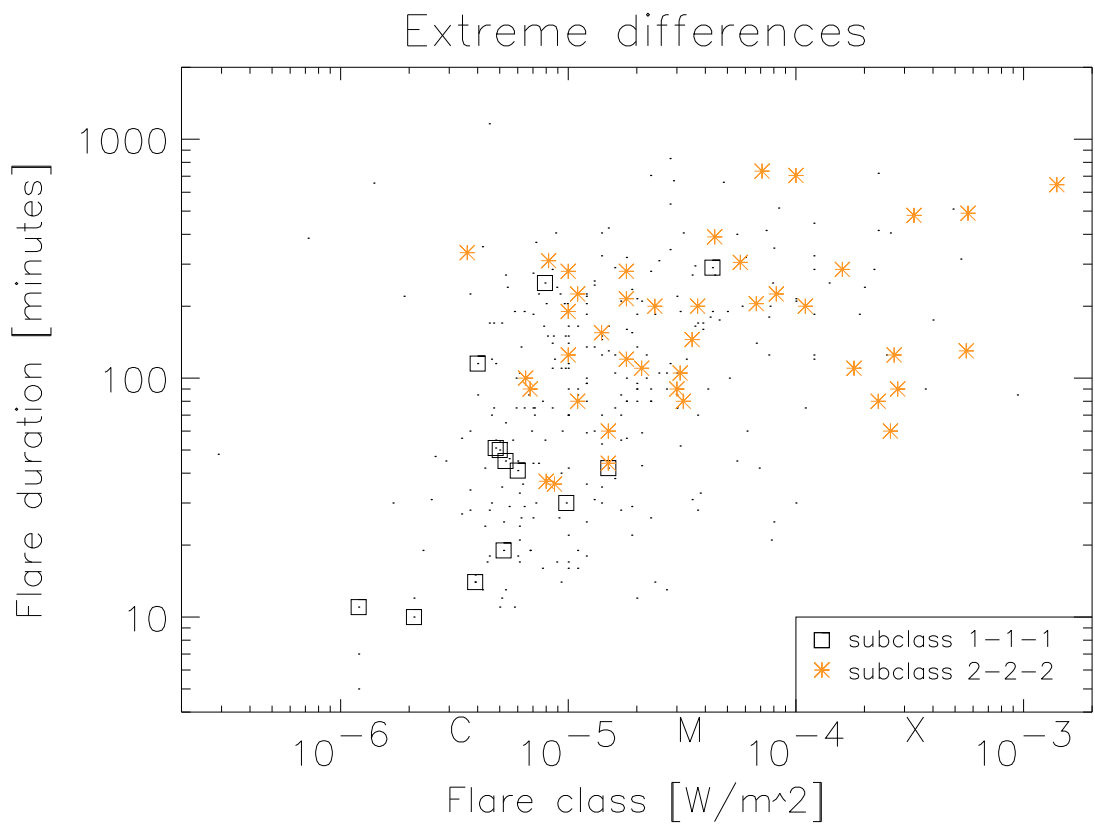}
\caption{Scatter plot of flare X-ray class versus flare total
duration. This plot compares flares associated with XPEs classified
according to different criteria that are employed simultaneously.
All points are marked with dots. A subset of well-observed (see
text) flares are additionally marked with boxes and stars for
collimated, confined, single (1,1,1) and loop-like, eruptive,
recurrent (2,2,2) XPEs of high quality (A--B),
respectively.\label{xpefl4}}
\end{figure}

One can expect that a difference between characteristics describing
associated flares should be even higher for two subclasses of XPEs
defined by combining our three criteria simultaneously. Indeed,
medians in Table~\ref{flare} for flares associated with the subclass
(1,1,1) -- collimated, confined, single XPEs -- and the subclass
(2,2,2) -- loop-like, eruptive, recurrent XPEs -- show extreme
differences (a factor 6.0 and 3.7 for X-ray class and duration,
respectively). As it is seen in Fig.~\ref{xpefl4}, in the diagram
X-ray class versus duration, flares associated with the subclasses
(1,1,1) and (2,2,2) of well-observed XPEs are almost separated.

In unbold-faced columns in Table~\ref{flare} we present medians for
flares associated with different subclasses of XPEs that were
defined less strictly, i.e. by including quality C events and
without excluding any doubtful examples. Ratios of medians for
subclasses 2 and subclasses 1 that were constituted more liberally
are usually lower in comparison with the more strictly defined
bold-faced values. It shows how some physical differences can be
masked by observational constrains.

\subsection{Hard X-rays} \label{hxr}

We included in the catalogue hard X-ray light curves of associated
flares, recorded by {\sl Yohkoh} HXT, for investigating the relation
between XPEs and non-thermal electron signatures. We considered
light curves in energy band M1 (23-33 keV) and interpreted a signal
above the doubled value of the background as the proof that in a
particular flare an acceleration of an appropriate number of
non-thermal electrons occurred. For 353 flares which were associated
with XPEs we found that 235 events, i.e. 66.6\% showed this
signature.

In the second and the third columns of Table\,\ref{flare} we present
detailed results of this relation for both populations of events and
for particular subclasses of XPEs. A percentage of associated flares
showing non-thermal electrons depends on how energetic is the
subclass, with higher values (75\%-82\%) for subclasses 2 and lower
values (57\%-75\%) for subclasses 1. The difference is the highest
for the kinematical criterion, but for the recurrence criterion
percentages for both subclasses are almost the same. After applying
all the three criteria we found that the difference between the
least energetic subclass (1,1,1) -- collimated, confined, single --
and the most energetic (2,2,2) -- loop-like, eruptive, recurrent --
is maximal: 57\% and 86\% of associated flares indicating
non-thermal electrons, respectively.

\begin{figure}
\includegraphics[angle=-90,scale=.65]{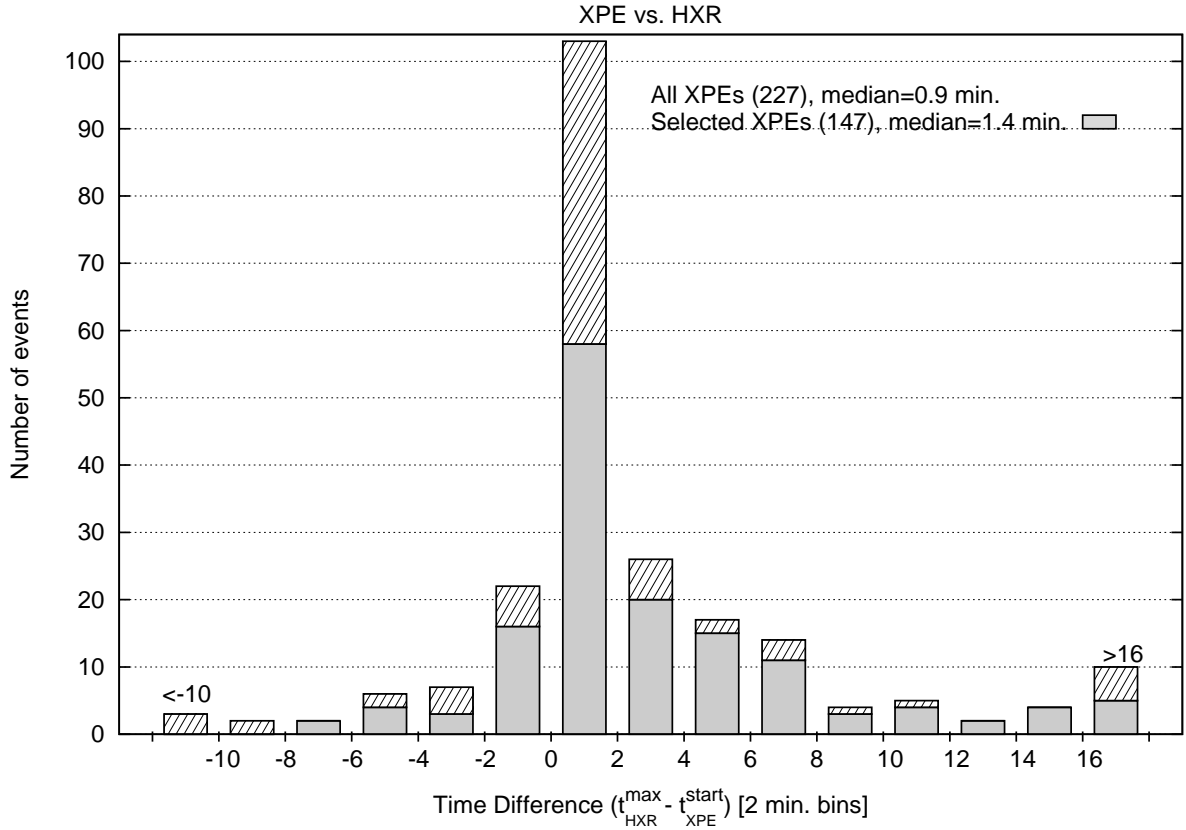}
\caption{Histogram of time differences between the HXT/M1 light
curve maximum and the XPE start. Gray and hatched bins represent the
better observed XPEs (quality A-B, so-called selected XPEs) and the
rest of XPEs (quality C), respectively. The size of bins is 2
minutes with exception of outermost ones. Numbers of all the
considered XPEs and the selected XPEs as well as their medians are
given. \label{hxr-xpe}}
\end{figure}

We also investigated time coincidence between macroscopic X-ray
plasma motions (XPEs) and non-thermal electron signatures (HXRs) in
detail. In this aim, we measured time differences between the XPE
start and the HXT/M1 flare peak. The results are presented as a
histogram in Fig.~\ref{hxr-xpe}. In 185 from 227 cases (81.5\%) XPEs
started before the HXR flare peak, in 18.5\% of cases the chronology
was opposite. However, the most frequent bin: 0--2 minutes in about
45\% of cases, suggests that both considered processes, i.e. soft
X-ray plasma motion and non-thermal electron acceleration, are
strongly coupled.

Similar investigation was performed by \citet{kim05a}. At first
glance our Fig.~\ref{hxr-xpe} and their Fig.~5 are different.
However, it is needed to know that the values in our histogram have
opposite sign and we used the HXR peak time for higher energy band
M1, 23-33 keV, than Kim et al. who used energy band L (14-23 keV).
In flares with a strong contribution of the non-thermal component,
the peak time in those energy bands are close, but in flares with a
stronger contribution of the superhot component in L band, the M1
peaks tend to occur earlier than the L peaks. Keeping in mind the
above mentioned differences in data organization we can conclude
that our results are consistent.

We also prepared variants of Fig.~\ref{hxr-xpe} for particular
subclasses of XPEs. However, we did not find any evident differences
between the considered distributions, namely, each of them shares
the common peak bin.

\section{XPEs association with Coronal Mass Ejections} \label{ass-cme}

In order to associate out XPEs with CMEs, we used the {\sl SOHO}
LASCO CME Catalog \citep{gop09}. Only 275 XPEs occurred when the
LASCO coronagraphs were operational. We found that 182 XPEs (66.2\%)
were associated with CMEs. This is slightly less than 69\% (95 from
137 events) obtained by \citet{kim05a}.  For particular subclasses
of XPEs the association was between 44\% and 88\% (see
Table~\ref{cmes}). According to \citet{yas08}, we consider a XPE-CME
pair as physically connected if the XPE occurred within the position
angle range defined by the CME angular width increased by
10$^{\circ}$ from either side. Moreover, the time of the XPE had to
fall within 3-hours-interval centered around the extrapolated time
of the CME front start at $h=1 R_{\sun}$. For the extrapolation we
used the time of the first appearance in the LASCO/C2 field of view
and the linear velocity taken from the CME catalog.

\subsection{Time coincidence} \label{cme-tc}

\begin{figure}
\includegraphics[angle=-90,scale=.65]{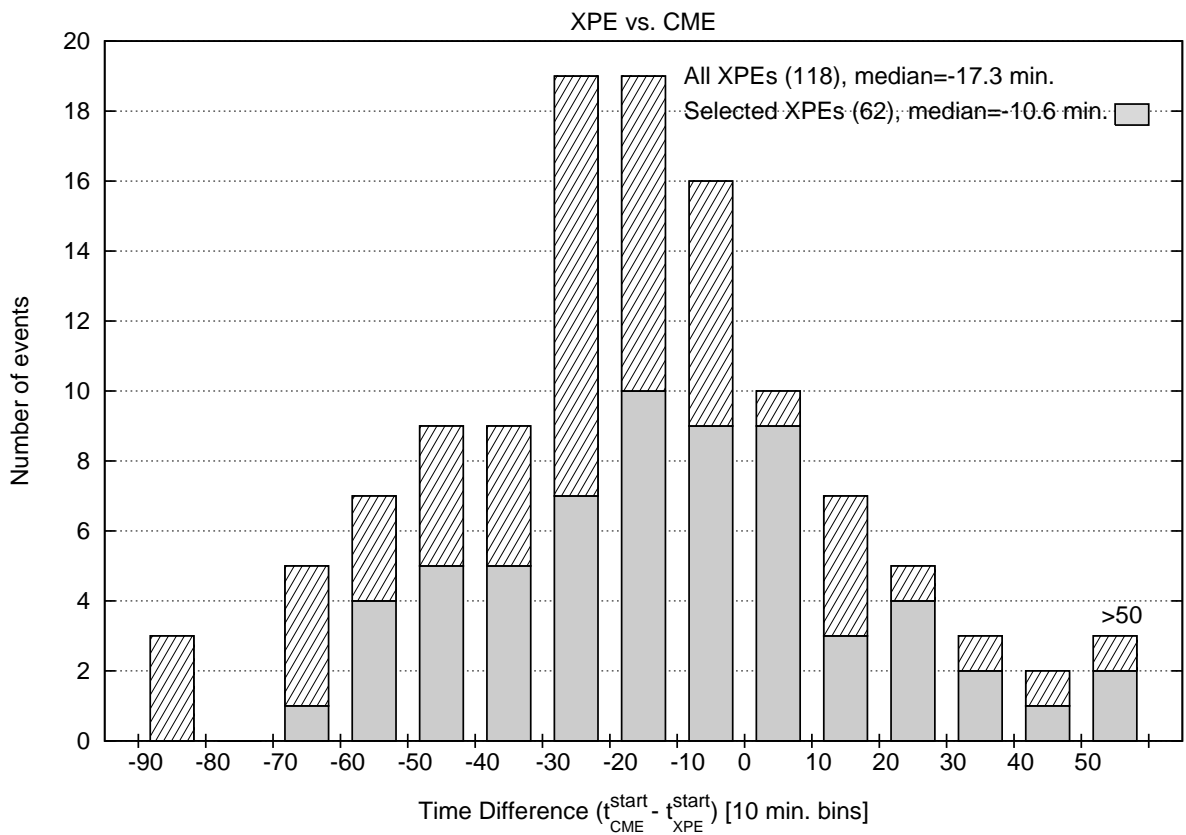}
\caption{Histogram of time differences between the extrapolated CME
front onset and the XPE start. Gray and hatched bins represent the
selected subgroup of better observed XPEs (quality A-B, $|\lambda| >
60^{\circ}$) and the rest of XPEs, respectively. The size of bins is
10 minutes with exception of outermost right one. Numbers of all the
considered XPEs and the selected XPEs as well as their medians are
given. \label{cme-xpe}}
\end{figure}

The histogram of the time differences between the extrapolated CME
front onset and the XPE start is presented in Fig.~\ref{cme-xpe}. As
we can see, there are more events with negative values, i.e., those
in which the CME starts before the XPE, than those with the opposite
chronology. The frequencies are 73.7\% (87 from 118) and 26.3\% (31
from 118), respectively. The carefully selected subgroup
(well-observed XPEs of quality A or B that occurred close to the
solar limb, $|\lambda| > 60^{\circ}$) shows slightly different
proportions: 66,1\% (41 from 62) and 33.9\% (21 from 62),
respectively. Both distributions: all the XPEs and the selected
XPEs, are quite gradual with slightly different medians: -17.3 min.
and -10.6 min., respectively.

\citet{kim05a} performed similar analysis for XPEs from the
two-years interval 1999-2001. Their Fig.~6 containing 43 events was
made under slightly different assumptions: (1) the CME-front times
were extrapolated at individual locations of XPEs in the {\sl
Yohkoh} field of view, (2) the CME speed was determined from the
first two observing times and heights. Despite these differences,
our histogram looks quite similar to those of Kim et al. Therefore
we conclude that, at least in a statistical sense, different ways of
extrapolating the CME onset time do not seriously affect the
temporal relation between XPEs and CMEs.

\begin{figure}
\includegraphics[angle=-90,scale=.65]{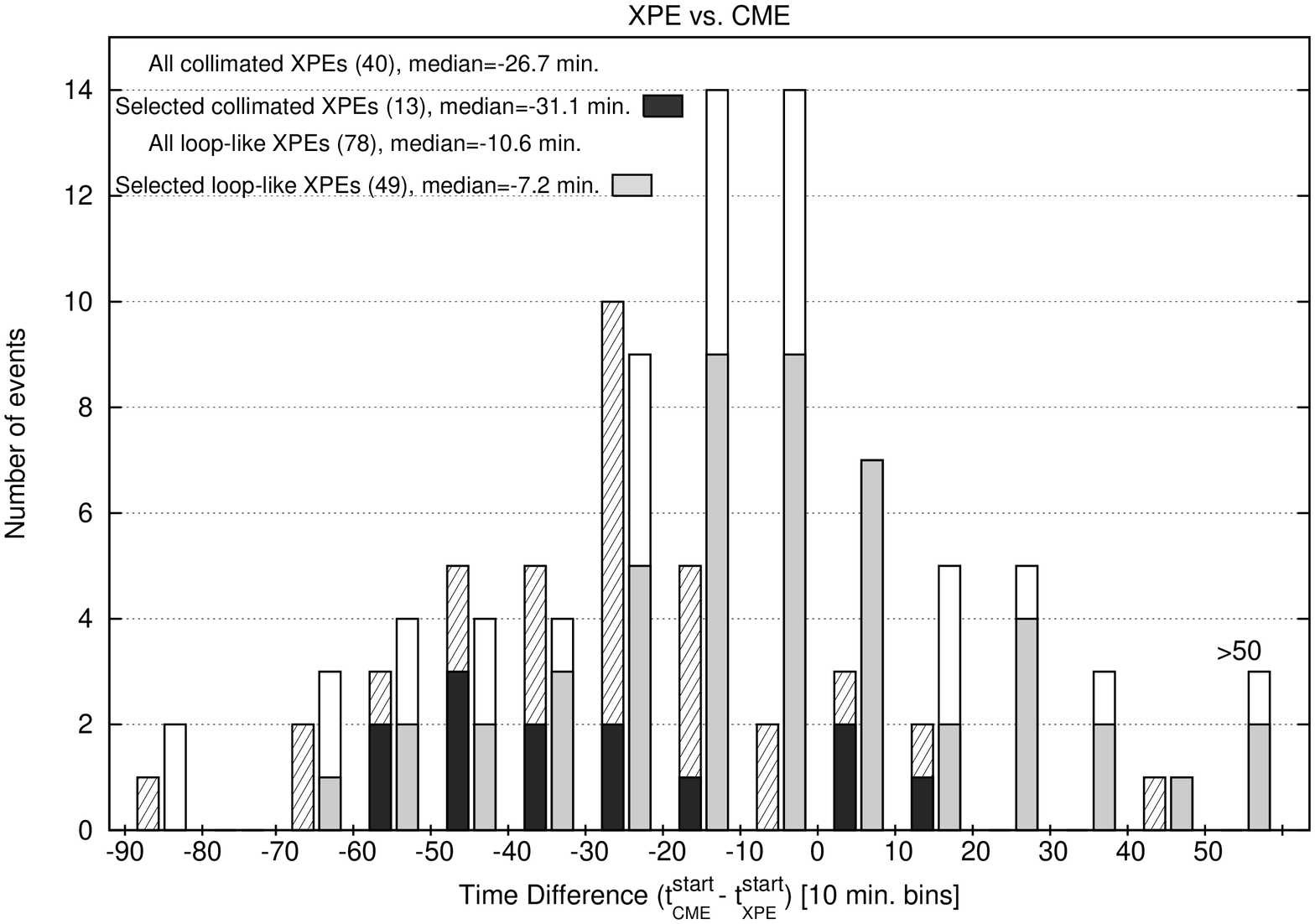}
\caption{Histogram of time differences between the extrapolated CME
front onset and the XPE start made for collimated and loop-like XPEs
separately. Black or gray and hatched or white bins represent the
selected subgroup of better observed XPEs (quality A-B, $|\lambda| >
60^{\circ}$) and the rest of XPEs, respectively. The size of bins is
10 minutes with exception of outermost right one. Numbers of all the
considered XPEs and the selected XPEs as well as their medians are
given. \label{cme-xpe-cl}}
\end{figure}

\begin{figure}
\includegraphics[angle=-90,scale=.65]{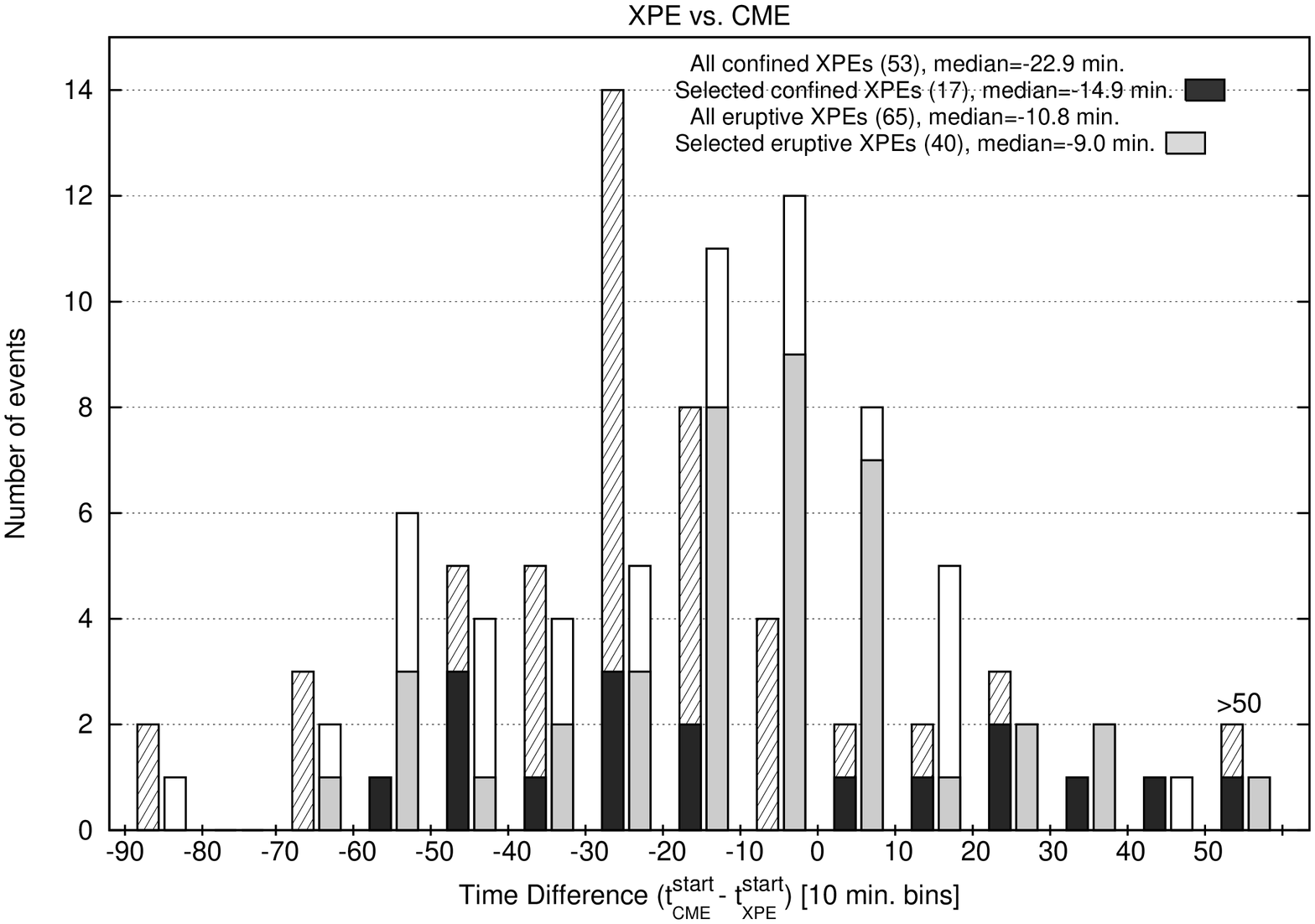}
\caption{Histogram of time differences between the extrapolated CME
front onset and the XPE start made for confined and eruptive XPEs
separately. Black or gray and hatched or white bins represent the
selected subgroup of better observed XPEs (quality A-B, $|\lambda| >
60^{\circ}$) and the rest of XPEs, respectively. The size of bins is
10 minutes with exception of outermost right one. Numbers of all the
considered XPEs and the selected XPEs as well as their medians are
given. \label{cme-xpe-ce}}
\end{figure}

\begin{figure}
\includegraphics[angle=-90,scale=.65]{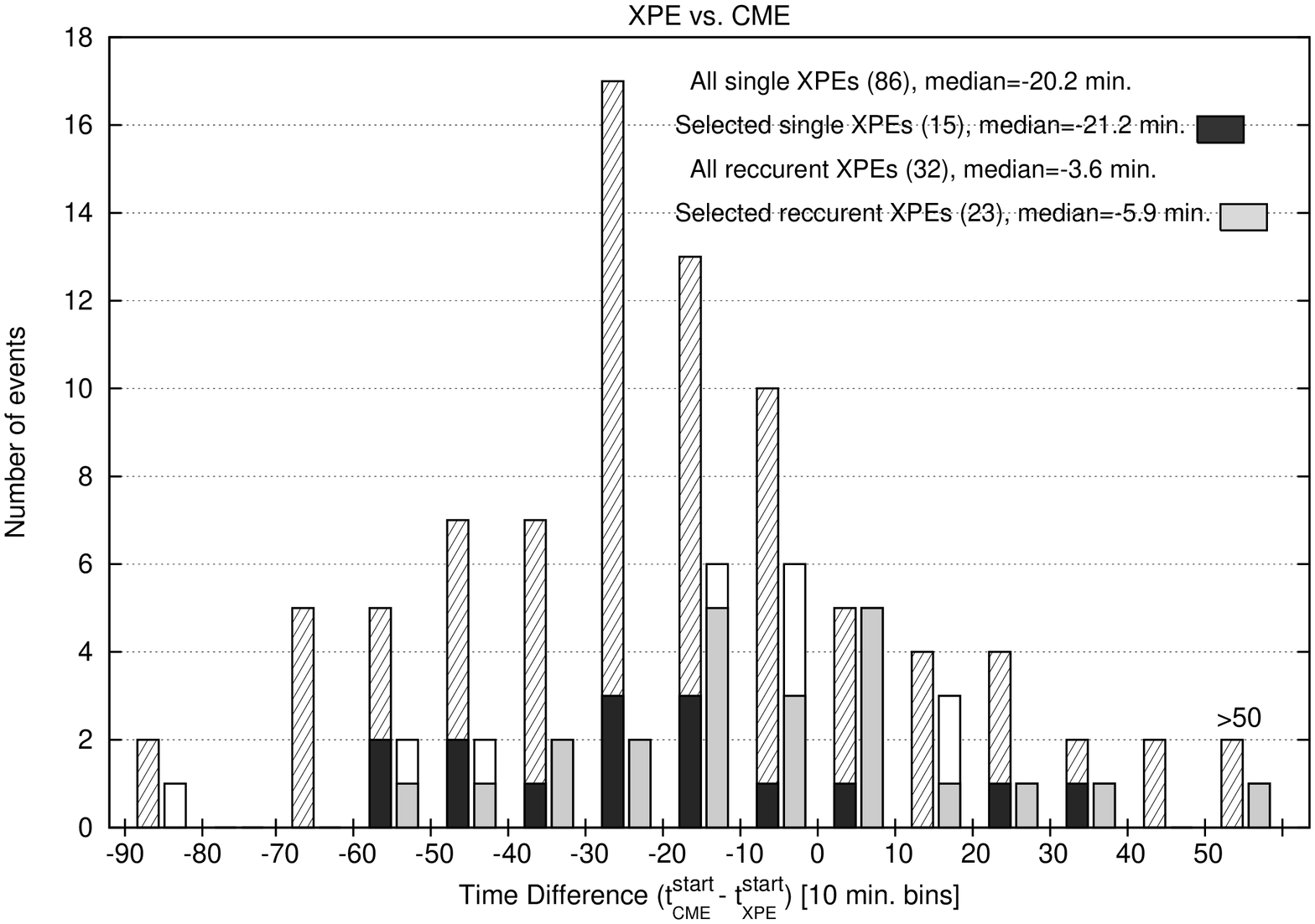}
\caption{Histogram of time differences between the extrapolated CME
front onset and the XPE start made for single and recurrent XPEs
separately. Black or gray and hatched or white bins represent the
selected subgroup of better observed XPEs (quality A-B, $|\lambda| >
60^{\circ}$) and the rest of XPEs, respectively. The size of bins is
10 minutes with exception of outermost right one. Numbers of all the
considered XPEs and the selected XPEs as well as their medians are
given. \label{cme-xpe-sr}}
\end{figure}

As in the analysis in Section~\ref{fl-tc}, we give further versions
of Fig.~\ref{cme-xpe}, in which particular subclasses of XPEs:
collimated and loop-like, confined and eruptive, single and
recurrent, are separated in Figs.~\ref{cme-xpe-cl},
\ref{cme-xpe-ce}, and \ref{cme-xpe-sr}, respectively. In
Fig.~\ref{cme-xpe-cl}, the histogram for loop-like XPEs shows a
relatively narrow maximum located close to the zero point. It means
that a large fraction of XPEs ($\sim$50\%) starts almost
simultaneously with the CME onset. Collimated XPEs are shifted
towards negative values in this figure and their maximum is
distinctly broader. The histogram made for the selected subgroup of
better observed events (gray and black bins in
Fig.~\ref{cme-xpe-cl}) show a similar trend.

A similar pattern can be seen in Fig.~\ref{cme-xpe-ce}. In this
Figure the histogram made for eruptive XPEs is narrower and centered
closer to the zero point than the histogram made for confined XPEs.
The distribution for confined XPEs is much broader, especially in
the plot for the selected subgroup of better observed events (black
bins), which makes an impression of a random occurrence within
almost the whole time window of the CME onset.

In Fig.~\ref{cme-xpe-sr} both histograms made for single and
recurrent XPEs show a similar width, but recurrent XPEs tend to
start earlier (almost simultaneously with the CME) than single ones.
The difference between medians is about 15 minutes for all events as
well as for the selected subgroup of better observed events.

\subsection{CME angular width and velocity} \label{aw+vel}

\begin{figure}
\includegraphics[scale=.90]{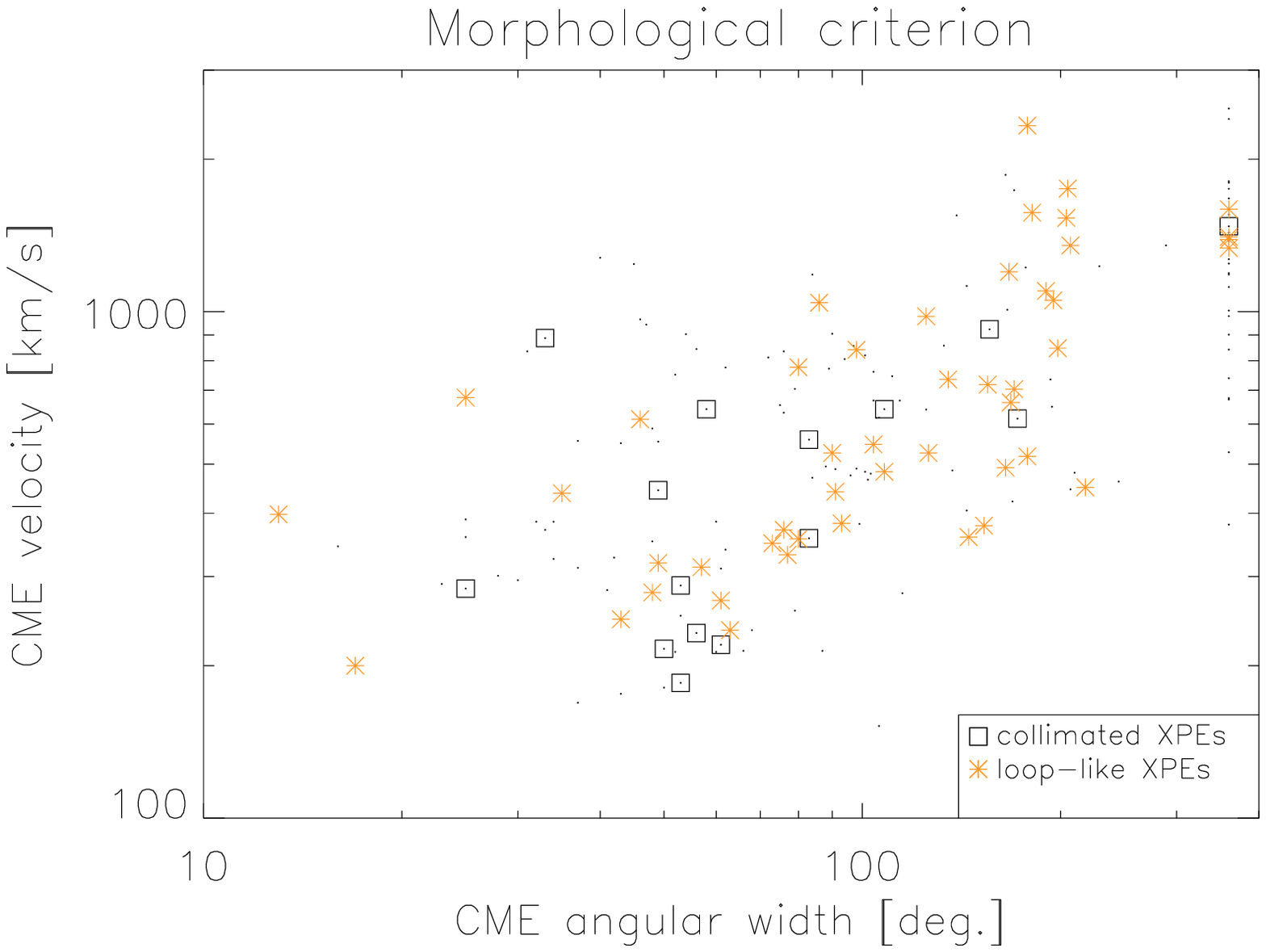}
\caption{Scatter plot of CME angular width versus CME linear
velocity. This plot compares CMEs associated with XPEs classified
according to the morphological criterion. All points are marked with
dots. A subset of well-observed (see text) CMEs are additionally
marked with boxes and stars for collimated and loop-like XPEs of
high quality (A--B), respectively. Values are taken from the {\sl
SOHO} LASCO CME catalog \citep{gop09}.\label{xpecme1}}
\end{figure}

In Fig.~\ref{xpecme1} we present a scatter plot of an angular width
versus a linear velocity for CMEs associated with morphological
subclasses of XPEs, i.e., collimated and loop-like XPEs. All points
are marked with dots. Additionally we emphasized well-observed XPEs
(quality A or B) that occurred close to the solar limb ($|\lambda| >
60^{\circ}$). CMEs associated with well-observed collimated and
loop-like XPEs are marked with boxes and stars, respectively. Both
groups of CMEs are mixed in the plot, however some shifts toward
wider and faster events is seen for CMEs associated with loop-like
XPEs.

\begin{figure}
\includegraphics[scale=.90]{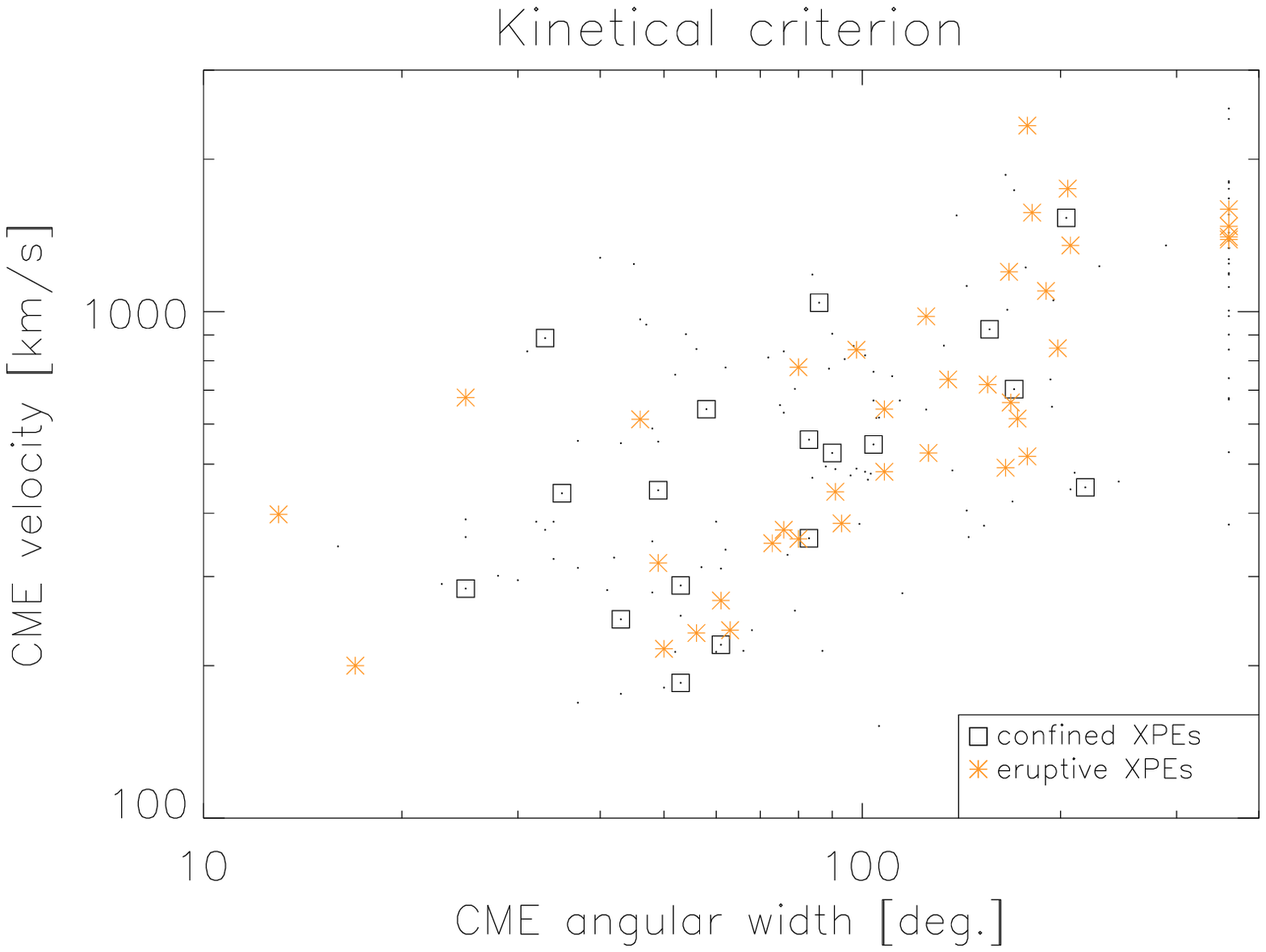}
\caption{Scatter plot of CME angular width versus CME linear
velocity. This plot compares CMEs associated with XPEs classified
according to the kinematical criterion. All points are marked with
dots. A subset of well-observed (see text) CMEs are additionally
marked with boxes and stars for confined and eruptive XPEs of high
quality (A--B), respectively. Values are taken from the {\sl SOHO}
LASCO CME catalog \citep{gop09}.\label{xpecme2}}
\end{figure}

\begin{figure}
\includegraphics[scale=.90]{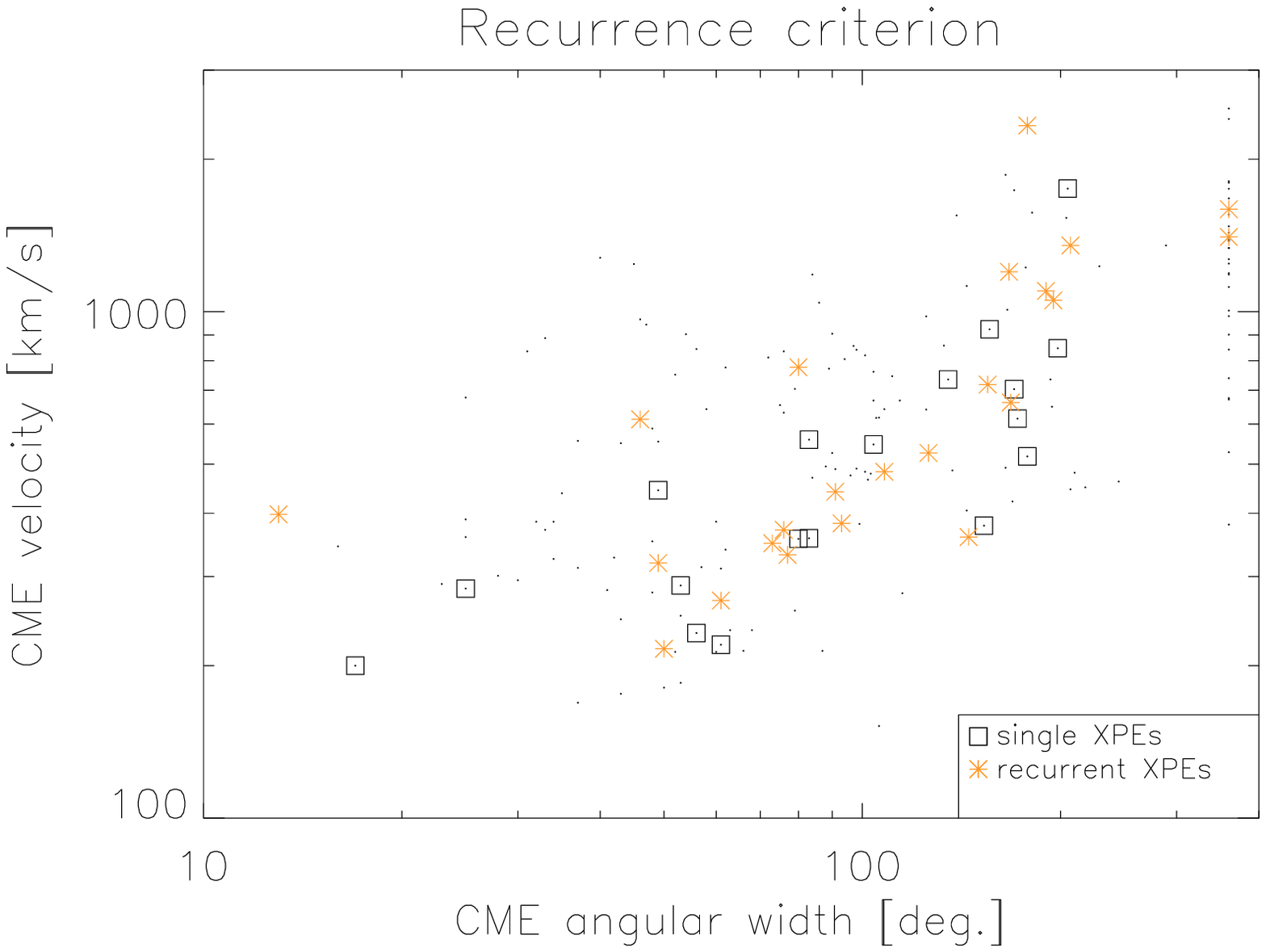}
\caption{Scatter plot of CME angular width versus CME linear
velocity. This plot compares CMEs associated with XPEs classified
according to the recurrence criterion. All points are marked with
dots. A subset of well-observed (see text) CMEs are additionally
marked with boxes and stars for single and recurrent XPEs of high
quality (A--B), respectively. Values are taken from the {\sl SOHO}
LASCO CME catalog \citep{gop09}.\label{xpecme3}}
\end{figure}

Similar scatter plots of an angular width versus a linear velocity
for CMEs associated with kinematical and recurrence subclasses of
XPEs are given in Figs.~\ref{xpecme2}-\ref{xpecme3}. All points are
marked with dots. Again, additionally we emphasized well-observed
XPEs (quality A or B) that occurred close to the solar limb
($|\lambda| > 60^{\circ}$). Moreover, we excluded events for which
the assignment of kinematical subclasses for XPEs was uncertain
(Fig.~\ref{xpecme2}) and events associated with XPEs that were
classified as single in case of partial time-coverage of
observations (Fig.~\ref{xpecme3}). In both figures, CMEs associated
with well-observed XPEs classified as the subclass 1 (confined and
single, respectively) are marked with boxes, whereas CMEs associated
with well-observed XPEs classified as the subclass 2 (eruptive and
recurrent, respectively) are marked with stars. Similarly to
Fig.~\ref{xpecme1}, both groups of CMEs are mixed in the plots and
some shifts toward wider and faster events is seen for CMEs
associated with XPEs of the subclasses 2.

The shifts seen in Figs.~\ref{xpecme1}-\ref{xpecme3} are confirmed
by medians calculated separately for both groups of CMEs for each
classification criterion. As it is seen in Table~\ref{cmes}
(bold-faced columns for well-observed events), the medians for CMEs
associated with XPEs of the subclass 2 are 1.2--2.0 times and
1.2--1.4 times greater than those for CMEs associated with XPEs of
the subclass 1 for CME angular width and linear velocity,
respectively. A higher angular width and velocity mean a more
energetic CME, thus we can conclude that more energetic XPEs are, on
average, associated with more energetic CMEs and less-energetic XPEs
rather prefer less-energetic CMEs.

\begin{figure}
\includegraphics[scale=.90]{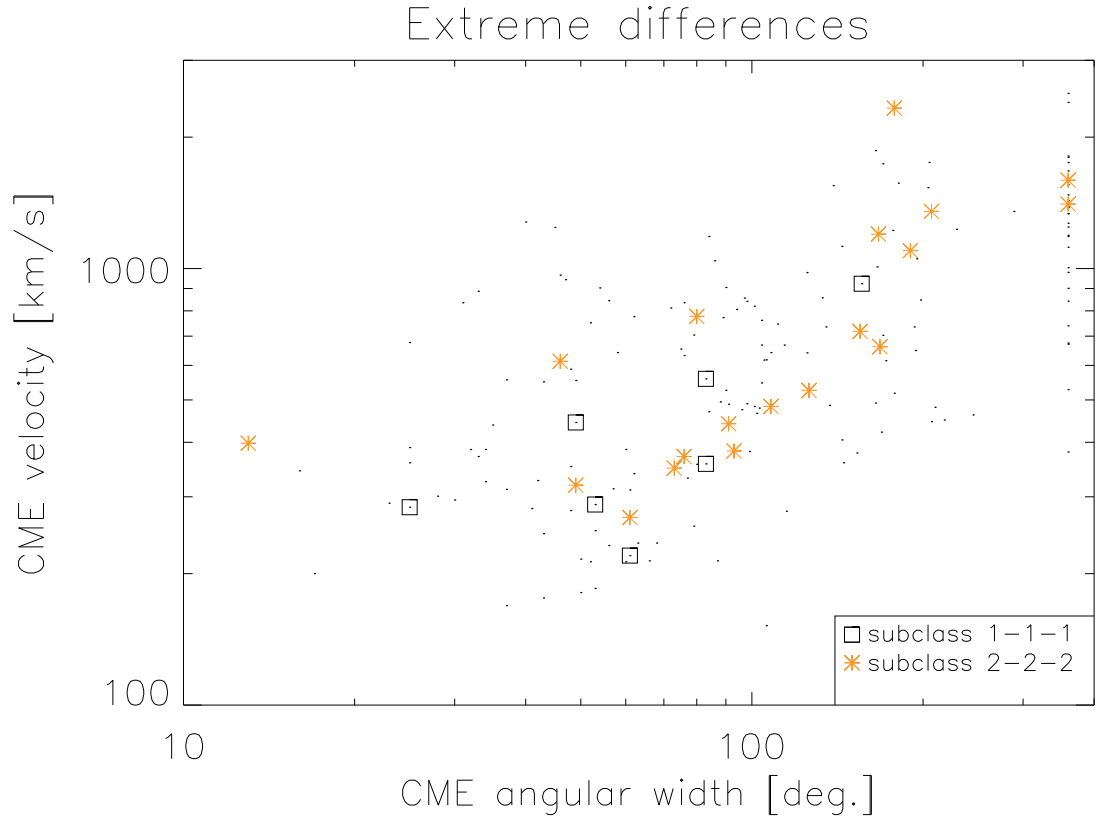}
\caption{Scatter plot of CME angular width versus CME linear
velocity. This plot compares CMEs associated with XPEs classified
according to different criteria that are employed simultaneously.
All points are marked with dots. A subset of well-observed (see
text) CMEs are additionally marked with boxes and stars for
collimated, confined, single (1,1,1) and loop-like, eruptive,
recurrent (2,2,2) XPEs of high quality (A--B), respectively. Values
are taken from the {\sl SOHO} LASCO CME catalog
\citep{gop09}.\label{xpecme4}}
\end{figure}

One can expect that a difference between characteristics describing
associated CMEs should be even higher for two subclasses of XPEs
that we define by applying our three criteria simultaneously.
Indeed, medians in Table~\ref{cmes} for CMEs associated with the
subclass (1,1,1) -- collimated, confined, single XPEs -- and the
subclass (2,2,2) -- loop-like, eruptive, recurrent XPEs -- show
extreme differences (factors 2.1 and 1.7 for angular width and
linear velocity, respectively). The difference between
characteristics describing associated CMEs is also seen in
Fig.~\ref{xpecme4}.

In unbold-faced columns in Table~\ref{cmes} we present medians for
CMEs associated with different subclasses of XPEs that were defined
less strictly, i.e. by including quality C, for $|\lambda| \le
60^{\circ}$ and without excluding any doubtful examples. Ratios of
medians for the subclasses 2 to medians for the subclasses 1 that
were constituted more liberally are often comparable to the more
strictly defined bold-faced values. It is opposite to flares for
which bigger differences between unbold-faced and bold-faced values
are evident (see Table~\ref{flare}). We suggest that the main reason
for this is the condition $|\lambda| > 60^{\circ}$, that we
constituted for specially selected (bold-faced) events. It excludes
the majority of halo CMEs being systematically wider and faster than
ordinary CMEs \citep{mic03}. In other words, more strict selection
criteria undoubtedly limit a scatter of values in two physically
different groups, however it is compensated with the bias introduced
by halo CMEs.

\section{Discussion} \label{disc}

The XPEs collected in the catalogue confirm the strong association
with flares. Starts of XPEs observed since their very beginning fall
usually within the rising phase of associated flares
(Figs.\,\ref{flb-xpeb} and \ref{flbn-xpebn}) and well coincide with
the HXR peaks (Fig.~\ref{hxr-xpe}). It means that symptoms of SXR
plasma motions occur when magnetic energy conversion in flares --
via reconnection -- is most vigorous \citep{ben08}. A small number
of exceptions is connected mainly with complex events in which X-ray
enhancements, recorded by {\sl GOES} and {\sl Yohkoh}/HXT, are
accumulated from at least two different positions on the Sun.

We would like to stress a lower correlation between XPEs and HXR
flares than between XPEs and SXR ones. As approximately one third of
flares associated with XPEs did not show any clear signatures of
non-thermal electrons, we can conclude that some macroscopic motion
of SXR plasma is more obvious characteristics of reconnection than
acceleration of non-thermal electrons. It can be caused by the the
limited sensitivity of the HXT. As we can see in Table~\ref{flare},
the more energetic subclasses 2 in our classification scheme of XPEs
show a stronger correlation with HXR flares. Thus, under the
assumption that the more energetic XPEs are associated with more
energetic flares, we can expect a larger fraction of them to be able
to produce the HXR emission above the threshold of the HXT.

Another proof of close association between processes responsible for
XPEs and flares is similarity between their durations. A comparison
between the medians in Figs.~\ref{flbn-xpebn} and \ref{flmn-xpeen}
shows that an XPE lasts, on average, as long as the rising phase of
an associated flare. Histograms presented for particular subclasses
of XPEs (Figs.~\ref{flbn-xpebn-cl}-\ref{flbn-xpebn-sr} and
\ref{flmn-xpeen-cl}-\ref{flmn-xpeen-sr}) show that the more
energetic subclasses 2 occur earlier and last shorter than the less
energetic subclasses 1. This difference probably reflects some
differences in reconnection processes occurring in both subclasses.
At first sight, this result is in contradiction with
Figs.~\ref{xpefl1}-\ref{xpefl3}, in which XPEs from subclasses 2
seem to prefer flares of longer duration, however we should remember
that in Figs.~\ref{flbn-xpebn-cl}-\ref{flbn-xpebn-sr} and
\ref{flmn-xpeen-cl}-\ref{flmn-xpeen-sr} the time is normalized with
the flare rising-phase duration.

Some interesting hints concerning hierarchy and chronology of
processes occurring in restructuring active regions can be found in
histograms of time differences between the XPE start and the
extrapolated CME onset for particular subclasses of XPEs
(Figs.~\ref{cme-xpe-cl}-\ref{cme-xpe-sr}). The more energetic
subclasses 2 of XPEs show close relationship with their associated
CMEs. They seem to start almost simultaneously, and small deviations
from the zero value are probably caused by unrealistic extrapolation
of the CME onsets. The start of XPEs of the less energetic
subclasses 1 shows a much looser connection with the CME start. Very
often the CME seems to occur first. Keeping in mind that XPEs are
usually caused by magnetic reconnection, we suggest that in the case
of the subclasses 2, the reconnection and loss-of-equilibrium of
magnetic structure, thus a CME development, occur almost
simultaneously. On the other hand, in the case of the subclasses 1,
the reconnection is usually a consequence of destabilization of
magnetic structure, which may occur earlier.

The results summarized in Tables~\ref{flare} and \ref{cmes} strongly
suggest that total amount of energy, converted from the magnetic
field in an active region during its magnetic reconfiguration,
determines characteristics of events including flares, CMEs, and
XPEs, which are thought to be consequences of this common
reconfiguration. Thus, more energetic XPEs are associated with more
energetic flares and CMEs, while less energetic ones -- seem to
occur commonly. This statistically averaged picture does not
exclude, for sure, exceptions in partitioning of magnetic energy.
For example, there are X-class confined flares completely devoid of
any CME \citep{w+z07,che11}. These flares are probably also devoid
of XPEs but this research is beyond the scope of this work.

Our investigation shows that characteristics of flares and CMEs
associated with particular subclasses of XPEs are different. We
found that a scale of differences is higher for flares than for
CMEs. We also found that the recurrence criterion proposed in our
XPE classification scheme does not separate the associated events as
strongly as the morphological and kinematical criteria.

\section{Conclusions}
\label{concl}

In our catalogue we have collected the most extensive database of
XPEs so far. Images from the SXT onboard {\sl Yohkoh} have been
organized into movies in the MPEG format. The events have been
classified on the basis of elementary and uniform criteria. The
catalogue also gives a piece of information concerning the
associated flares and CMEs by using entries to the {\it Yohkoh}
Flare Catalogue (HXT/SXT/SXS/HXS) and the {\it SOHO} LASCO CME
Catalog, respectively. The collected data allow us to study XPEs
more comprehensively as a separate solar activity phenomena and also
as elements of more complex processes occurring in the solar corona.

XPEs constitute a strongly inhomogeneous group of events. Their
appearances include expanding loop structures, moving blobs, rising
columns, and so on. Their strong inhomogeneity is responded by wide
range of values of basic parameters: altitude ($10^8-10^{10}$~cm),
volume ($10^{26}-10^{30}$~cm$^{-3}$), duration ($10^1-10^3$~s),
velocity ($10^0-10^3$~km\,s$^{-1}$), acceleration
($-10^3-10^4$~m\,s$^{-2}$), mass ($10^{12}-10^{15}$~g), energy
($10^{25}-10^{31}$~ergs).

It is difficult to point out a universal mechanism responsible for
all the events presented in the catalogue. There is no doubt that
the majority of XPEs is connected somehow with magnetic
reconnection. However, in many events, the evolution is far from
what may be expected from the canonical CSHKP model, suggesting the
existence of more complex 3D quadrupolar reconnection \citep{nit10}.
We often observe an XPE as a result of magnetic reconnection that
leads to chromospheric evaporation as a hydrodynamic response of
intensified plasma heating or non-thermal electron beams in a flare
magnetic structure.

On the other hand, the close morphological and kinematical
connection of some XPEs with CMEs, together with the similar start
time, suggests a mechanism of loss-of-equilibrium type common for
CMEs. [Indeed, movies illustrating evolution of some XPEs resemble
cartoons presenting the tether release model or the tether straining
model leading to the magnetic breakout model.] Finally, some movies
in the catalogue give an impression that SXR plasma leaks out from
the magnetic structure probably under low-$\beta$-plasma conditions.

For proper interpretation of the data, we need to identify the
mechanism responsible for the observed XPE. An inappropriate choice
of the mechanism can lead to meaningless and erroneous conclusions
regarding processes occurring in the solar corona. In the context of
strong inhomogeneity of XPEs and several possible mechanisms of
their origin, it is not advised to routinely interpret all XPEs in
terms of a single and same mechanism. The similar conclusion were
given by \citet{nit10} who criticized the tendency to employ the
CSHKP model for description all ``Masuda-type'' flares
\citep{mas94}.

If we consider a one particular XPE, it is basically difficult to
decide which mechanism is responsible for its occurrence without the
complete quantitative analysis including plasma diagnostics and a
modeling of magnetic field structure. These conditions were
unreachable in practice for the majority of events in the catalogue.
Therefore, in advent of new observations of XPEs derived by modern
instruments onboard {\sl Hinode}, {\sl Solar-Terrestrial Relations
Observatory} ({\sl STEREO}), and the {\sl Solar Dynamics
Observatory} ({\sl SDO}), we have been trying to give some solutions
that would be correct at least in statistical sense.

We have shown that the subclasses of XPEs separated on the basis of
our simple observational criteria have different levels of
correlation with other solar-activity phenomena. The difference is
also seen if we consider basic parameters describing these flares
and CMEs. However, the association of XPEs with different flares or
CMEs does not mean a specific physical mechanism as far as these
flares or CMEs represent physically different groups. In the
meantime, discussions concerning a difference in observational
characteristics to justify separate physical mechanisms responsible
for flares or CMEs are still open. Are there two different classes
of flares \citep{pal77} or all flares can be explained by only one
mechanism \citep{shi95}? Are there two kinematically different
classes of CMEs \citep{she99} or the division is artificial
\citep{vrs05}?

We have found that more energetic XPEs are better correlated with
flares and CMEs and that more energetic XPEs correlate with more
energetic flares and CMEs. Virtually the effect of observational
conditions works in the same way and we cannot resolve correctly the
influence of the effect on our conclusion.

The most promising way in the investigation of XPEs is to deal them
as an element of a larger ensemble. The usage of observations made
in temporal, spatial, and spectral ranges broader than those needed
for direct monitoring of XPEs allows the better understanding of
processes in which XPEs participate. Recently, a similar picture of
flares as global events was presented by \citet{hud11}. Our
experience is that XPEs are strongly coupled with flare HXR
quasi-periodic oscillations \citep{n+m09}, probably because the
reconnection rate is controlled by plasmoid generation
\citep{nis09}. We also found that XPEs are somehow associated with
progressive spectral hardening in HXRs \citep{tom08}, thus with
Solar Energetic Particles \citep{kip95, gra09}.

In the future we are going to upgrade the XPEs catalogue by adding
entries devoted to associated prominences and radio bursts.
Moreover, the {\sl TRACE} movies will be added, if available. We
also are going to perform the comprehensive analysis of several,
very interesting events from the catalogue that have been omitted by
other {\sl Yohkoh} researchers.

\acknowledgments

{\it Yohkoh} is a project of the Institute for Space and
Astronautical Sciences, Japan, with substantial participation from
other institutions within Japan and with important contributions
from the research groups in the US and the UK under the support of
NASA and SERC. The SXT instrument was jointly developed by the
Lockheed Palo Alto Research Laboratory and the National Astronomical
Observatory of Japan. Collaborators include the University of Tokyo,
Stanford University, the University of California at Berkeley, and
the University of Hawaii. The CME catalog used in this work is
generated and maintained at the CDAW Data Center by NASA and The
Catholic University of America in cooperation with the Naval
Research Laboratory. SOHO is a project of international cooperation
between ESA and NASA. The {\it Yohkoh} Flare Catalogue
(HXT/SXT/SXS/HXS) used in this work was developed by J. Sato, K.
Yoshimura, T. Watanabe, M. Sawa, M. Yoshimori, Y. Matsumoto, S.
Masuda and T. Kosugi. We acknowledge solar data collected and
distributed by U. S. National Geophysical Data Center. We thank Dr.
Masamitsu Ohyama for providing us his list of XPEs from years
1991-1998. We appreciate the valuable remarks of the referee, which
helped us to improve this paper. This work was supported by Polish
Ministry of Science and High Education grant No. N\,N203\,1937\,33.

\clearpage

\clearpage

\clearpage

\clearpage

\clearpage

\clearpage

\clearpage

\clearpage

\clearpage

\clearpage

\clearpage

\clearpage

\clearpage

\clearpage

\clearpage

\clearpage

\clearpage

\clearpage

\clearpage

\clearpage

\clearpage

\clearpage

\clearpage

\clearpage

\clearpage

\clearpage

\begin{deluxetable}{rcccclllcl}
\tabletypesize{\footnotesize} \tablewidth{0pt} \tablecaption{XPEs
presented in the catalogue\label{tb1}} \tablehead{ \colhead{No.} &
\colhead{Date} & \colhead{Time} & \colhead{Class.} & \colhead{Q.} &
\colhead{AR} & \colhead{GOES} & \colhead{Coordinates} &
\colhead{CME} & \colhead{References}} \startdata
001 & 91/10/22 & 06:42.2-07:28.8\tablenotemark{b} & 2,1,2 & B & 6891 & M1.2 & S11\,E85 & \nodata & 19 \\
002 & 91/11/02 & 16:31.2-16:57.3 & 2,2,1 & B & 6891 & M4.8 & S10\,W84 & \nodata & 19 \\
003 & 91/11/17 & 18:32.9-18:41.8\tablenotemark{b} & 1,1,2 & C & 6929 & M1.9 & S12\,E78 & \nodata & 19 \\
004 & 91/12/02 & 04:50.6-05:21.0 & 2,2,1 & A & 6952 & M3.6 & N18\,E92 & \nodata & 18,19,27,28,54\\
005 & 91/12/03 & 16:35.0-17:04.5 & 2,1,1 & B & 6952 & X2.2 & N17\,E72 & \nodata & 19 \\
006 & 91/12/09 & 02:02.7-02:06.5\tablenotemark{b} & 2,1,1 & C & 6966 & M1.4 & S06\,E91 & \nodata & 19 \\
007 & 91/12/10 & 04:03.0-04:10.1\tablenotemark{b} & 1,1,1 & B & 6968 & C9.3 & S14\,E93 & \nodata & 19 \\
008 & 92/01/13 & 17:27.9-17:35.2\tablenotemark{b} & 2,2,2 & C & 6994 & M2.0 & (S15\,W89) & \nodata & 18,19,27,28 \\
009 & 92/01/13 & 19:04.1-19:13.8\tablenotemark{b} & 1,2,1 & C & 7012 & M1.3 & S10\,E95 & \nodata & 19 \\
010 & 92/01/14 & 19:29.0\tablenotemark{a}-19:32.6\tablenotemark{b} & 2,2,1 & B & 7012 & M1.7 & S11\,E89 & \nodata & 19 \\
011 & 92/01/15 & 18:56.1-19:04.5 & 2,2,1 & B & 7012 & M2.0 & (S09\,E72)& \nodata & 19 \\
012 & 92/01/30 & 17:07.6-17:17.6\tablenotemark{b} & 2,1,1 & C & 7042 & M1.6 & S13\,E84 & \nodata & 19 \\
013 & 92/02/06 & 03:17.4-03:36.6\tablenotemark{b} & 2,2,1 & B & 7030 & M7.6 & N05\,W82 & \nodata & 18,19,27,28 \\
014 & 92/02/06 & 20:52.7-21:24.8\tablenotemark{b} & 2,2,1 & B & 7030 & M4.1 & N05\,W94 & \nodata & 19 \\
015 & 92/02/09 & 03:01.0-03:10.7\tablenotemark{b} & 2,1,1 & C & 7035 & M1.2 & S17\,W74 & \nodata & 19 \\
016 & 92/02/17 & 15:41.8-16:25.0 & 1,2,2 & C & 7050 & M1.9 & N16\,W81 & \nodata & 18,19,27,28 \\
017 & 92/02/18 & 18:00.1-18:29.9 & 2,1,1 & C & 7067 & +\tablenotemark{d} & (N05\,E89) & \nodata & 13 \\
018 & 92/02/19 & 14:45.4-15:39.1\tablenotemark{b} & 2,2,1 & B & 7067 & M1.2 & N06\,E94 & \nodata & 19 \\
019 & 92/02/21 & 03:11.9-03:19.2\tablenotemark{b} & 2,1,1 & C & 7070 & M3.2 & (N09\,E80) & \nodata & 19 \\
020 & 92/02/21 & 22:04.6-22:08.0\tablenotemark{b} & 2,1?,1 & C & 7070 & M2.2 & N05\,E65 & \nodata & 19 \\
021 & 92/04/01 & 10:12.8-10:22.5 & 2,2,1 & B & 7123 & M2.3 & (S03\,E89) & \nodata & 18,27,28 \\
022 & 92/06/05 & 18:08.7-19:08.9 & 2,1,1,& C & 7186 & C2.6 & N07\,E28 & \nodata & 13 \\
023 & 92/06/07 & 01:41.0-01:50.0\tablenotemark{b} & 2,2,1 & B & 7186 & M2.7 & N09\,E10 & \nodata & 4 \\
024 & 92/07/20 & 17:18.1\tablenotemark{a}-17:49.6 & 1,1,1 & C & 7222 & -- & (S06\,W88) & \nodata & 13 \\
025 & 92/07/29 & 20:19.7-21:09.8 & 2,2,2 & B & 7236 & -- & (N19\,W88) & \nodata & 13 \\
026 & 92/08/25 & 19:02.6-19:36.4\tablenotemark{b} & 1,1,1 & B & 7260 & C8.7 & N13\,W98 & \nodata & 19 \\
027 & 92/09/09 & 02:06.2-02:18.6 & 1,1,1 & C & 7270 & M3.1 & S10\,W72 & \nodata & 19 \\
028 & 92/09/09 & 17:57.9-18:07.2 & 1,2?,1 & C & 7270 & M1.9 & S11\,W78 & \nodata & 19 \\
029 & 92/10/04 & 22:14.0-22:32.4 & 2,2,2 & B & 7293 & M2.4 & S05\,W90 & \nodata & 18,19,24,27,28,53 \\
030 & 92/10/05 & 09:24.3-09:52.0 & 1,2,1 & B & 7293 & M2.0 & S08\,W90 & \nodata & 10,12,18,19,21,24,53 \\
031 & 92/11/05 & 06:19.0-06:40.7\tablenotemark{c} & \nodata & D & 7323 & M2.0 & S16\,W90 & \nodata & 18,19,27,28 \\
032 & 92/11/05 & 20:30.1-21:08.6\tablenotemark{c} & \nodata & D & 7323 & C8.7 & S17\,W92 & \nodata & 19 \\
033 & 93/02/14 & 12:51.9-12:59.4 & 2,2,1 & B & 7427 & M2.0 & S22\,E78 & \nodata & 19 \\
034 & 93/02/17 & 10:35.4-10:53.5\tablenotemark{b} & 1,2,1 & B & 7420 & M5.8 & S07\,W87 & \nodata & 14,18,19,24,27,28,53 \\
035 & 93/02/21 & 00:31.2-00:45.2\tablenotemark{b} & 2,2,1 & B & 7433 & M1.4 & N13\,E75 & \nodata & 19 \\
036 & 93/03/15 & 20:31.9-21:15.2 & 2,2,1 & B & 7440 & M2.9 & S03\,W93 & \nodata & 4 \\
037 & 93/03/23 & 01:21.0\tablenotemark{a}-01:29.5\tablenotemark{b} & 2,2,1 & B & 7448 & M2.3 & N18\,W78 & \nodata & 4 \\
038 & 93/05/07 & 20:56.6-21:30.6\tablenotemark{b} & 2,2,1 & B & 7500 & M1.6 & N14\,E41 & \nodata & 4 \\
039 & 93/05/14 & 22:00.1-22:10.3 & 2,2,1 & A & 7500 & M4.4 & N19\,W48 & \nodata & 23 \\
040 & 93/06/25 & 03:13.8\tablenotemark{a}-03:40.3 & 2,2,1 & B & 7530 & M5.1 & S09\,E88 & \nodata & 19 \\
041 & 93/06/28 & 01:06.7-01:23.0\tablenotemark{b} & 2,2,2 & B & 7535 & C6.5 & N03\,E69 & \nodata & 6 \\
042 & 93/09/26 & 17:26.2-17:28.3\tablenotemark{b} & 2,1,1 & B & 7590 & C3.4 & N14\,E94 & \nodata & 19 \\
043 & 93/09/27 & 12:07.5-12:17.8 & 2,2,1 & B & 7590 & M1.8 & N08\,E90 & \nodata & 18,19 \\
044 & 93/10/01 & 23:51.3-00:01.3\tablenotemark{b} & 2,1,1 & C & 7592 & C8.5 & S14\,E69 & \nodata & 19 \\
045 & 93/11/11 & 11:15.4-11:31.8 & 1,2,1 & B & 7618 & C9.7 & N10\,E95 & \nodata & 10,14,20 \\
046 & 93/11/13 & 06:38.6-06:48.9 & 1,2?,1 & B & 7618 & M2.1 & N08\,E73 & \nodata & 19 \\
047 & 94/01/05 & 06:49.4-06:59.6\tablenotemark{b} & 2,2?,1 & B & 7647 & M1.0 & S13\,W23 & \nodata & 4 \\
048 & 94/01/16 & 23:09.6-23:22.9\tablenotemark{b} & 2,2,1 & B & 7654 & M6.1 & N05\,E71 & \nodata & 19 \\
049 & 94/01/27 & 03:47.8-03:59.0\tablenotemark{b} & 1,1,1 & B & 7654 & C4.6 & N08\,W68 & \nodata & 4 \\
050 & 94/01/28 & 16:53.2-17:26.5\tablenotemark{b} & 2,2,2 & B & 7654 & M1.8 & N08\,W85 & \nodata & 19 \\
051 & 94/02/27 & 09:02.7-09:18.9\tablenotemark{b} & 2,2,1 & B & 7671 & M2.8 & N08\,W98 & \nodata & 4 \\
052 & 94/08/30 & 08:20.5-08:41.4\tablenotemark{b} & 1,1,1 & B & 7773 & M1.1 & S06\,E82 & \nodata & 19 \\
053 & 96/04/20 & 06:51.7-07:04.4 & 1,2,1 & A & 7956 & B2.9 & N04\,W68 & \nodata & 24,36,53 \\
054 & 96/08/22 & 07:42.5-07:52.1\tablenotemark{b} & 2,2,1 & B & 7986 & C4.5 & S14\,E107 & + & 3,50,52 \\
055 & 97/02/23 & 01:30.2-02:15.2\tablenotemark{b} & 2,2?,1 & B & 8019 & B7.2 & (N31\,E90+) & -- & 2 \\
056 & 97/02/23 & 02:58.2-03:31.0\tablenotemark{b} & 2,1,1 & C & 8019 & B7.2 & (N33\,E81) & + & 4 \\
057 & 97/05/16 & 11:44.0-12:10.7\tablenotemark{b} & 1,2,1 & C & 8038 & -- & (N20\,W73) & + &  4 \\
058 & 97/08/09 & 16:32.9-16:35.3\tablenotemark{b} & 1,2?,1 & C & 8069 & C8.5 & N19\,W85 & + & 16,19 \\
059 & 97/09/17 & 11:38.4-11:49.2 & 1,2,1 & B & 8084 & M1.7 & N21\,W82 & + & 16,19 \\
060 & 97/09/17 & 17:48.4-18:24.6\tablenotemark{b} & 2,2,2 & B & 8084 & M1.0 & N21\,W84 & + & 19 \\
061 & 97/11/06 & 11:50.8-12:06.3 & 2,2,1 & C & 8100 & X9.4 & S18\,W63 & + & 17,48 \\
062 & 97/11/14 & 09:11.0-09:20.8\tablenotemark{b} & 2,1,1 & B & 8108 & C2.5 & N21\,E70 & + & 24,48,53 \\
063 & 97/11/27 & 13:10.4-13:25.2 & 2,2,2 & A & 8113 & X2.6 & N17\,E63 & + & 29,48 \\
064 & 98/03/23 & 02:45.9-03:12.1 & 2,2,2 & B & 8179 & M2.3 & S22\,W99 & + & 16,19,50 \\
065 & 98/03/25 & 13:04.1-13:17.1 & 2,2,1 & C & 8180 & C5.3 & (S37\,W90+) & + & 2 \\
066 & 98/04/20 & 09:43.8-10:12.5\tablenotemark{b} & 2,2?,2 & C & 8194 & M1.4 & S30\,W90 & + & 17 \\
067 & 98/04/23 & 05:29.4-05:46.8 & 2,2,2 & A & 8210 & X1.2 & S18\,E104 & + & 1,10,16,18,19,24,26, \\
    &          &                 &       &   &      &      &           &   & 50,51,52,53 \\
068 & 98/04/24 & 08:47.6-08:55.2 & 2,1,1 & C & 8210 & C8.9 & S20\,E91 & + & 19 \\
069 & 98/04/25 & 14:21.9-14:49.2 & 2,2,2 & B & 8210 & C3.6 & S19\,E73 & + & 2 \\
070 & 98/04/27 & 08:50.0-08:55.3\tablenotemark{b} & 2,2?,1 & B & 8210 & X1.0 & S16\,E50 & + & 34 \\
071 & 98/05/03 & 21:17.4-21:25.5 & 2,2,2 & A & 8210 & M1.4 & S13\,W34 & + & 30 \\
072 & 98/05/06 & 07:54.3-08:11.0 & 2,2,2 & A & 8210 & X2.7 & S11\,W65 & + & 17,24,35,48,53 \\
073 & 98/05/08 & 01:50.4-02:20.1 & 2,2,2 & A & 8210 & M3.1 & (S16\,W90+) & + & 16,18,19 \\
074 & 98/05/08 & 14:21.2-14:49.4 & 2,2,2 & B & 8210 & M1.8 & S17\,W95 & + & 19 \\
075 & 98/05/09 & 02:04.7-02:20.1 & 1,2,1 & C & 8210 & C7.0 & (S15\,W90+) & -- & 16 \\
076 & 98/05/09 & 03:17.7-03:37.3 & 2,2,2 & A & 8210 & M7.7 & S17\,W102 & + & 16,18,19,50 \\
077 & 98/05/10 & 13:25.4-13:33.2 & 2,2?,1 & B & 8220 & M3.9 & S27\,E89 & -- & 19 \\
078 & 98/05/28 & 06:22.5-06:28.2 & 1,1,1 & B & 8226 & C1.2 & N20\,W78 & -- & 19 \\
079 & 98/05/28 & 19:02.4-19:17.6 & 2,2,2 & B & 8226 & C8.7 & (N16\,W87) & + & 16,19 \\
080 & 98/05/29 & 01:02.8-01:12.0 & 1,2,1 & C & 8226 & M6.7 & (N15\,W89) & + & 19 \\
081 & 98/06/11 & 10:00.8-10:18.3\tablenotemark{b} & 2,2,2 & C & 8243 & M1.4 & (N21\,E89) & + & 16,50 \\
082 & 98/06/15 & 07:26.8-07:34.3\tablenotemark{b} & 2,2,1 & C & 8232 & C1.4 & (S16\,W87) & + & 2 \\
083 & 98/07/03 & 01:06.5-01:12.7 & 2,2?,1 & C & 8256 & M1.2 & S26\,W14 & \nodata & 4 \\
084 & 98/08/14 & 08:25.5-08:31.7 & 2,2,1 & B & 8293 & M3.1 & S23\,W74 & \nodata & 19 \\
085 & 98/08/18 & 08:17.5-08:26.5 & 2,2,2 & B & 8307 & X2.8 & N33\,E68 & \nodata & 18,19,48 \\
086 & 98/08/18 & 22:15.1-22:20.7 & 2,2,1 & B & 8307 & X4.9 & N33\,E87 & \nodata & 18,19,31,48\\
087 & 98/09/06 & 02:12.3-02:18.5 & 1,1?,1 & B & 8323 & C2.3 & S22\,W35 & \nodata & 4 \\
088 & 98/09/20 & 02:38.5-03:19.1 & 2,2,2 & B & 8340 & M1.8 & N22\,E62 & \nodata & 4 \\
089 & 98/09/23 & 00:30.2-00:47.6 & 2,2,1 & B & 8344 & C9.3 & S20\,E22 & \nodata & 4 \\
090 & 98/09/23 & 06:45.3-07:05.2 & 2,2,2 & B & 8340 & M7.1 & N18\,E09 & \nodata & 40 \\
091 & 98/09/28 & 16:07.9-16:11.7 & 2,2,1 & C & 8339 & C6.8 & (S15\,W87) & \nodata & 4 \\
092 & 98/09/30 & 13:16.4-13:40.5 & 1,2,2 & B & 8340 & M2.8 & N23\,W81 & \nodata & 4 \\
093 & 98/10/07 & 17:08.6-17:44.8\tablenotemark{b} & 2,2,1 & A & 8355 & M2.3 & S23\,E68 & \nodata &  32 \\
094 & 98/10/20 & 20:34.5-21:16.1 & 2,2,2 & C & 8360 & C7.4 & (S21\,W86) & \nodata & 4 \\
095 & 98/11/03 & 19:14.3-19:36.3 & 2,1?,2 & B & 8375 & M1.0 & N21\,E02 & + & 4 \\
096 & 98/11/05 & 01:03.4-01:12.2 & 2,1,2 & C & 8375 & C7.1 & (N20\,W14) & + & 33 \\
097 & 98/11/06 & 02:42.1-02:51.8\tablenotemark{b} & 1,1,1 & C & 8375 & C4.4 & N19\,W24 & -- & 4 \\
098 & 98/11/06 & 09:09.9-09:16.6 & 1,1?,1 & C & 8375 & C4.2 & N19\,W27 & + & 4 \\
099 & 98/11/07 & 17:52.1-18:01.1 & 1,2?,1 & B & 8375 & C5.3 & N19\,W48 & -- & 4 \\
100 & 98/11/16 & 23:11.4-23:18.7 & 2,2,2 & B & 8385 & C6.8 & (N24\,W83) & \nodata & 4 \\
101 & 98/11/22 & 06:39.6-06:47.2 & 2,2,2 & C & 8384 & X3.7 & S27\,W82 & \nodata & 28,34 \\
102 & 98/11/22 & 16:21.5-16:31.1 & 1,2,2 & B & 8384 & X2.5 & S30\,W89 & \nodata & 28,34 \\
103 & 98/11/24 & 02:19.6-02:28.3 & 2,2,1 & C & 8384 & X1.0 & S30\,W103 & + & 4 \\
104 & 98/11/25 & 14:00.8\tablenotemark{a}-14:07.5 & 2,2,1 & C & 8395 & C6.4 & N18\,E68 & + & 4 \\
105 & 98/11/28 & 05:29.5-05:49.6 & 2,2,2 & B & 8395 & X3.3 & N17\,E32 & + & 34,48 \\
106 & 98/12/18 & 17:17.5-17:27.9\tablenotemark{b} & 2,2,1 & C & 8414 & M8.0 & (N34\,E40) & + & 4 \\
107 & 98/12/23 & 05:53.5\tablenotemark{a}-06:02.9\tablenotemark{b}& 2,1,1 & C & 8414 & M2.3 & (N28\,E70) & \nodata & 4 \\
108 & 98/12/24 & 01:22.2-01:29.1\tablenotemark{b} & 2,1,1 & C & 8421 & C6.2 & N29\,E85 & \nodata & 4 \\
109 & 98/12/25 & 06:18.4-06:35.0\tablenotemark{b} & 2,1?,2 & B & 8421 & M1.2 & N30\,E66 & \nodata & 4 \\
110 & 98/12/28 & 05:16.2-05:31.9 & 2,1,2 & B & 8416 & M1.4 & N28\,E26 & \nodata & 4 \\
111 & 99/01/03 & 08:15.4-09:03.8\tablenotemark{b} & 2,2,2 & A & 8420 & C8.2 & N16\,W70 & \nodata & 4 \\
112 & 99/01/06 & 23:59.7-00:05.4\tablenotemark{b} & 1,2,1 & B & 8422 & C8.0 & (S23\,W90) & \nodata & 4 \\
113 & 99/01/14 & 10:06.9-10:17.0 & 2,2,1 & B & 8439 & M3.0 & N18\,E64 & \nodata & 4 \\
114 & 99/02/10 & 23:09.1-23:25.3\tablenotemark{b} & 2,2?,1 & C & 8457 & C3.7 & (N18\,E57) & + & 4 \\
115 & 99/02/12 & 15:23.7-15:32.1\tablenotemark{b} & 2,2?,1 & B & 8456 & C5.7 & N12\,E27 & -- & 4 \\
116 & 99/02/16 & 21:19.7-21:27.8& 2,2,2 & C & 8462 & C5.3 & N19\,W12 & \nodata & 4 \\
117 & 99/02/21 & 13:16.0-13:41.5\tablenotemark{b} & 1,1?,2 & B & 8462 & M1.3 & N24\,W81 & -- & 4 \\
118 & 99/03/07 & 04:05.5-04:11.9 & 2,2,1 & C & 8477 & C2.9 & N24\,W03 & + & 4 \\
119 & 99/04/03 & 23:02.0-23:10.5 & 1,1,1 & B & 8508 & M4.3 & N29\,E81 & + & 9 \\
120 & 99/05/07 & 04:31.5-04:45.0 & 2,2,2 & A & 8535 & M3.2 & N20\,E87 & + & 9,10 \\
121 & 99/05/08 & 14:33.0-14:52.1 & 1,1,1 & C & 8526 & M4.6 & N23\,W75 & + & 9 \\
122 & 99/05/09 & 12:34.7-13:05.5 & 2,1?,2 & C & 8526 & M1.0 & (N22\,W84) & + & 9 \\
123 & 99/05/09 & 18:04.2-18:17.3 & 1,2,1 & B & 8526 & M7.6 & (N22\,W86) & + & 8,9 \\
124 & 99/05/11 & 21:46.9-22:02.7 & 2,1,1 & C & 8542 & C4.7 & S19\,E79 & + & 9 \\
125 & 99/05/16 & 17:26.2-17:32.9\tablenotemark{b} & 1,1,1 & B & 8534 & M1.1 & S17\,W76 & -- & 9 \\
126 & 99/05/17 & 03:43.2-03:51.6 & 1,1,2 & B & 8534 & C3.9 & (S16\,W82) & -- & 8,9,24,37,53 \\
127 & 99/05/17 & 04:57.5-05:01.0\tablenotemark{b} & 1,1,1 & C & 8534 & M2.3 & (S15\,W82) & -- & 9 \\
128 & 99/05/29 & 03:08.1-03:14.2\tablenotemark{b} & 2,2,1 & B & 8557 & M1.6 & (S21\,E65) & + & 9 \\
129 & 99/06/11 & 11:25.4-11:38.4\tablenotemark{b} & 2,2,1 & B & 8585 & C8.8 & (N46\,E90) & + & 9 \\
130 & 99/06/19 & 22:31.8-22:42.8\tablenotemark{b} & 2,1,1 & C & 8592 & C4.0 & (N25\,E86) & + & 9 \\
131 & 99/06/23 & 00:41.2-00:51.0\tablenotemark{b} & 2,1,1 & B & 8583 & C7.9 & S12\,W78 & + & 9 \\
132 & 99/06/30 & 03:11.0-03:18.3 & 1,1,1 & C & 8613 & C4.5 & (N19\,E90) & + & 9 \\
133 & 99/06/30 & 04:36.2-05:00.8\tablenotemark{b} & 2,2,2 & B & 8611 & M2.1 & S26\,E28 & + & 4 \\
134 & 99/07/07 & 06:20.7-06:30.3 & 1,2?,1 & C & 8611 & C1.2 & S26\,W76 & + & 9 \\
135 & 99/07/07 & 09:19.1\tablenotemark{a}-09:24.4\tablenotemark{b} & 1,1,1 & C & 8611 & C1.2 & (S26\,W76) & + & 9 \\
136 & 99/07/09 & 22:38.5-22:48.2 & 1,1,1 & C & 8629 & C6.9 & N22\,W69 & -- & 9 \\
137 & 99/07/12 & 23:31.9-23:38.0 & 1,1,1 & B & 8626 & C2.1 & (S19\,W76) & -- & 9 \\
138 & 99/07/16 & 15:50.8-15:55.3\tablenotemark{b} & 1,1,1 & C & 8635 & M3.1 & N43\,W71 & + & 9 \\
139 & 99/07/23 & 04:45.9-05:03.8\tablenotemark{b} & 2,1,1 & B & 8645 & C9.3 & S23\,E97 & -- & 9 \\
140 & 99/07/23 & 15:56.1-16:26.3\tablenotemark{b} & 1,1,1 & C & 8645 & M1.0 & S26\,W92 & -- & 9 \\
141 & 99/07/23 & 22:58.1-23:10.9 & 2,1,1 & C & 8644 & M1.2 & S26\,E87 & -- & 9 \\
142 & 99/07/24 & 03:31.7-04:06.6 & 2,1,1 & C & 8636 & M1.7 & S29\,E87 & -- & 9 \\
143 & 99/07/24 & 08:00.4-08:16.7\tablenotemark{b} & 2,1?,1 & B & 8636 & M3.3 & S28\,E78 & + & 9 \\
144 & 99/07/25 & 13:08.8\tablenotemark{a}-13:14.7\tablenotemark{b} & 2,2,1 & B & 8639 & M2.4 & N38\,W81 & + & 8,9,10,24,38,53 \\
145 & 99/08/04 & 05:58.5-06:11.5 & 2,1,1 & C & 8647 & M6.0 & S16\,W64 & + & 9 \\
146 & 99/08/04 & 21:54.2-22:03.2\tablenotemark{b} & 1,1,1 & C & 8647 & C3.3 & (S18\,W80) & -- & 9 \\
147 & 99/08/06 & 10:05.3-10:18.0\tablenotemark{b} & 2,1,1 & C & 8647 & C7.1 & S28\,W81 & + & 9 \\
148 & 99/08/07 & 20:52.0-21:10.9 & 2,2,2 & B & 8645 & M1.7 & S29\,W104 & + & 9 \\
149 & 99/08/10 & 06:02.4-06:13.2\tablenotemark{b} & 1,1,1 & C & 8656 & C4.8 & N15\,W74 & -- & 9 \\
150 & 99/08/13 & 15:07.9-15:16.2\tablenotemark{b} & 1,1,1 & B & 8668 & C5.3 & (N17\,E90) & -- & 9 \\
151 & 99/08/14 & 12:06.0-12:16.9\tablenotemark{b} & 1,1,1 & C & 8668 & C7.2 & N23\,E72 & -- & 9 \\
152 & 99/08/20 & 12:40.4-13:36.6\tablenotemark{b} & 2,2?,2 & C & 8674 & M1.8 & S28\,E76 & + & 9 \\
153 & 99/08/28 & 17:54.1-18:04.6\tablenotemark{b} & 2,2,2 & B & 8674 & X1.1 & S26\,W14 & + & 48 \\
154 & 99/09/08 & 12:13.6-12:14.5\tablenotemark{b} & 2,2?,1 & C & 8690 & M1.4 & N12\,E53 & + & 25 \\
155 & 99/09/21 & 10:16.0-10:34.6\tablenotemark{b} & 2,2,1 & B & 8692 & C6.4 & (S25\,W84) & + & 9 \\
156 & 99/10/25 & 19:37.8-19:44.1 & 2,2,1 & C & 8737 & C4.8 & S19\,W69 & + & 9 \\
157 & 99/10/26 & 21:12.0-21:27.6\tablenotemark{b} & 2,2,2 & B & 8737 & M3.7 & (S16\,W86) & + & 9,10,11,18,28,39 \\
158 & 99/10/27 & 09:11.2-09:13.9\tablenotemark{b} & 2,2,1 & B & 8737 & M1.0 & S12\,W88 & -- & 9 \\
159 & 99/10/27 & 13:35.2-14:04.9\tablenotemark{b} & 2,2,1 & B & 8737 & M1.8 & S15\,W90 & + & 9 \\
160 & 99/10/27 & 15:21.1-15:43.6\tablenotemark{b} & 2,1,1 & B & 8737 & M1.4 & S14\,W92 & + & 9 \\
161 & 99/11/05 & 18:22.5-18:48.6 & 2,2?,1 & B & 8759 & M3.0 & N12\,E96 & + & 9,10 \\
162 & 99/11/06 & 06:32.5-06:35.1 & 1,1,1 & C & 8759 & C4.6 & (N12\,E86) & -- & 8,9 \\
163 & 99/11/06 & 17:04.4-17:14.4 & 1,1,1 & B & 8759 & C5.0 & (N12\,E86) & + & 9 \\
164 & 99/11/08 & 06:06.0-06:16.0\tablenotemark{c} & \nodata & D & 8749 & C5.9 & S18\,W82 & + & 9 \\
165 & 99/11/13 & 02:22.7-03:09.4 & 2,2?,1 & C & 8763 & M1.3 & S15\,E44 & + & 9 \\
166 & 99/11/27 & 12:12.1-12:23.3 & 1,1,1 & C & 8771 & X1.4 & S15\,W68 & + & 9,48,49 \\
167 & 99/12/18 & 01:26.8-01:39.1 & 2,2,1 & C & 8806 & C9.4 & N19\,E82 & -- & 9 \\
168 & 00/01/12 & 20:50.0-20:56.9 & 2,2,2 & B & 8829 & M1.1 & N13\,E67 & + & 9 \\
169 & 00/01/18 & 09:36.0-09:58.8\tablenotemark{b} & 1,2,1 & A & 8827 & M1.2 & S15\,W106 & + & 9,10,24,53 \\
170 & 00/01/18 & 17:12.1-17:18.1 & 2,2?,1 & C & 8831 & M3.9 & S19\,E11 & + & 4 \\
171 & 00/01/22 & 17:58.9\tablenotemark{a}-18:08.8\tablenotemark{b} & 2,1,2 & C & 8831 & M1.0 & S23\,W50 & + & 4 \\
172 & 00/02/04 & 09:14.6-09:57.4 & 2,1,1 & C & 8858 & M3.0 & N25\,E71 & -- & 28 \\
173 & 00/02/04 & 19:28.3-19:43.5\tablenotemark{b} & 1,2,2 & B & 8858 & C7.0 & N25\,E71 & + & 9 \\
174 & 00/02/05 & 19:33.9-19:43.9 & 2,1?,2 & B & 8858 & X1.2 & N26\,E52 & + & 48 \\
175 & 00/02/22 & 20:26.2-21:24.0 & 2,1,1 & C & 8882 & C9.2 & (S18\,E90+) & + & 9 \\
176 & 00/02/26 & 23:38.5-23:43.8\tablenotemark{b} & 2,2,2 & B & 8889 & M1.0 & N29\,E50 & -- & 4 \\
177 & 00/03/02 & 08:23.7-08:28.7 & 2,2,1 & B & 8882 & X1.1 & (S18\,W55) & + & 48,49 \\
178 & 00/03/02 & 13:12.5-13:20.5 & 1,2,1 & C & 8882 & C5.5 & S19\,W60 & + & 4 \\
179 & 00/03/03 & 02:11.6-02:14.6\tablenotemark{b} & 2,2,1 & B & 8882 & M3.8 & S15\,W60 & + & 4 \\
180 & 00/03/06 & 10:47.9-10:50.0\tablenotemark{b} & 1,1,1 & B & 8889 & C4.5 & (N20\,W75) & -- & 9 \\
181 & 00/03/06 & 16:20.0-16:28.4\tablenotemark{b} & 2,1,1 & B & 8889 & C3.9 & N20\,W78 & -- & 9 \\
182 & 00/03/18 & 20:50.7-20:57.3\tablenotemark{b} & 2,2,1 & B & 8906 & M2.1 & (S15\,W68) & + & 4 \\
183 & 00/03/18 & 21:53.5-21:59.0 & 1,1,1 & B & 8906 & C4.2 & S19\,W67 & + & 4 \\
184 & 00/03/18 & 23:16.9-23:27.4\tablenotemark{b}& 2,1,1 & B & 8902 & +\tablenotemark{d} & (S18\,E90+) & + & 9 \\
185 & 00/03/27 & 13:59.1-14:25.5\tablenotemark{c}& \nodata & D & 8926 & C8.4 & S09\,W69 & -- & 9 \\
186 & 00/03/27 & 15:32.8\tablenotemark{a}-15:41.9 & 2,1,1 & C & 8926 & C8.9 & S10\,W69 & -- & 24 \\
187 & 00/03/31 & 06:27.8-06:50.8 & 2,2?,1 & B & 8936 & M1.2 & S15\,E55 & + & 9 \\
188 & 00/04/06 & 02:22.4-02:31.9 & 2,2,1 & B & 8948 & M1.8 & S15\,E53 & + & 4 \\
189 & 00/04/08 & 02:37.5-02:47.5 & 2,2,1 & C & 8948 & M2.0 & S15\,E26 & + & 4 \\
190 & 00/05/02 & 14:53.2-15:03.0 & 1,1,2 & C & 8971 & M2.8 & N22\,W68 & + & 9 \\
191 & 00/05/05 & 15:18.3-15:47.5 & 2,2,2 & B & 8977 & M1.5 & S18\,W110 & + & 9 \\
192 & 00/05/12 & 08:40.9-08:50.7\tablenotemark{c} & \nodata & D & 8998 & C8.1 & S14\,E90 & -- & 9,41 \\
193 & 00/05/12 & 21:32.8-21:39.8\tablenotemark{c} & \nodata & D & 8998 & C9.8 & S15\,E81 & -- & 9 \\
194 & 00/05/13 & 01:46.6-02:10.1 & 2,2,1 & C & 9002 & M1.1 & N22\,E109 & -- & 9 \\
195 & 00/05/13 & 23:12.4-23:21.8\tablenotemark{b} & 2,1,1 & C & 9002 & C7.4 & N22\,E96 & + & 9 \\
196 & 00/05/15 & 08:26.8-08:53.1 & 2,2,1 & C & 9002 & M4.4 & (N23\,E87) & + & 9 \\
197 & 00/05/18 & 15:55.6-16:05.0 & 2,2?,1 & C & 9002 & M2.7 & N23\,E30 & + & 4 \\
198 & 00/05/23 & 17:50.1-18:00.1 & 1,1,1 & C & 8996 & C4.3 & S22\,W80 & + & 9 \\
199 & 00/05/24 & 00:09.2-00:19.1 & 1,1,1 & C & 9017 & C6.8 & (S12\,E90+) & -- & 9 \\
200 & 00/05/24 & 03:14.3-04:02.8 & 1,1,1 & C & 9017 & C7.0 & S12\,E93 & -- & 9 \\
201 & 00/05/24 & 11:43.3-12:13.0 & 1,1,1 & C & 9017 & M1.1 & (S12\,E90) & -- & 9 \\
202 & 00/05/24 & 21:05.6-21:51.1\tablenotemark{b} & 1,1,1 & C & 9017 & C9.7 & (S12\,E90) & -- & 9 \\
203 & 00/05/26 & 11:33.1-11.37.6\tablenotemark{b} & 2,2?,1 & B & 8998 & C6.1 & S13\,W90 & + & 9 \\
204 & 00/05/28 & 10:23.6-10:34.6 & 1,1,1 & C & 9002 & C8.6 & (N21\,W88) & + & 9 \\
205 & 00/06/02 & 03:44.6-03:50.8\tablenotemark{b} & 2,1,1 & B & 9026 & M1.2 & (N22\,E77) & + & 9 \\
206 & 00/06/02 & 20:32.7-21:10.3 & 2,1,1 & B & 9026 & M3.1 & (N21\,E67) & + & 9 \\
207 & 00/06/06 & 15:35.6-15:45.6 & 1,1,1 & C & 9026 & X2.3 & (N20\,E18) & + & 4 \\
208 & 00/06/07 & 15:42.9-15:52.9\tablenotemark{b} & 2,2?,1 & B & 9026 & X1.2 & N23\,E03 & + & 4 \\
209 & 00/06/12 & 12:24.4-12:57.5 &  2,1,1 & C & 9042 & C6.4 & N16\,E87 & -- & 9 \\
210 & 00/06/17 & 02:29.6-02:45.3 & 1,1,2 & C & 9033 & M3.5 & N22\,W72 & + & 9 \\
211 & 00/06/18 & 02:04.4-02:14.4\tablenotemark{b} & 1,2,1 & C & 9033 & X1.0 & N23\,W85 & + & 9 \\
212 & 00/06/19 & 04:18.1-04:23.0 & 1,1,1 & C & 9033 & C1.7 & (N21\,W89) & + & 9 \\
213 & 00/06/21 & 09:24.5-09:34.1 & 1,1,1 & C & 9042 & M1.3 & N24\,W42 & + & 4 \\
214 & 00/06/23 & 14:22.7-14:37.9 & 2,2,1 & B & 9042 & M3.0 & N26\,W72 & + & 9 \\
215 & 00/06/23 & 22:14.9-22:24.8 & 1,1?,2 & C & 9042 & C7.7 & N22\,W74 & + & 9 \\
216 & 00/06/28 & 12:17.4-12:19.5\tablenotemark{b} & 2,1?,1 & B & 9064 & C6.1 & (S22\,W73) & -- & 9 \\
217 & 00/06/29 & 10:28.6-10:38.7 & 1,1,1 & B & 9064 & C4.8 & (S17\,W82) & -- & 9 \\
218 & 00/07/01 & 12:37.6-12:43.1\tablenotemark{b} & 1,1,1 & B & 9054 & C6.0 & (N15\,W75) & -- & 9 \\
219 & 00/07/01 & 23:21.4-23:31.3 & 2,1,1 & C & 9054 & M1.5 & N07\,W88 & -- & 9 \\
220 & 00/07/10 & 21:14.7-21:22.7 & 2,2,2 & B & 9077 & M5.7 & N18\,E49 & + & 4 \\
221 & 00/07/12 & 03:32.6-03:37.5 & 1,1,1 & C & 9078 & M1.4 & (S11\,E87) & + & 9 \\
222 & 00/07/12 & 10:26.4-10:38.7 & 2,2,1 & C & 9077 & X1.9 & N17\,E27 & + & 48 \\
223 & 00/07/12 & 16:24.9-16:32.7 & 2,1,1 & C & 9070 & M1.0 & N17\,W68 & + & 9 \\
224 & 00/07/12 & 19:46.1-20:18.1 & 2,2,1 & C & 9077 & M1.5 & (N17\,W72) & + & 9 \\
225 & 00/07/13 & 02:04.5-02:14.2\tablenotemark{b} & 1,1,1 & B & 9069 & C6.1 & S16\,W71 & -- & 9 \\
226 & 00/07/13 & 06:59.8-07:09.3 & 2,1,1 & C & 9070 & C6.8 & (N16\,W80) & -- & 9 \\
227 & 00/07/13 & 18:14.1-19:04.8\tablenotemark{c} & \nodata & D & 9070 & M1.3 & (N16\,W83) & -- & 9 \\
228 & 00/07/14 & 00:42.0-00:44.6 & 1,1,1 & B & 9070 & M1.5 & (N17\,W86) & -- & 9 \\
229 & 00/07/14 & 10:20.0-10:27.2 & 2,2,2 & B & 9077 & X5.7 & N22\,W07 & + & 48,49 \\
230 & 00/07/14 & 13:45.1-13:55.0 & 2,2,1 & B & 9077 & M3.7 & N20\,W08 & -- & 4 \\
231 & 00/07/16 & 01:24.5-01:32.5 & 2,1,1 & B & 9087 & C6.3 & S11\,E53 & -- & 4 \\
232 & 00/07/18 & 05:02.9-05:10.5\tablenotemark{b} & 2,2,2 & B & 9077 & M1.8 & N17\,W58 & \nodata & 4 \\
233 & 00/07/20 & 09:46.1-10:18.7 & 2,2?,2 & C & 9087 & M3.6 & S12\,W08 & \nodata & 24 \\
234 & 00/07/21 & 14:33.3-14:40.6 & 2,1,2 & C & 9090 & M5.5 & (N10\,E12) & -- & 4 \\
235 & 00/07/22 & 11:21.1-11:27.1 & 2,2,1 & B & 9085 & M3.7 & N14\,W56 & + & 4 \\
236 & 00/07/25 & 02:46.8-02:52.4 & 2,2,1 & B & 9097 & M8.0 & N06\,W08 & + & 42 \\
237 & 00/07/26 & 03:54.6-04:13.6 & 1,1,1 & C & 9087 & C8.9 & S13\,W89 & -- & 9 \\
238 & 00/07/27 & 04:08.1-04:17.9 & 2,1,1 & B & 9090 & M2.4 & N10\,W72 & -- & 9 \\
239 & 00/07/27 & 16:46.0-16:52.6\tablenotemark{b} & 2,1,1 & B & 9087 & M1.5 & S09\,W105 & -- & 9 \\
240 & 00/08/02 & 08:17.8-08:27.1 & 1,1,1 & B & 9114 & C7.9 & (N10\,E85) & + & 9 \\
241 & 00/08/12 & 09:49.6-10:00.5 & 2,2,2 & B & 9119 & M1.1 & (S15\,W84) & + & 9 \\
242 & 00/08/14 & 05:01.1-05:10.8\tablenotemark{b} & 1,1,1 & C & 9126 & C8.1 & N06\,W75 & + & 9 \\
243 & 00/08/24 & 09:01.8-09:11.6\tablenotemark{b} & 2,1,1 & C & ? & C6.2 & (N27\,W88) & -- & 9 \\
244 & 00/08/25 & 14:27.7-14:32.5 & 2,2,1 & B & 9143 & M1.4 & S15\,E67 & + & 7,9,25,43 \\
245 & 00/09/07 & 20:38.0-20:47.6 & 1,1,2 & B & 9151 & C7.2 & N06\,W47 & + & 44 \\
246 & 00/09/15 & 14:41.2-14:43.9\tablenotemark{b} & 1,2?,1 & C & 9165 & M2.0 & N12\,E07 & + & 24 \\
247 & 00/09/22 & 23:46.9-23:56.5\tablenotemark{b} & 1,1,1 & C & 9165 & C8.5 & N14\,W94 & -- & 9 \\
248 & 00/09/30 & 17:53.4-18:21.1 & 2,1,1 & B & ? & M1.0 & (S29\,E85) & + & 9 \\
249 & 00/09/30 & 23:17.3-23:29.0 & 2,2,1 & B & 9169 & X1.2 & N07\,W90+ & -- & 9,18,28,45,48 \\
250 & 00/10/01 & 07:01.7-07:14.2\tablenotemark{b} & 2,1,1 & C & 9169 & M5.0 & N08\,W97 & -- & 9 \\
251 & 00/10/01 & 13:59.2-14:06.5\tablenotemark{b} & 2,1,1 & A & 9169 & M2.2 & N09\,W101 & -- & 9 \\
252 & 00/10/16 & 05:35.1-06:03.9 & 2,2,1 & A & 9182 & C7.0 & N04\,W107 & + & 9,24,53 \\
253 & 00/10/16 & 06:42.5\tablenotemark{a}-06:46.9\tablenotemark{b} & 2,2?,1 & B & 9182 & M2.5 & N03\,W108 & + & 9 \\
254 & 00/10/26 & 05:10.2-05:18.2\tablenotemark{b} & 1,1,1 & B & 9199 & C3.7 & (N14\,W79) & + & 8,9 \\
255 & 00/10/26 & 06:11.2-06:20.7\tablenotemark{b} & 2,1,1 & C & 9209 & C4.3 & S25\,E71 & -- & 9 \\
256 & 00/10/26 & 11:43.9-11:46.9\tablenotemark{c} & \nodata & D & 9203 & C6.9 & N17\,W77 & + & 9 \\
257 & 00/10/26 & 15:59.1-16:35.9 & 2,1?,2 & B & 9209 & C8.5 & S20\,E64 & + & 9 \\
258 & 00/10/29 & 01:32.8-01:57.4 & 2,2,2 & B & 9209 & M4.4 & S25\,E35 & \nodata & 4 \\
259 & 00/11/01 & 12:04.5\tablenotemark{a}-12:21.1 & 1,1,1 & C & 9212 & -- & (N10\,E29) & -- & 24 \\
260 & 00/11/08 & 23:18.2-23:44.0 & 2,1,1 & C & 9213 & M7.4 & (N10\,W77) & + & 9,17 \\
261 & 00/11/09 & 03:03.2-03:12.9\tablenotemark{b} & 2,1?,1 & C & 9213 & M1.2 & (N08\,W86) & -- & 9 \\
262 & 00/11/09 & 06:28.9-06:38.6\tablenotemark{b} & 1,1,1 & C & 9213 & M1.2 & (N12\,W88) & -- & 9 \\
263 & 00/11/14 & 16:26.2-16:35.9 & 2,2,2 & B & 9232 & M1.0 & (N14\,E90) & + & 9 \\
264 & 00/11/18 & 16:48.3-16:57.9\tablenotemark{b} & 2,1,1 & C & 9227 & C5.9 & (S16\,W79) & -- & 9 \\
265 & 00/11/24 & 14:55.0-15:20.5 & 2,2,2 & A & 9236 & X2.3 & N22\,W07 & + & 46,48 \\
266 & 00/11/24 & 21:50.2-22:03.5 & 2,2,2 & B & 9236 & X1.8 & N21\,W14 & + & 15,48 \\
267 & 00/11/25 & 01:00.1-01:10.4 & 2,2,2 & A & 9240 & M8.2 & N07\,E50 & + & 4 \\
268 & 00/11/25 & 09:10.0-09:21.0 & 2,2,2 & B & 9236 & M3.5 & N18\,W24 & + & 4 \\
269 & 00/11/25 & 18:36.5-18:46.4 & 2,1,1 & C & 9236 & X1.9 & N20\,W23 & + & 48 \\
270 & 00/11/26 & 16:38.3-16:47.4 & 2,1,2 & B & 9236 & X4.0 & N18\,W38 & + & 48 \\
271 & 00/11/30 & 08:58.0-09:23.5\tablenotemark{b} & 2,2,1 & A & 9236 & M1.0 & (N17\,W88) & \nodata & 4 \\
272 & 00/12/06 & 22:22.0-22:47.5\tablenotemark{b} & 2,2?,1 & C & 9246 & M1.6 & S10\,W66 & + & 9 \\
273 & 00/12/18 & 08:58.6-09:06.2\tablenotemark{b} & 1,1,1 & C & 9276 & C5.2 & S14\,W76 & -- & 9 \\
274 & 00/12/19 & 10:23.5-11:05.1 & 1,1,1 & C & 9276 & +\tablenotemark{d} & (S12\,W85) & + & 9 \\
275 & 00/12/23 & 04:57.9-05:07.5\tablenotemark{b} & 1,1,1 & C & 9283 & C3.4 & S12\,E76 & -- & 9 \\
276 & 00/12/23 & 08:20.2-08:23.1\tablenotemark{b} & 1,1,1 & C & 9283 & C3.7 & S13\,E75 & -- & 9 \\
277 & 00/12/24 & 01:06.7-01:15.0 & 2,1,1 & C & 9283 & C7.0 & S15\,E66 & + & 9 \\
278 & 00/12/26 & 23:47.9-23:56.1 & 1,1,1 & B & 9289 & C4.0 & (S06\,E87) & -- & 9 \\
279 & 00/12/27 & 15:52.0-16:05.3 & 2,1,1 & C & 9289 & M4.3 & S07\,E73 & -- & 9 \\
280 & 01/01/04 & 08:56.9-09:06.0\tablenotemark{b} & 2,1,1 & B & 9302 & C4.5 & N25\,E87 & -- & 9 \\
281 & 01/01/05 & 18:29.5-18:39.3\tablenotemark{b} & 2,1,1 & A & 9302 & C5.8 & N20\,E72 & -- & 9 \\
282 & 01/01/08 & 10:36.5-10:53.3\tablenotemark{b} & 2,1?,1 & B & 9302 & C5.1 & (N21\,E36) & + & 24 \\
283 & 01/01/09 & 08:50.0-08:59.5\tablenotemark{b} & 2,1,1 & B & 9297 & C5.1 & (N21\,W87) & -- & 9 \\
284 & 01/01/19 & 17:09.2-18:08.4 & 1,1,1 & C & 9313 & M1.0 & S07\,E61 & + & 9 \\
285 & 01/01/24 & 14:42.7-15:11.4\tablenotemark{c} & \nodata & D & 9311 & M1.0 & N06\,W77 & -- & 9 \\
286 & 01/01/25 & 07:10.8-07:18.5\tablenotemark{b} & 2,2,1 & C & 9325 & C7.4 & (N10\,E74) & + & 9 \\
287 & 01/01/30 & 00:57.1-01:03.2\tablenotemark{b} & 2,2,1 & C & 9313 & C3.7 & (S08\,W89) & + & 9 \\
288 & 01/02/19 & 20:55.4-21:05.1\tablenotemark{b} & 2,1,1 & B & 9360 & C5.4 & (S09\,E89) & + & 9 \\
289 & 01/03/06 & 10:10.1-10:19.7 & 2,2,1 & B & 9364 & C6.7 & (S11\,W89) & + & 9 \\
290 & 01/03/07 & 14:49.2-14:59.1\tablenotemark{b} & 2,1,1 & C & 9371 & C5.8 & N23\,W75 & + & 9 \\
291 & 01/03/09 & 19:54.4-20:00.5 & 1,1,1 & C & 9371 & C4.2 & (N22\,W84) & + & 9 \\
292 & 01/03/10 & 17:14.3-17:23.8\tablenotemark{b} & 1,1,1 & C & 9365 & C5.9 & (S12\,W90) & + & 9 \\
293 & 01/03/11 & 08:47.4-08:57.2\tablenotemark{b} & 2,1,1 & B & 9376 & C5.0 & S14\,E87 & + & 9,24 \\
294 & 01/03/21 & 02:36.6-02:44.4 & 2,2?,2 & B & 9373 & M1.8 & S05\,W65 & + & 9 \\
295 & 01/03/21 & 11:25.2-11:31.5 & 1,1,1 & B & 9373 & C9.8 & S05\,W70 & + & 9 \\
296 & 01/03/24 & 01:36.5-01:40.5 & 2,2?,1 & C & 9376 & M1.2 & S14\,W82 & + & 9 \\
297 & 01/03/24 & 23:32.6-00:08.8 & 1,1,1 & C & 9393 & M1.1 & N19\,E60 & -- & 9 \\
298 & 01/03/28 & 10:56.0-11:05.7\tablenotemark{b} & 2,1,1 & B & 9397 & M4.3 & S09\,E29 & + & 24 \\
299 & 01/03/29 & 01:37.4-01:58.0\tablenotemark{b} & 1,1,1 & C & 9393 & -- & (N15\,W09) & -- & 24 \\
300 & 91/03/29 & 12:47.1-13:02.3 & 2,1,1 & C & 9393 & C7.6 & (N16\,W11) & -- & 24 \\
301 & 01/04/01 & 10:59.8-11:03.5\tablenotemark{b} & 1,2,1 & B & 9415 & M5.5 & S21\,E107 & + & 4 \\
302 & 01/04/03 & 03:37.6-03:49.3 & 1,1?,1 & C & 9415 & X1.2 & S21\,E83 & + & 4 \\
303 & 01/04/05 & 08:30.7-08:49.5 & 2,2,1 & B & 9393 & M8.4 & N14\,W103 & + & 24,53 \\
304 & 01/04/06 & 19:12.5-19:23.5\tablenotemark{b} & 2,2,2 & B & 9415 & X5.6 & S21\,E31 & + & 48 \\
305 & 01/04/09 & 15:23.6-15:24.8\tablenotemark{b} & 2,1,1 & C & 9415 & M7.9 & S21\,W04 & + & 24 \\
306 & 01/04/10 & 05:19.2-05:33.9 & 1,1,1 & C & 9415 & X2.3 & S23\,W09 & + & 4 \\
307 & 01/04/12 & 10:12.8-10:22.0 & 2,2,1 & B & 9415 & X2.0 & S19\,W43 & + & 47,48 \\
308 & 01/04/15 & 13:29.7-13:53.0 & 2,2,2 & B & 9415 & Y1.4 & S20\,W85 & + & 17,18,48 \\
309 & 01/04/17 & 12:15.4-12:20.7\tablenotemark{b} & 2,2?,1 & C & 9415 & C1.9 & (S21\,W90) & -- & 4 \\
310 & 01/04/20 & 21:30.7-21:35.5\tablenotemark{b} & 2,2,2 & B & 9433 & C8.0 & (N17\,E45) & \nodata & 4 \\
311 & 01/04/23 & 10:15.7-10:23.8\tablenotemark{b} & 2,1?,1 & B & 9433 & C9.1 & N17\,E12 & -- & 4 \\
312 & 01/04/24 & 05:36.1-05:52.0 & 2,1,2 & B & 9433 & M2.1 & N18\,E01 & -- & 4 \\
313 & 01/04/25 & 13:43.5-13:56.7 & 1,1,2 & B & 9433 & M2.7 & N18\,W09 & + & 4 \\
314 & 01/04/26 & 13:10.7\tablenotemark{a}-13:18.2\tablenotemark{b} & 1,1,1 & C & 9433 & M7.8 & N17\,W31 & + & 4 \\
315 & 01/05/08 & 00:41.8-01:09.9\tablenotemark{b} & 2,1,1 & B & 9445 & C9.9 & N23\,W43 & -- & 4 \\
316 & 01/05/12 & 23:26.8-23:39.7 & 2,2,2 & B & 9455 & M3.0 & S17\,E00 & -- & 4 \\
317 & 01/05/15 & 02:57.4-03:04.4 & 2,2,1 & B & 9455 & M1.0 & S17\,W29 & + & 4 \\
318 & 01/06/19 & 23:19.8-23:29.5\tablenotemark{b} & 2,1,1 & B & 9501 & C4.2 & S10\,W37 & + & 4 \\
319 & 01/06/22 & 20:25.5-20:35.1 & 2,2,1 & B & 9511 & C5.5 & N09\,E28 & -- & 4 \\
320 & 01/07/16 & 03:17.0-03:23.3\tablenotemark{b} & 2,2?,1 & C & 9539 & M1.2 & S18\,W20 & + & 4 \\
321 & 01/07/23 & 06:22.7-06:29.5\tablenotemark{b} & 2,1,1 & B & 9545 & C5.0 & N10\,W65 & \nodata & 4 \\
322 & 01/07/30 & 20:40.6-20:47.7 & 2,2,1 & B & 9562 & C6.0 & (N05\,E79) & \nodata & 4 \\
323 & 01/07/31 & 04:02.5-04:09.4 & 1,1,1 & B & 9562 & C6.0 & (N05\,E76) & \nodata & 4 \\
324 & 01/08/08 & 07:09.4-07:17.7 & 1,1,1 & B & 9557 & C3.9 & (S18\,W86) & + & 4 \\
325 & 01/08/09 & 18:27.7-18:37.4 & 2,2?,2 & B & 9570 & C7.8 & S17\,E19 & -- & 4 \\
326 & 01/08/11 & 01:19.3-01:28.0 & 1,1,1 & B & 9563 & C5.2 & (N20\,W83) & + & 4 \\
327 & 01/08/25 & 16:25.8\tablenotemark{a}-16:34.6 & 1,1,2 & B & 9591 & X5.3 & S17\,E34 & + & 5 \\
328 & 01/08/26 & 13:18.9\tablenotemark{a}-13:58.9\tablenotemark{b} & 2,1,1 & C & ? & M1.3 & (N16\,E89) & + & 4 \\
329 & 01/08/31 & 10:38.1-10:47.8\tablenotemark{b} & 2,2,1 & B & 9601 & M1.6 & N15\,E37 & + & 4 \\
330 & 01/09/02 & 13:44.8-14:04.5\tablenotemark{b} & 2,1,2 & B & 9591 & M3.0 & S20\,W53 & -- & 24,53 \\
331 & 01/09/03 & 18:19.1-18:34.0 & 2,2,2 & B & 9608 & M2.5 & S22\,E96 & + & 4 \\
332 & 01/09/07 & 15:28.9-15:33.8\tablenotemark{b} & 1,1,1 & C & 9601 & M1.2 & N19\,W65 & + & 4 \\
333 & 01/09/08 & 16:43.6-16:49.7 & 2,1,1 & B & 9608 & C5.1 & S23\,E32 & -- & 4 \\
334 & 01/09/12 & 21:44.0-21:48.6\tablenotemark{b} & 1,2,1 & C & 9606 & C9.6 & (S17\,W63) & + & 4 \\
335 & 01/09/13 & 19:50.7\tablenotemark{a}-19:54.2 & 1,2?,1 & B & 9606 & C5.8 & (S18\,W79) & -- & 4 \\
336 & 01/09/14 & 21:44.6-21:45.9\tablenotemark{b}& 2,2,1 & B & 9616 & M3.7 & (S14\,E39) & + & 4 \\
337 & 01/09/17 & 08:21.0-08:29.8 & 2,2,2 & B & 9616 & M1.5 & S14\,E04 & + & 4 \\
338 & 01/09/17 & 21:04.7-21:10.2 & 2,2,1 & C & 9616 & M1.0 & S11\,W06 & -- & 4 \\
339 & 01/09/20 & 18:15.3-18:18.7 & 2,2,2 & B & 9631 & M1.5 & N09\,W11 & + & 4 \\
340 & 01/09/21 & 04:52.8-05:01.5 & 2,2,1 & B & 9620 & C4.1 & N10\,E12 & -- & 4 \\
341 & 01/09/22 & 18:09.9-18:15.6\tablenotemark{b} & 2,2?,1 & C & 9633 & C5.4 & (N14\,E88) & -- & 4 \\
342 & 01/09/24 & 09:50.1-10:13.7\tablenotemark{b} & 2,1,1, & C & 9632 & X2.6 & S18\,E27 & + & 17 \\
343 & 01/09/30 & 11:34.1-11:41.0\tablenotemark{b} & 1,1,1 & B & 9628 & M1.0 & S20\,W75 & -- & 28 \\
344 & 01/10/01 & 04:45.8-04:50.8\tablenotemark{b} & 1,1,2 & B & -- & C4.7 & (N19\,W89) & -- & 28 \\
345 & 01/10/01 & 04:57.9-05:17.9\tablenotemark{b} & 2,2,2 & B & 9628 & M9.1 & (S21\,W84) & + & 28 \\
346 & 01/10/02 & 17:11.5-17:50.5 & 2,1,2 & A & 9628 & C4.7 & (S21\,W85) & -- & 4 \\
347 & 01/10/03 & 06:42.5-06:47.4 & 1,2,2 & B & 9636 & C6.1 & N19\,W46 & + & 4 \\
348 & 01/10/09 & 07:36.8-07:46.7 & 2,2,2 & B & 9645 & C7.0 & (S19\,W86) & + & 4 \\
349 & 01/10/19 & 16:25.2-16:35.9 & 2,2,2 & B & 9661 & X1.6 & N15\,W29 & + & 4 \\
350 & 01/10/20 & 21:11.2-21:17.6 & 2,2,1 & B & 9674 & C4.6 & S09\,E24 & -- & 4 \\
351 & 01/10/22 & 00:38.0-00:43.4 & 1,1,1 & C & 9658 & M1.0 & S16\,W86 & + & 4 \\
352 & 01/10/22 & 14:40.6-14:58.5 & 2,2,2 & B & 9672 & M6.7 & S21\,E18 & + & 4 \\
353 & 01/10/22 & 17:52.1-18:01.8 & 2,2,2 & C & 9672 & X1.2 & S18\,E16 & + & 4 \\
354 & 01/10/25 & 22:52.6-23:00.9 & 1,2,2 & B & 9678 & C6.1 & (N08\,E24) & -- & 4 \\
355 & 01/10/29 & 08:14.6-08:21.7 & 1,1,1 & C & 9672 & M1.0 & S18\,W82 & + & 4 \\
356 & 01/11/01 & 14:03.6-14:29.6 & 2,2?,2 & B & 9687 & M1.7 & S19\,E77 & + & 2 \\
357 & 01/11/04 & 16:05.3-16:19.8 & 2,2,2 & B & 9684 & X1.0 & N06\,W18 & + & 17 \\
358 & 01/11/06 & 03:01.3\tablenotemark{a}-03:03.6\tablenotemark{b} & 2,1,1 & B & 9687 & M2.0 & S19\,E10 & -- & 4 \\
359 & 01/11/08 & 15:12.9-15:45.1 & 2,1?,1 & B & 9690 & M4.2 & S17\,E36 & \nodata & 4 \\
360 & 01/11/09 & 18:32.3-18:42.1 & 1,1,1 & C & 9687 & M1.9 & S21\,W42 & \nodata & 4 \\
361 & 01/11/17 & 05:17.3-05:49.3 & 2,2,1 & B & 9704 & M2.8 & S13\,E42 & + & 4 \\
362 & 01/11/28 & 15:41.4-15:44.4\tablenotemark{b} & 1,2,1 & C & 9715 & C7.7 & (N05\,E17) & + & 4 \\
363 & 01/11/28 & 15:51.9-16:03.1 & 1,1,1 & C & 9715 & C2.1 & (N05\,E17) & + & 4 \\
364 & 01/11/28 & 16:14.9-16:22.3\tablenotemark{b} & 1,1,1 & C & 9715 & M6.9 & N04\,E16 & + & 4 \\
365 & 01/11/29 & 01:45.7-01:49.0\tablenotemark{b} & 2,2,1 & C & 9715 & M1.1 & N04\,E12 & -- & 4 \\
366 & 01/11/30 & 01:03.7-01:08.8 & 2,1,2 & B & 9718 & M3.5 & S06\,E57 & -- & 4 \\
367 & 01/12/02 & 21:38.0-22:25.5 & 2,2,2 & B & 9714 & M2.0 & (S09\,W88) & + & 4 \\
368 & 01/12/10 & 22:46.1-22:55.9 & 2,1,2 & B & 9733 & C7.0 & N10\,E52 & -- & 4 \\
\enddata
\tablenotetext{a}{earlier event start} \tablenotetext{b}{later event
end} \tablenotetext{c}{no identification, time interval of available
SXT observations} \tablenotetext{d}{{\sl GOES} class disturbed by a
flare in another active region}
%% You can append references to a table using the \tablerefs command.
\tablerefs{(1) Alexander, Metcalf \& Nitta 2002; (2)
Bak-Ste$\acute{\rm{s}}$licka 2010, private communication; (3)
Chertok 2000; (4) Chmielewska 2010; (5) Falewicz, Tomczak \&
Siarkowski 2002; (6) Hori 1999; (7) Khan et al. 2002; (8) Kim et al.
2004; (9) Kim et al. 2005a; (10) Kim et al. 2005b; (11) Kim et al.
2009; (12) Kliem, Karlicky \& Benz 2000; (13) Klimchuk et al. 1993;
(14) Kundu et al. 2001; (15) Nishizuka et al. 2010; (16) Nitta \&
Akiyama 1999; (17) Nitta, Cliver \& Tylka 2003; (18) Nitta, Freeland
\& Liu; (19) Ohyama 2009, private communication; (20) Ohyama \&
Shibata 1997; (21) Ohyama \& Shibata 1998; (22) Ohyama \& Shibata
2000; (23) Ohyama \& Shibata 2008; (24) Ronowicz 2007; (25)
Saint-Hilaire \& Benz 2003; (26) Shanmugaraju et al. 2003; (27)
Shibata et al. 1995; (28) Shimizu et al. 2008; (29) SXT SN
1997/11/28 (Nitta); (30) SXT SN 1998/05/08 (McKenzie \& Hudson);
(31) SXT SN 1998/08/22 (Alexander); (32) SXT SN 1998/10/09
(McKenzie); (33) SXT SN 1998/11/06 (Hudson); (34) SXT SN 1998/11/27
(Hudson); (35) SXT SN 1998/12/25 (Hudson \& Akiyama); (36) SXT SN
1999/04/04 (Akiyama); (37) SXT SN 1999/05/21 (McKenzie); (38) SXT SN
1999/09/11 (Nitta); (39) SXT SN 1999/10/29 (McKenzie \& Fletcher);
(40) SXT SN 2000/04/18 (Hudson); (41) SXT SN 2000/05/12 (Hudson);
(42) SXT SN 2000/07/28 (Hudson); (43) SXT SN 2000/09/01 (Fletcher \&
Hudson); (44) SXT SN 2000/09/08 (Hudson); (45) SXT SN 2000/10/06
(Handy); (46) SXT SN 2000/12/22 (Nitta); (47) SXT SN 2001/04/13
(Hudson); (48) SXT SN 2001/08/17 (Nitta); (49) SXT SN 2002/10/04
(Nitta); (50) Tomczak 2003; (51) Tomczak 2004; (52) Tomczak 2005;
(53) Tomczak \& Ronowicz 2007; (54) Tsuneta 1997}
\end{deluxetable}

\clearpage

\begin{deluxetable}{ll}
\tablewidth{0pt} \tablecaption{Quality of the observed XPEs
\label{tb2}} \tablehead{ \colhead{Quality} & \colhead{Quantity} }
\startdata A (excellent) & 20/368 (5.4\%) \\
B (good) & 190/368 (51.6\%) \\
C (poor) & 149/368 (40.5\%) \\
D (problematic) & 9/368 (2.5\%) \\
\enddata
\end{deluxetable}

\clearpage

\begin{deluxetable}{ll}
\tablewidth{0pt} \tablecaption{Location of the observed XPEs
\label{tb3}} \tablehead{ \colhead{Heliographic} & \colhead{Quantity}
\\ \colhead{longitude ($|\lambda|$)} & \colhead{} }
\startdata $< 60^{\circ}$ & 106/368 (28.8\%) \\
60$^{\circ}$-90$^{\circ}$ & 218/368 (59.2\%) \\
$>90^{\circ}$ & 44/368 (12.0\%) \\
\enddata
\end{deluxetable}

\clearpage

\begin{deluxetable}{ll}
\tablewidth{0pt} \tablecaption{Time coverage of the observed XPEs
\label{tb4}} \tablehead{ \colhead{}  & \colhead{Quantity} }
\startdata Full & 206/359 (57.4\%) \\
Partial & 153/359 (42.6\%) \\
\enddata
\end{deluxetable}

\clearpage

\begin{deluxetable}{llrr}
\tablewidth{0pt} \tablecaption{Population of particular subclasses
of XPEs \label{tb5}}
\tablehead{ \colhead{} & \colhead{} & \colhead{} & \colhead{Quantity} \\
\cline{3-4}
\colhead{Subclass} & \colhead{Description} & \colhead{total} & \colhead{special selection} \\
\colhead{} & \colhead{} & \colhead{(T)} & \colhead{(SS)} }
\startdata
1,1,1 & collimated, confined, single & 74 & 14 \\
1,1,2 & collimated, confined, recurrent & 10 & 5 \\
1,2,1 & collimated, eruptive, single & 24 & 6 \\
1,2,2 & collimated, eruptive, recurrent & 6 & 5 \\
2,1,1 & loop-like, confined, single & 72 & 7 \\
2,1,2 & loop-like, confined, recurrent & 15 & 7 \\
2,2,1 & loop-like, eruptive, single & 94 & 30 \\
2,2,2 & loop-like, eruptive, recurrent & 64 & 52 \\
\cline{2-4}
 & total & 359 & 126 \\
\enddata
\end{deluxetable}

\clearpage

\begin{deluxetable}{lcccccc}
\tablewidth{0pt} \tablecaption{Properties of flares associated with
particular subclasses of XPEs \label{flare}} \tablehead{ \colhead{}
 & \colhead{Number} & \colhead{Number} & \colhead{Flare}
 & \colhead{Flare} & \colhead{Flare} & \colhead{Flare} \\
 \colhead{XPE} & \colhead{of HXR} & \colhead{of HXR} & \colhead{class} & \colhead{class} & \colhead{duration}
 & \colhead{duration} \\
 \colhead{subclass} & \colhead{events} & \colhead{events} & \colhead{(median)} & \colhead{(median)} & \colhead{(median)}
 & \colhead{(median)} \\
 \colhead{} & \colhead{} & \colhead{} &
 \colhead{[W\,m$^{-2}$]} & \colhead{[W\,m$^{-2}$]} & \colhead{[min.]} &
 \colhead{[min.]} \\
 \cline{2-7}
 \colhead{} & \colhead{tot.} & \colhead{ss} & \colhead{tot.}
 & \colhead{ss} & \colhead{tot.} & \colhead{ss} }
\startdata \multicolumn{7}{c}{Morphological criterion:} \\
 1 (collimated) & 67/114& {\bf 24/42} & C8.7 & {\bf C6.1} & 70 & {\bf 51} \\
 2 (loop-like) & 168/245 & {\bf 102/136} & M1.6 & {\bf M1.8} & 120 & {\bf 110} \\
 2 to 1 ratio & & & 1.8 & {\bf 3.0} & 1.7 & {\bf 2.2} \\
 \multicolumn{7}{c}{Kinematical criterion:} \\
 1 (confined) & 97/171 & {\bf 32/56} & C9.3 & {\bf C6.1} & 75 & {\bf 45} \\
 2 (eruptive) & 138/188 & {\bf 80/98} &  M1.8 & {\bf M2.3} & 125 & {\bf 120} \\
 2 to 1 ratio & & & 1.9 & {\bf 3.8} & 1.7 & {\bf 2.7} \\
 \multicolumn{7}{c}{Recurrence criterion:} \\
 1 (single) & 165/264 & {\bf 43/57} & M1.1 & {\bf M1.4} & 90 & {\bf 75} \\
 2 (recurrent) & 70/95 & {\bf 53/69} &  M1.9 & {\bf M2.1} & 135 & {\bf 155} \\
 2 to 1 ratio & & & 1.7 & {\bf 1.5} & 1.5 & {\bf 2.1} \\
 \multicolumn{7}{c}{Extreme differences:} \\
 (1,1,1) & 42/74 & {\bf 8/14} & C7.9 & {\bf C5.2} & 45 & {\bf 42} \\
 (2,2,2) & 50/64 & {\bf 38/44} & M2.1 & {\bf M3.1} & 160 & {\bf 155} \\
 (2,2,2) to & & & 2.7 & {\bf 6.0} & 3.6 & {\bf 3.7} \\
 (1,1,1) ratio & & & & & & \\
\enddata
\end{deluxetable}

\clearpage

\begin{deluxetable}{lcccccc}
\tablewidth{0pt} \tablecaption{Properties of CMEs associated with
particular subclasses of XPEs \label{cmes}} \tablehead{ \colhead{}
 & \colhead{} & \colhead{} & \colhead{CME} & \colhead{CME} & \colhead{CME}
 & \colhead{CME} \\
 \colhead{XPE} & \colhead{Number} & \colhead{Number} &
 \colhead{angular} & \colhead{angular} & \colhead{velocity} & \colhead{velocity} \\
 \colhead{subclass} & \colhead{of events} & \colhead{of events} &
 \colhead{width} & \colhead{width} & \colhead{(median)} & \colhead{(median)} \\
 \colhead{} & \colhead{} & \colhead{} &
 \colhead{(median)} & \colhead{(median)} & \colhead{[km\,s$^{-1}$]} &
 \colhead{[km\,s$^{-1}$]} \\
 \cline{2-7}
 \colhead{} & \colhead{tot.} & \colhead{ss} & \colhead{tot.}
 & \colhead{ss} & \colhead{tot.} & \colhead{ss} }
\startdata \multicolumn{7}{c}{Morphological criterion:} \\
 1 (collimated) & 54/90 & {\bf 13/26} & 89$^{\circ}$ & {\bf 61$^{\circ}$} & 522 & {\bf 444} \\
 2 (loop-like) & 128/185 & {\bf 49/67} & 113$^{\circ}$ & {\bf 125$^{\circ}$} & 649 & {\bf 547} \\
 2 to 1 ratio & & & 1.3 & {\bf 2.0} & 1.2 & {\bf 1.2} \\
 \multicolumn{7}{c}{Kinematical criterion:} \\
 1 (confined) & 75/142 & {\bf 17/39} & 90$^{\circ}$ & {\bf 83$^{\circ}$} & 522 & {\bf 450} \\
 2 (eruptive) & 107/133 & {\bf 38/43} &  125$^{\circ}$ & {\bf 126$^{\circ}$} & 629 & {\bf 642} \\
 2 to 1 ratio & & & 1.4 & {\bf 1.5} & 1.2 & {\bf 1.4} \\
 \multicolumn{7}{c}{Recurrence criterion:} \\
 1 (single) & 125/200 & {\bf 18/27} &  93$^{\circ}$ & {\bf 104$^{\circ}$} & 528 & {\bf 518} \\
 2 (recurrent) & 57/75 & {\bf 24/30} &  155$^{\circ}$ & {\bf 126$^{\circ}$} & 662 & {\bf 602} \\
 2 to 1 ratio & & & 1.7 & {\bf 1.2} & 1.3 & {\bf 1.2} \\
 \multicolumn{7}{c}{Extreme differences:} \\
 (1,1,1) & 33/63 & {\bf 7/12} & 83$^{\circ}$ & {\bf 61$^{\circ}$} & 500 & {\bf 357} \\
 (2,2,2) & 42/52 & {\bf 20/23} & 168$^{\circ}$ & {\bf 126$^{\circ}$} & 718 & {\bf 613} \\
 (2,2,2) to (1,1,1) & & & 2.0 & {\bf 2.1} & 1.4 & {\bf 1.7} \\
 ratio & & & & & & \\
\enddata
\end{deluxetable}

\end{document}